# A universal strategy for decoupling stiffness and extensibility of polymer networks

Baiqiang Huang[1], Shifeng Nian[1], Li-Heng Cai[1,2,3,]*

**Affiliations:**

[1]Soft Biomatter Laboratory, Department of Materials Science and Engineering, University of Virginia, Charlottesville, VA 22904, USA

[2]Department of Chemical Engineering, University of Virginia, Charlottesville, VA 22904, USA

[3]Department of Biomedical Engineering, University of Virginia, Charlottesville, VA 22904, USA

*Corresponding author. Email: liheng.cai@virginia.edu

**One-Sentence Summary:** Foldable bottlebrush polymer network strands release stored length upon elongation to decouple the inherent stiffness-extensibility trade-off of unentangled single-network elastomers.

**Supplementary Videos:**

https://figshare.com/s/2561a90f2426c5abba06

**Abstract:** Since the invention of polymer networks in the 19th century (e.g., crosslinked natural rubber by Goodyear), it has been a dogma that stiffer networks are less stretchable, a trade-off inherent to the molecular nature of polymer network strands. Here, we report a *universal* strategy for decoupling the stiffness and extensibility of single-network elastomers. Instead of using linear polymers as network strands, we use foldable bottlebrush polymers, which feature a collapsed backbone grafted with many linear side chains. Upon elongation, the collapsed backbone unfolds to release stored length, enabling remarkable extensibility. By contrast, the network elastic modulus is inversely proportional to the network strand mass and is determined by the side chains. We validate this concept by creating a series of unentangled single-network elastomers with nearly constant Young's modulus (30 kPa) while increasing tensile breaking strain by 40-fold, from 20% to 800%. We show that this strategy applies to networks of different polymer species and topologies. Our discovery opens an avenue for developing polymer networks with extraordinary mechanical properties.

**Main Text**

Stiffness and extensibility are two fundamental mechanical properties of polymer networks. Although these properties seem distinct, they share a common microscopic origin. For an unentangled single-network elastomer, the basic component of all kinds of polymer networks, the stiffness (Young's modulus $E$) is about the thermal energy $k_BT$ per volume $V$ of a network strand, $E \approx 3k_BT/V$.[1] By contrast, the extensibility $\varepsilon_{max}$, or tensile strain at break, increases with the network strand size (**Fig. 1a**). Thus, the network stiffness and extensibility are correlated: $E \propto (\varepsilon_{max})^{-\alpha}$, where $\alpha=2$ for a flexible linear network strand, $\alpha>2$ if the network strand is pre-strained[2,3], or $\alpha<2$ if the network strand is a semiflexible brush-like polymer[4] (**Fig. 1b**; **Supplementary Note 1**, eqs. S4, S18, and S28). Nevertheless, $\alpha$ must be positive for single-network elastomers.

A widely accepted strategy to mitigate the stiffness-extensibility trade-off is incorporating a weak structure within a strong network; examples include clusters of nanoparticles in filled rubber[5,6], reversible bonds in dual-crosslinked polymer networks[7–10], and brittle networks in interpenetrating-network hydrogels[11–13] and elastomers[14]. When subjected to deformation, the weak structure undergoes fracture to prevent localized, amplified stress near network defects[15] or along network strands, avoiding premature failure of the strong network[16]. An alternative strategy to prevent premature network facture is introducing mobile crosslinkers such as entanglements[17–19] and slide-rings[20], which move along network strands to redistribute stress throughout the polymer network. These two strategies, however, do not change the nature of network strands and cannot break the inherent stiffness-extensibility trade-off of single-network elastomers. An emerging strategy to extend a network strand beyond its nominal stretching limit is through mechanochemistry, where mechano-sensitive monomers release stored length upon force-triggered cycloreversion, which converts a cyclic compound to its acyclic constituents[21,22]. However, this process is irreversible and often results in impaired network mechanical properties. Nevertheless, it represents a fundamental challenge to decouple the stiffness and extensibility of unentangled single-network elastomers.

We seek to develop a strategy to split the inherent stiffness-extensibility trade-off of unentangled single-network elastomers. Instead of using linear polymers as network strands, we propose to use our recently discovered hybrid bottlebrush polymers, which consist of many linear side chains randomly separated by small spacer monomers[23]. The design criteria require that the side chains have a relatively high molecular weight (MW) and a low glass transition temperature ($T_g$); by contrast, the spacer monomer is low MW and highly incompatible with the side chains (**Fig. 1c, i**). Reminiscent of oil droplets in water, the spacer monomers are prone to aggregate to minimize interfacial free energy. However, because of chain connectivity, the spacer monomers cannot form spherical droplets; instead, they collapse into a cylindrical core with its surface densely grafted with side chains (**Fig. 1c, ii**). Yet the folded bottlebrush polymer remains elastic at room temperature (RT) because of its low $T_g$ side chains. Upon elongation, the collapsed backbone unfolds to release the stored length, enabling remarkable network extensibility (**Fig. 1c, iii**). By contrast, the network stiffness, or the MW of this so-called foldable bottlebrush polymer (fBB), is not much affected by the backbone but is determined by the side chains. Thus, we hypothesize that using fBB as network strands enables independent control over polymer stiffness and extensibility.

To test this hypothesis, we design a fBB polymer using linear poly(dimethyl siloxane) (PDMS) as the side chain and benzyl methacrylate (BnMA) as the spacer monomer (**Fig. 2a**). Poly(benzyl methacrylate) (PBnMA) and PDMS are highly incompatible with the Flory-Huggins interaction parameter $\chi \approx 0.2$, and have dramatically different $T_g$ of 54 °C and -100 °C, respectively[24]. We fix the degree of polymerization (DP) ($N_{sc}$=14) of the PDMS side chain (MW~1000 g/mol) while changing the number of side chains ($n_{sc}$) and the number ratio between spacers and side chains ($r_{sp}$) within the bottlebrush polymer. This approach allows us to reduce the four design parameters of fBB polymers, [$n_{sc}$, $r_{sp}$, $N_{sc}$, $\chi$], to two, [$n_{sc}$, $r_{sp}$].

We exploit the self-assembly of ABA triblock copolymers[25] to crosslink the fBB polymers to create networks. We synthesize an fBB polymer[23] and then grow onto its two ends a high $T_g$ linear polymer (PBnMA)[24], forming a linear-fBB-linear triblock copolymer (**Fig. 2a**). We start with two fBB polymers consisting of nearly the same number of side chains ($n_{sc}$≈550) but different spacer ratios ($r_{sp}$=0, 0.84) (**Extended Data Table 1**). Simultaneously, we fix the DP of a linear end block ($N_l$≈0.3$n_{sc}$≈175) to reach a volume fraction of 10% (**Supplementary Data Set 1**). At RT, the linear blocks aggregate into spherical glassy nodules that crosslink the fBB polymers, as evidenced by the hollow-cone dark-field transmission electron microscopy (TEM) (**Fig. 2b**). This microstructure is further confirmed by small-angle X-ray scattering (SAXS), which reveals a pronounced primary scattering peak, $q^*$, that corresponds to the average inter-domain distance, $d=2\pi/q^*$ (left arrow, **Fig. 2c**; **Fig. 2d**). These results demonstrate the formation of end-crosslinked fBB polymer networks encoded by three molecular architecture parameters [$n_{sc}$, $N_l$, $r_{sp}$] (**Fig. 2d**).

As the spacer ratio increases from 0 to 0.84, the network shear storage modulus $G'$, measured at 1 rad/sec, remains nearly constant at 3 kPa. By contrast, the yield strain increases by more than three-fold from 161% to 515% (**Fig. 2e**). At large deformations, the control network ($r_{sp}$=0) exhibits strain-stiffening characterized by a rapid increase of $G'$ with strain (solid circles, **Fig. 2e**). This phenomenon is classical to unentangled polymer networks, attributed to extending the network strand to its stretching limit. Surprisingly, for the network with spacer monomers ($r_{sp}$=0.84) the strain-stiffening occurs after a remarkable strain-softening regime with a reduction of 14% in shear modulus (left arrow, **Fig. 2e**). This strain-softening diminishes for fBB polymer networks with fewer side chains ($n_{sc}$<360) (**Extended Data Fig. 1**). Nevertheless, it has never been observed in any existing unentangled single-network elastomers. The remarkable strain-softening strongly suggests the strain-triggered unfolding of the collapsed bottlebrush backbone: As the fBB polymer unfolds, it becomes unable to sustain stress efficiently, resulting in reduced network stiffness (**Fig. 1c**). These results indicate the potential of using fBB polymers as network strands to increase network extensibility without altering stiffness.

To identify the parameter space ([$n_{sc}$, $N_l$, $r_{sp}$]) within which fBB polymer networks allow for decoupled stiffness and extensibility, we fix the number of side chains ($n_{sc}$≈200) while increasing the spacer ratio within a wide range ($r_{sp}$=0 to 3.46) (**Extended Data Table 1**; **Extended Data Fig. 2a, c**; **Supplementary Data Set 1**). Simultaneously, we fix the end block volume fraction relative to the side chains at ~6% ($N_f$≈33). All polymers self-assemble to end-crosslinked networks, as confirmed by the presence of primary SAXS scattering peaks (**Extended Data Fig. 3a**). As the average DP of the spacer segment, $N_g$=$r_{sp}$+1, increases from 1 to 4.46, $d$ increases less than twice from ~30 nm to ~50 nm (green circles, **Fig. 2f**). By contrast, the contour length of the fBB polymer $L_{max}$=$N_g n_{sc} l$ ($l$=2.56 Å is the main-chain length of a chemical monomer[23]), the maximum extent to which the fBB polymer can be stretched, increases by more than four times from ~50 nm to ~220 nm (red squares, **Fig. 2f**). The dramatic difference between the values of $d$ and $L_{max}$ highlights the ability of fBB polymers to store length as network strands.

In the self-assembled polymer networks, however, the glassy nodules are incompatible with the elastic bottlebrush network strands, resulting in interfacial repulsion that generates tension along the bottlebrush backbone[2] (**Supplementary Note 1.2.3**). To determine whether the tension unfolds the collapsed bottlebrush backbone, we compare the molecular structure of unperturbed fBB polymers in the melt to that in the self-assembled networks. Using wide-angle X-ray scattering (WAXS), we observe a characteristic peak, $q_{bb}$, associated with the inter-backbone distance between two neighboring fBB polymers, $D_{bb}=2\pi/q_{bb}$ (right arrow, **Fig. 2c**; **Extended Data Fig. 3a**). As $N_g$ increases from 1 to 4.46, $D_{bb}$ monotonically increases by 60% from 3.54 nm to 5.51 nm (circles, **Fig. 2g**). Notably, in the self-assembled networks, the dependence of $D_{bb}$ on $N_g$ quantitatively agrees with that observed in the melt (filled squares in **Fig. 2g**). These findings show that the fBB polymers remain folded in the self-assembled networks.

We emphasize that the observed increase of $D_{bb}$ with $N_g$ contradicts the understanding of conventional bottlebrush polymer melts (dashed line, **Fig. 2g**; **Supplementary Note 1**, eq. S10)[26–28]. However, this discrepancy is well-explained by our recent theory that accounts for the incompatibility between side chains and backbone within a bottlebrush polymer[23]. The theory predicts that the backbone folds into a cylindrical core, with all grafting sites located on its surface, to minimize the interfacial free energy between the side chains and the bottlebrush backbone (**Fig. 1c**). As the grafting density decreases, the backbone polymer collapses, leading to an increase in the cylindrical core diameter. Simultaneously, the distance between grafting sites in space reduces, enhancing steric repulsion among the side chains that further extends the side chains. Consequently, the bottlebrush diameter increases with the decrease of grafting density (solid lines, **Fig. 2g**; **Supplementary Note 2**, eq. S39).

After confirming that fBB polymers remain folded in the self-assembled networks, we quantify the network mechanical properties using uniaxial tensile tests (**Supplementary Videos 1-5**). The stress-strain curves exhibit *three* distinct behaviors depending on the spacer ratio. For low spacer ratios (Regime I, $0<r_{sp}<r_{sp,t}\approx1.5$), the stress-strain curves nearly overlap at low strains, yet the networks with more spacers become more stretchable (**Fig. 3a, i**). Moreover, the tensile fracture strain, $\varepsilon_f$, is the same as $\varepsilon_{max}$, the strain at which the nominal stress is maximum. Quantitatively, the network Young's moduli $E$ remain nearly the same of ~30 kPa; by contrast, $\varepsilon_{max}$ increases by a remarkable 20-fold from $(21\pm6)\%$ to $(428\pm59)\%$ (light green regions, **Figs. 3c, d**). The Young's moduli are consistent with network shear moduli, $G$, following the classic relation $E=3G$ (**Extended Data Fig. 4**).[1] Moreover, the dependence of network modulus on the spacer ratio can be well explained by the theoretical prediction for unentangled single-network elastomers (**Extended Data Fig. 4e**). These results show that using fBB polymers as network strands enables truly decoupled network stiffness and extensibility ($E\propto(\varepsilon_{max})^{-\alpha}$, $\alpha=0$) (light green region, **Fig. 3e**).

Remarkably, for intermediate spacer ratios (Regime II, $r_{sp,t}<r_{sp}<r_{sp,m}\approx2.3$), the networks are extremely stretchable with $\varepsilon_f$ up to ~2800% (grey line, **Fig. 3a, ii**; **Supplementary Video 3**). Yet, these networks exhibit plastic deformation above a critical strain ($\varepsilon_{max}$~900%) with the nominal stress decreasing dramatically with strain (**Fig. 3a, ii**). This plastic deformation is likely because of pulling the linear block out from the glassy nodules, which occurs for relatively weak ones. Consistent with this understanding, the plastic deformation disappears if glassy nodules become strong (**Extended Data Fig. 5**) or the crosslinkers are permanent covalent bonds (**Extended Data Fig. 9**). At high spacer ratios (Regime III, $r_{sp}>r_{sp,m}$), there is no plastic deformation and $\varepsilon_f=\varepsilon_{max}$. (**Fig. 3a, iii**). Thus, we use $\varepsilon_{max}$ to describe the extensibility attributed to fBB polymer network strands (arrow, **Fig. 3a, ii**).

Plotting $E$ against $\varepsilon_{max}$ reveals two distinct correlations at relatively high spacer ratios ($r_{sp}>r_{sp,l}$). There exists a small window ($r_{sp,l}<r_{sp}<r_{sp,m}$) in which stiffer networks are more stretchable ($\alpha<0$) (light red region, **Fig. 3e**). This behavior highlights the potential of exploiting fBB polymers to simultaneously enhance network stiffness and extensibility, a capability inaccessible by conventional single-network elastomers. At high spacer ratios ($r_{sp}>r_{sp,m}$), the networks resume the classical stiffness-extensibility trade-off ($\alpha>0$) yet remain quite stretchable ($\varepsilon_{max}>400\%$) (light grey region, **Fig. 3e**). Notably, the network stiffness increases exponentially with the spacer ratio (red dashed line, **Fig. 3d**). This polymer stiffening is caused by elevated $T_g$ of fBB polymers, such that the fBB polymers themselves become viscoelastic solids (**Extended Data Fig. 6a-f**). Indeed, at RT, the shear modulus of fBB polymers with high spacer ratios dominates the elastic contribution of fBB polymers as network strands and increases exponentially with the spacer ratio (**Extended Data Fig. 6g**). These results highlight the importance of keeping fBB polymers elastic (of low $T_g$) to increase network extensibility without altering stiffness.

Despite macroscopic evidence from the remarkable strain-softening (**Fig. 2e**) and truly decoupled network stiffness and extensibility (light green region, **Fig. 3e**), it has yet to be microscopically validated the unfolding of a collapsed bottlebrush backbone upon elongation (**Fig. 1c**). We perform *in situ* SAXS/WAXS measurements on a network undergoing uniaxial extension (**Fig. 4a**). We choose the network with a relatively high spacer ratio ($r_{sp}$=2.9), so that the inter-backbone distance ($D_{bb}$=5.39 nm) is 50% greater than that of the control network (**Fig. 4b**). Along the stretching direction, the inter-domain distance increases linearly with strain (**Fig. 4c, d**), yet the inter-backbone distance remains constant (**Extended Data Fig. 7c**). Remarkably, perpendicular to the stretching direction, the inter-backbone distance decreases with strain (**Fig. 4e, f**).

These contrasting behaviors originate from two distinct phenomena that resulted from unfolding the collapsed bottlebrush backbone. *First,* along the stretching direction, the network strand unfolds, resulting in an increased inter-domain distance (**Extended Data Fig. 8a, b**) and more aligned bottlebrush backbone (**Extended Data Fig. 8c**). *Second,* the unfolding process reduces the diameter of the cylindrical core of the collapsed bottlebrush backbone. Simultaneously, it decreases the grafting distance of side chains in space, such that the side chains become less crowded and their size decreases (**Extended Data Fig. 8d**). Consequently, as the strain increases, perpendicular to the stretching direction, not only the value of inter-backbone spacing decreases but also its orientation is more ordered, as evidenced by the enhanced orientation factor (**Fig. 4g, Extended Data Fig. 7a, b**). Collectively, our results show that a collapsed bottlebrush backbone unfolds upon extension.

Finally, we examine the universality of our design strategy by using another spacer monomer methyl methacrylate (MMA). MMA fulfills the design criteria as it has a low MW (100 g/mol) and is highly incompatible with PDMS[29]. We synthesize a series of fBB polymer networks with a wide range of MMA spacer ratios, [~200, ~60, 0-3.62] (**Extended Data Table 1; Extended Data Fig. 2b, c; Supplementary Data Set 2**). We observe similar behavior from the network microstructure (**Extended Data Fig. 3b**) to the mechanical properties (**Fig. 3b**; blue squares, **Fig. 3c, d**; **Extended Data Figs. 4, 6**). Moreover, for both MMA and BnMA spacers, $E$ versus $\varepsilon_{max}$ collapses to a universal relation, where $\varepsilon_{max}$ can be increased by nearly 40-fold, from ~20% to ~800% while keeping $E$ constant of ~30 kPa ($\alpha$=0) (light green region, **Fig. 3e**).

In summary, we have discovered that employing fBB polymers as network strands offers a general strategy to decouple the inherent stiffness-extensibility tradeoff of single-network elastomers. The unfolding of a fBB polymer is a reversible physical process, reminiscent of unzipping titin protein in muscle[30]. Importantly, our design approach is not limited to network topology; it applies to both end-crosslinked (**Fig. 3**) and randomly crosslinked single-network elastomers (**Extended Data Fig. 9**; **Supplementary Videos 6, 7**). Moreover, the extensibility of fBB polymer networks is significantly larger than that of conventional bottlebrush polymer networks[2–4,28,31] (**Extended Data Fig. 10**; **Supplementary Table S1**). Given that single-network polymers are the fundamental component of all kinds of networks, the universality of our design strategy and the resulting extremely stretchable polymer networks, our discovery opens an avenue for the development of polymer networks with extraordinary mechanical properties.

**Methods**

1. Materials. Monomethacryloxypropyl-terminated poly(dimethyl siloxane) (MCR-M11, MW~1000 g/mol) and methacryloxypropyl terminated poly(dimethyl siloxane) (DMS-R18, MW~5000 g/mol) are purchased from Gelest and purified using basic alumina columns to remove inhibitors. Benzyl methacrylate (BnMA, 98%) and methyl methacrylate (MMA, 98%) are purchased from Sigma Aldrich and purified using basic alumina columns to remove inhibitors. Copper (II) chloride ($CuCl_2$, 99.999%), tris[2-(dimethylamino)ethyl]amine ($Me_6TREN$), ethylene bis(2-bromoisobutyrate (2f-BiB, 97%), Tin(II) 2-ethylhexanoate ($Sn(EH)_2$, 92.5–100%), anisole (≥99.7%) and *p*-xylene (≥99.7%) are purchased from Sigma Aldrich and used as received. Tetrahydrofuran (THF, analytical reagent (AR)), purchased from Macron Fine Chemicals, and THF (High-performance liquid chromatography (HPLC) grade), methanol (Certified ACS), dichloromethane (DCM, Certified ACS), dimethylformamide (DMF, Certified ACS), purchased from Fisher, are used as received.

2. Polymer synthesis and characterization. In a fBB polymer, the side chains are randomly separated by spacer monomers. However, the side chains and the spacer monomers have different reactivity unless in a well-controlled condition. To this end, we use our previously established methods[23,31,32] to synthesize end-crosslinked and randomly crosslinked fBB polymer networks.

2.1 End-crosslinked fBB polymer networks. We present a detailed synthesis procedure using a sample with BnMA as the spacer ([195, 40, 1.08]) as an illustrative example.

Step I. Synthesis of a fBB middle block. A 50 mL Schlenk flask is charged with ethylene bis(2-bromoisobutyrate) (2f-BiB, 4.3 mg, 0.012 mmol), MCR-M11 (6g, 6 mmol), BnMA (705 mg, 4mmol), *p*-xylene (4 mL) and anisole (4 mL). Me$_6$TREN (46 mg, 0.2 mmol) and CuCl$_2$ (4.5 mg, 0.02 mmol) are dissolved in 1 mL DMF to prepare a catalyst solution. Then, 30 μL catalyst solution is added to the mixture. Followed by a 60-minute bubbling of the mixture with nitrogen, Sn(EH)$_2$ (14.6 mg, 0.036 mmol) in 100 μL *p*-xylene is quickly added into the reaction mixture while bubbling. The flask is then sealed with a rubber stopper and immersed in an oil bath at 60 °C. The reaction is monitored by proton nuclear magnetic resonance spectroscopy ($^1$H NMR) and stopped at 39% conversion. The reaction mixture is diluted with THF and passed through a neutral alumina column to remove the catalyst. The collected solution is concentrated using a rotary evaporator (Buchi R-205). To remove unreacted monomers and other impurities, the concentrated polymer solution is precipitated in methanol three times. After purification, the number of BnMA spacers is determined using $^1$H NMR, which is 211 for this fBB polymer. Additionally, the polymer contains 195 PDMS side chains, corresponding to a spacer/side chain ratio of $r_{sp}$=1.08 (**Supplementary Data, Fig. S5**).

Step II. Synthesis of a linear-fBB-linear triblock copolymer. A 25 mL Schlenk flask is charged with BnMA (282 mg, 1.6 mmol), macroinitiator (fBB PDMS, 232 kg/mol, 470 mg, 2 × 10$^{-3}$ mmol), *p*-xylene (1.3 mL), and anisole (1.3 mL). Me$_6$TREN (46 mg, 0.2 mmol) and CuCl$_2$ (2.7 mg, 0.02 mmol) are dissolved in 1 mL of DMF to prepare a catalyst solution. Then, 85 μL of the catalyst solution containing 1.7 × 10$^{-2}$ mmol Me$_6$TREN and 1.7 × 10$^{-3}$ mmol CuCl$_2$ is added to the mixture, and the resulting mixture is bubbled with nitrogen for 45 min to remove oxygen. Subsequently, the reducing agent, Sn(EH)$_2$ (27.5 mg, 6.8 × 10$^{-2}$ mmol) in 200 μL of *p*-xylene, is quickly added to the reaction mixture using a glass syringe. The flask is sealed and immersed in an oil bath at 60 °C. We stop the reaction at 10% conversion, purify the polymers following the same procedure as that in Step I, and use $^1$H NMR to confirm the DP of BnMA is 40 per end block (**Supplementary Data, Fig. S22**). Finally, the sample is dried in a vacuum oven (Thermo Fisher, Model 6258) at RT for 24 hours. The polymer is a transparent solid at RT.

2.2. Randomly crosslinked fBB polymer networks. These networks are synthesized following the procedure in Step I (**Methods, Section 2.1**) with the only difference that a di-functional crosslinking chain (DMS-R18) is added to the reaction mixture at a molar ratio 1:100 to the side chains (**Extended Data Fig. 9a, b**). Thus, the average number of side chains per network strand is $n_{sc}$=100. For each network, we use a co-solvent, DMF and THF with a volume ratio of 1:1, to remove unreacted monomers and catalyst, and then use THF to wash the network three times to remove DMF. The polymer networks are dried in a vacuum oven at RT overnight before being subjected to mechanical measurements.

3. $^1$H NMR characterization. We use $^1$H NMR to determine the average number of side chains per bottlebrush, the average number of spacer monomers, and the DP of each end block in a triblock copolymer polymer. The first one is calculated based on the conversion of linear PDMS macromonomers to bottlebrush PDMS. The number of spacer monomers is determined based on the NMR spectra of a purified bottlebrush polymer. The DP of each end block is determined based on the NMR spectra of purified triblock polymers. Chemical shifts for $^1$H NMR spectra are reported in parts per million compared to a singlet at 7.26 ppm in CDCl$_3$.

4. Gel permeation chromatography (GPC). GPC measurements are conducted using the TOSOH EcoSEC HLC-8320 GPC system equipped with two TOSOH Bioscience TSKgel GMHHR-M 5 μm columns in series. The GPC system includes a refractive index detector and operates at a temperature of 40 °C. HPLC grade THF is used as the eluent, and it is delivered at a flow rate of 1 mL/min. The samples for analysis are prepared by dissolving them in THF at a concentration of approximately 5 mg/mL.

5. SAXS/WAXS measurements. To prepare a sample for SAXS/WAXS characterization, we dissolve a triblock copolymer in toluene at a concentration of 100 mg/mL with a total volume of 3 mL in a glass vial and allow the solvent to slowly evaporate for 24 hours. Because toluene is a solvent close to being equally good for PBnMA and PDMS, it avoids the effects of solvent selectivity on the self-assembly process. Subsequently, we subject the sample to thermal annealing in a vacuum oven for 6 hours at 180 °C. Following the thermal annealing step, we slowly cool down the sample to RT at a rate of 0.5 °C/min. Throughout this cooling process, the microstructure of the self-assembled polymers does not change.

We use the Soft Matter Interfaces (12-ID) beamline at the Brookhaven National Laboratory to conduct SAXS/WAXS measurements on fBB polymers and networks. We perform measurements at multiple locations, thereby ensuring the consistency of the acquired two-dimensional (2D) scattering patterns. The distance between the sample and the detector is 8.3 meters, and the radiation wavelength used is $\lambda = 0.77$ Å. The scattered X-rays are captured using an in-vacuum Pilatus 1M detector, which consists of an array of 0.172 mm square pixels in a 941×1043 configuration. The raw SAXS images are converted into $q$-space, visualized in Xi-CAM software, and then radially integrated using customized Python code. The resulting one-dimensional intensity profile, denoted as $I(q)$, is plotted as a function of the scattering wave vector, $|\vec{q}| = q = 4\pi\lambda^{-1}\sin(\theta/2)$, where $\theta$ represents the scattering angle.

We perform *in situ* tensile tests using a Linkam TST-350 tensile stage equipped with a 2.5 N load cell. The Linkam stage is positioned in front of the X-ray beam with a horizontal orientation relative to the X-ray detector. To prepare the sample, we cut an annealed polymer into a rectangular shape with typical dimensions of 5-8 mm in length, 2-4 mm in width, and 0.5-1.0 mm in thickness. We load the sample to the Linkham tensile stage and stretch the elastomer at a strain rate of 0.01/sec while using SAXS/WAXS to characterize the microstructure of the network.

Because long-time exposure to X-ray may damage the elastomer, we limit the number of SAXS/WAXS measurements to 10 during target tensile strain. For each sample, data acquisition is conducted at a rate of one point per 35 seconds. Additionally, throughout the experiment, the beamline is maintained consistently passing through the central region of the sample. SAXS and WAXS patterns are recorded with a Pilatus 1M detector and a Pilatus 9K, respectively, with a pixel size of 0.172 mm. The distance between the sample and the WAXS detector is 2.5 meters, and the radiation wavelength used is $\lambda = 0.77$ Å.

In WAXS, the characteristic scattering peak corresponds to the inter-backbone distance of fBB polymers. The orientation of inter-backbone spacing, not the bottlebrush backbone, is determined from the azimuthal spread of the peak intensity from a 2D WAXS pattern:

$$\langle \cos^2 \phi \rangle = \frac{\int_0^{\pi/2} I(\phi) \cos^2 \phi \sin \phi \, d\phi}{\int_0^{\pi/2} I(\phi) \sin \phi \, d\phi} \tag{1}$$

Here, $\phi$ is the azimuthal angle and $I(\phi)$ is the intensity for the characteristic peak, as outlined by the region between two dashed circles in **Extended Data Fig. 7a**. The Herman's orientation parameter $S$ is defined as:

$$S = \frac{3\langle\cos^2\phi\rangle - 1}{2} \qquad (2)$$

The value of $S$ is 0 if the inter-backbone spacing exhibits no preferred direction, 1 when aligned parallel to the stretching direction, and -0.5 when aligned perpendicular to the stretching direction. Note that WAXS does not measure the alignment of the bottlebrush backbone but the inter-backbone spacing. Thus, when the bottlebrush backbones are more aligned on the stretching direction, the orientation of inter-backbone spacing is more ordered perpendicular to the stretching direction, resulting in the decrease of $S$. Consistent with this understanding, as the tensile strain increases from 0 to ~6, the value of $S$ decreases from 0 to ~-0.1 (**Fig. 4g**).

6. Transmission electron microscopy (TEM). We use our previously established method to prepare the samples for TEM imaging[24,33]. We use hollow-cone dark-field TEM (FEI Titan 80-300) at the electron energy of 300 keV with a tilt angle of 0.805° to characterize the annealed samples. This tilt angle allows for sharp contrast between PDMS and PBnMA domains without staining. The size of spherical domains is calculated using ImageJ, and >200 domains are used to ensure sufficiently powered statistics.

7. Rheological characterization. Rheological measurements are performed using a stress-controlled rheometer (Anton Paar MCR 302) with a plate-plate geometry of 8 mm in diameter. We exploit the solvent reprocessability of our polymers to prepare samples *in situ*. Specifically, we dissolve a polymer in DCM at a volume ratio of 1:2 to make a homogenous mixture.

Approximately 1 mL of the solution is pipetted onto the bottom plate of the rheometer and allowed to dry in ambient air at RT for one hour. Subsequently, the bottom plate is heated to 40 °C for 20 minutes. These procedures allow us to prepare a relatively thick film, ~0.3 mm, without the occurrence of cavities resulting from solvent evaporation. Then, we lower the upper plate and trim the excess sample at the edge of the geometry.

For frequency sweep, we fix the temperature at 20 °C and the oscillatory shear strain at 0.5% while varying the shear frequency from 0.1 rad/sec to 100 rad/sec. For strain sweep, we fix the temperature at 20 °C and the oscillatory shear frequency at 1 rad/sec while increasing the shear strain from 1% to 1000%. For the temperature sweep, we fix the oscillatory frequency at 1 rad/s and the shear strain at 5% while increasing the temperature from –20 °C to 80 °C, well above the $T_g$=54 °C of PBnMA[34]. As detailed in our previous work[32], we use a slow temperature ramping rate, 1 °C /minute, and wait for 20 minutes at each temperature point before collecting data; this ensures that the self-assembled microstructure is in equilibrium at each temperature point.

8. Tensile test. We prepare the self-assembled fBB polymer networks using a Teflon model described above (**Methods, Section 5**). We use a normalized cutter to cut the elastomers into dumbbell-shaped samples, which have a central part of 13 mm in length, 3 mm in width, and 0.5-1.0 mm in thickness. To load a sample, we use epoxy to glue the two ends of a sample to a hard cardboard, which is clamped to a tensile grip to avoid possible damage to the elastomers.

For uniaxial tensile test, use a Mark-10 ESM303 Motorized Test Stand equipped with a 2.5 N load cell or an Instron (6800-SC) equipped with a 10 N load cell. Nominal stress is defined as the tensile force per unit of the initial cross-section area of the central part. During the tensile test, we use a camera to record the strain by tracking the displacement of the central part of a dumbbell-shaped sample. Each measurement is performed at RT at a strain rate of 0.02/sec and repeated at least three times on parallel samples.

9. Differential scanning calorimetry. We determine the $T_g$ of fBB polymers using a differential scanning calorimeter (DSC Q20, TA instruments). All the samples are prepared using a combination of solvent and thermal annealing (**Methods, Section 5**) to ensure equilibrated state. Before characterization, the samples are further dried at RT (~293 K) under vacuum (30 mbar) for at least 24 hours. A standard aluminum DSC pan is used for all the measurements with approximately 10 mg of sample loaded. All the samples were annealed at 393K for 2 min to erase the thermal history, followed by cooling at 10 K/min to 213 K and then heating at 10 K/min to 393 K. The specific heat capacity, $C_p$, is determined following the second heating cycle. The $T_g$ values are determined as the midpoint of the specific heat capacity jump; these values are also consistent with those determined by the inflection point of the heat capacity under temperature sweep.

**Acknowledgements:** We thank Dr. Daniel A. Rau, Dr. Mikhail Zhernenkov, Dr. Guillaume Freychet, Dr. Lutz Wiegart, and Dr. Patryk Wasik for assistance with SAXS/WAXS measurements, and Dr. Shiwang Cheng for DSC measurements. We thank Dr. Michael Rubinstein and Dr. Edward Samulski for critical reading of the manuscript before submission.

**Funding:** This work is supported by the National Science Foundation CAREER Award (DMR-1310266).

**Author contributions:**

Conceptualization: LHC

Methodology: BH, NS, LHC

Investigation: BH, NS, LHC

Visualization: LHC, BH

Funding acquisition: LHC

Project administration: LHC

Supervision: LHC

Writing – original draft: LHC

Writing – review & editing: LHC, BH, SN

**Competing interests:** LHC, BH, and SN have filed a provisional patent application (63/517,846) regarding foldable bottlebrush polymers and networks.

**Data and materials availability:** All data are available in the manuscript or the supplementary information.

**Supplementary Information**

Supplementary Notes

Supplementary Figures

Supplementary Tables

Supplementary Data

Supplementary Videos

Supplementary References

# Figures

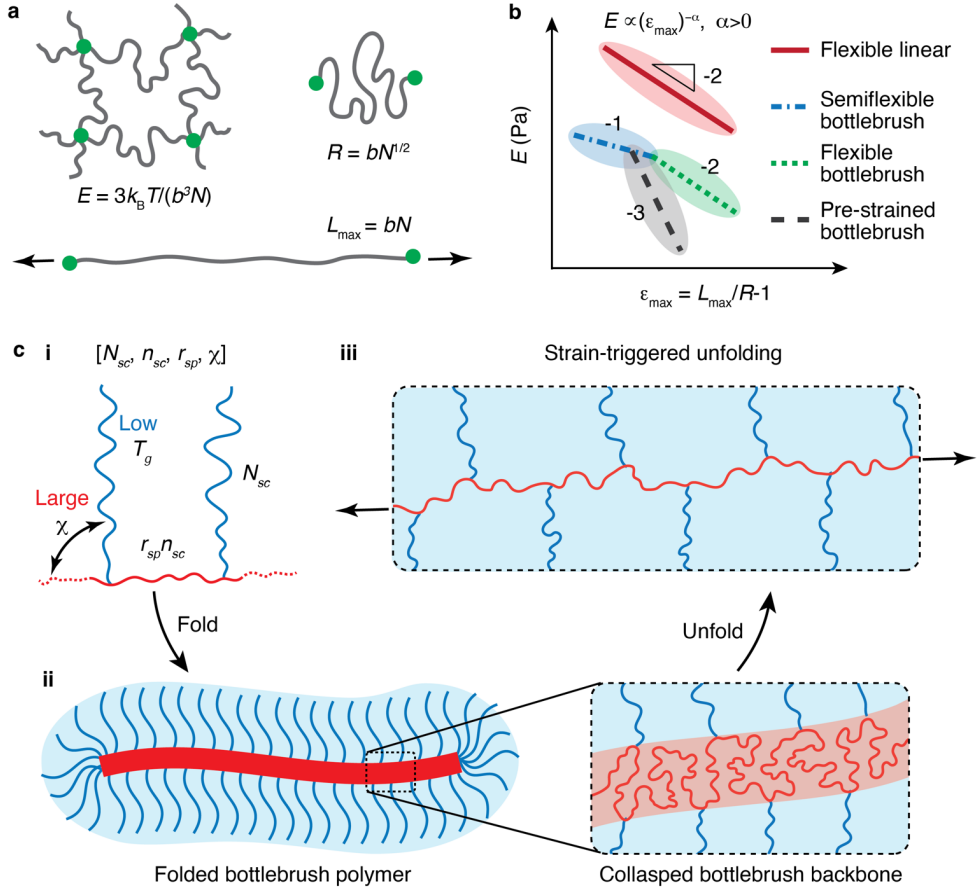

**Figure 1. Concept of foldable bottlebrush (fBB) polymer network strands. a**, A conventional single-network elastomer is comprised of crosslinked flexible linear polymers without solvents. The network extensibility is limited by the ratio of the maximum length, $L_{max}$, to the initial size, $R$, of the network strand: $\varepsilon_{max}=L_{max}/R-1\sim N^{1/2}$, where $R=bN^{1/2}$ is the random walk of $N$ Kuhn monomers of size $b$ per network strand, and $L_{max}=bN$ is the contour length of the network strand. The network Young's modulus is $E\approx 3k_BT/V\sim 1/N$, where $k_B$ is Boltzmann constant, $T$ is the absolute temperature, and $V\approx Nb^3$ is the volume of a network strand. Thus, the network stiffness and extensibility are negatively correlated: $E\propto(\varepsilon_{max})^{-2}$. **b**, Using conventional bottlebrush polymers increases the network strand molecular weight and, therefore, reduces network stiffness, but does not break the inherent stiffness-extensibility trade-off: $E\propto(\varepsilon_{max})^{-\alpha}$, where the value of $\alpha$ is 2, 1, or >2 if the bottlebrush polymer is flexible, semiflexible, or pre-strained (**Supplementary Note 1**). **c**, A fBB polymer consists of many linear side chains ($n_{sc}$) randomly separated by small spacer monomers at a molar ratio of $r_{sp}$. **(i)** The design criteria are that the side chain has a relatively high MW (DP $N_{sc}$) and a low glass transition temperature ($T_g$), whereas the spacer monomer has a low MW and is highly incompatible with the side chains (relatively large Flory-Huggins interaction parameter $\chi$). **(ii)** To minimize interfacial free energy, the backbone collapses into a cylindrical core with its surface densely grafted with side chains, despite the strong steric repulsion among the overlapped side chains in both the folded and unfolded states. The fBB polymer remains elastic because of its low $T_g$ side chains. **(iii)** Upon stretching, the collapsed bottlebrush backbone unfolds to release the stored length. By contrast, the molecular weight of this fBB polymer is dominated by the side chains. Thus, it is expected that using fBB polymers as network strands enables decoupling network stiffness and extensibility ($\alpha=0$).

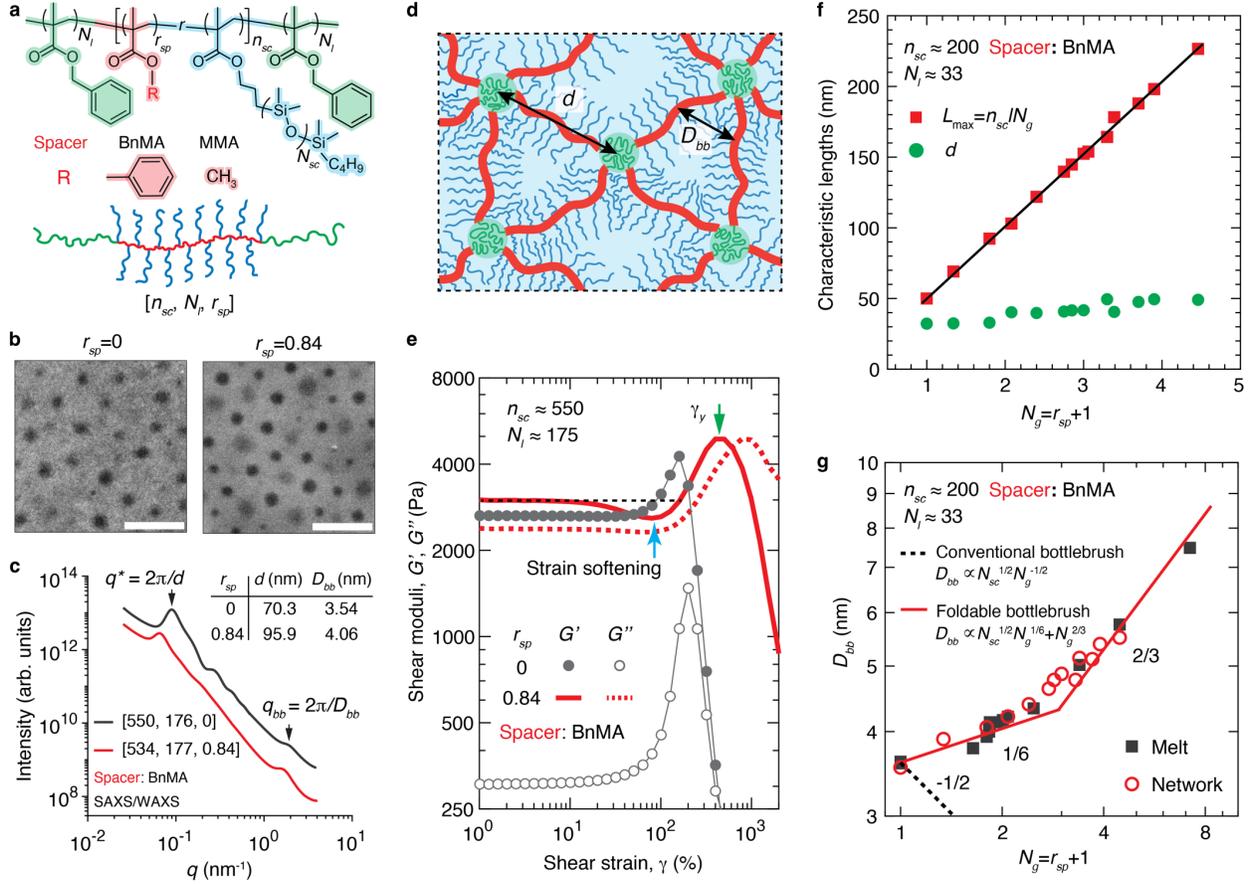

**Figure 2. Design and synthesis of foldable bottlebrush polymer networks. a,** A proof-of-concept design of fBB polymer networks by exploiting the classical ABA triblock copolymer self-assembly. Side chain: linear poly(dimethyl siloxane) (PDMS) of 1,000 g/mol; spacer: benzyl methacrylate (BnMA) of 176 g/mol or methyl methacrylate (MMA) of 100 g/mol; linear blocks: poly(benzyl methacrylate) (PBnMA) with DP $N_l$;. Since the side chain MW is constant, this linear-fBB-linear triblock copolymer has three design parameters [$n_{sc}$, $N_l$, $r_{sp}$]. **b,** Hollow-cone dark-field TEM images of the microstructures self-assembled by two triblock copolymers. Left: control polymer without spacers, [550, 176, 0]; right: polymer with spacers, [534, 177, 0.84]. The dark dots are the spherical nodules aggregated by linear PBnMA end blocks. The average inter-domain distances are $d$=75.7±13.1 nm ($r_{sp}$=0) and 85.2±10.8 nm ($r_{sp}$=0.84). Scale bars, 200 nm. **c,** Dependence of SAXS/WAXS intensity on the magnitude of wavevector, $q$, for the self-assembled networks. **d**, A schematic illustrating two characteristic lengths in the self-assembled networks: (1) the average inter-domain distance, $d=2\pi/q^*$, and (2) the inter-backbone distance, $D_{bb}=2\pi/q_{bb}$, of fBB polymers, as noted by the arrows in **(c)**. **e,** Large amplitude oscillatory shear (LAOS) measurements reveal that introducing spacer monomers increases the shear yield strain ($\gamma_y$) but nearly does not alter network stiffness. The network with spacers exhibits a remarkable strain-softening regime, with a reduction of 14% in shear modulus, followed by the classical strain-stiffening. $G'$ and $G''$ are, respectively, shear storage and loss moduli measured at 1 rad/sec at 20 °C. **f, g,** Characteristic length scales ($d$, $L_{max}$, $D_{bb}$) of fBB polymer networks with various BnMA spacer ratios, [~200, ~33, 0-3.46]. **(f)** As the average DP of the spacer segment ($N_g=r_{sp}+1$) increases, the inter-domain distance ($d$) remains nearly the same, but the bottlebrush backbone contour length ($L_{max}=n_{sc}lN_g$) increases dramatically. Here, $l$ is the main-chain bond length of a repeating unit in the backbone. **(g)** The inter-backbone distance ($D_{bb}$) increases with the decrease of grafting density ($1/N_g$) in both melts (squares) and self-assembled networks (circles). This behavior contradicts the understanding of conventional bottlebrush polymers (dashed line, $D_{bb} \propto N_{sc}^{1/2}N_g^{-1/2}$, **Supplementary Note 1**, eq. S10). Yet, it can be well-explained by our recent molecular theory that accounts for the incompatibility between spacer monomers and side chains within fBB polymers (solid line, $D_{bb} \propto N_{sc}^{1/2}N_g^{1/6}+N_g^{2/3}$; **Supplementary Note 2**, eq. S39).

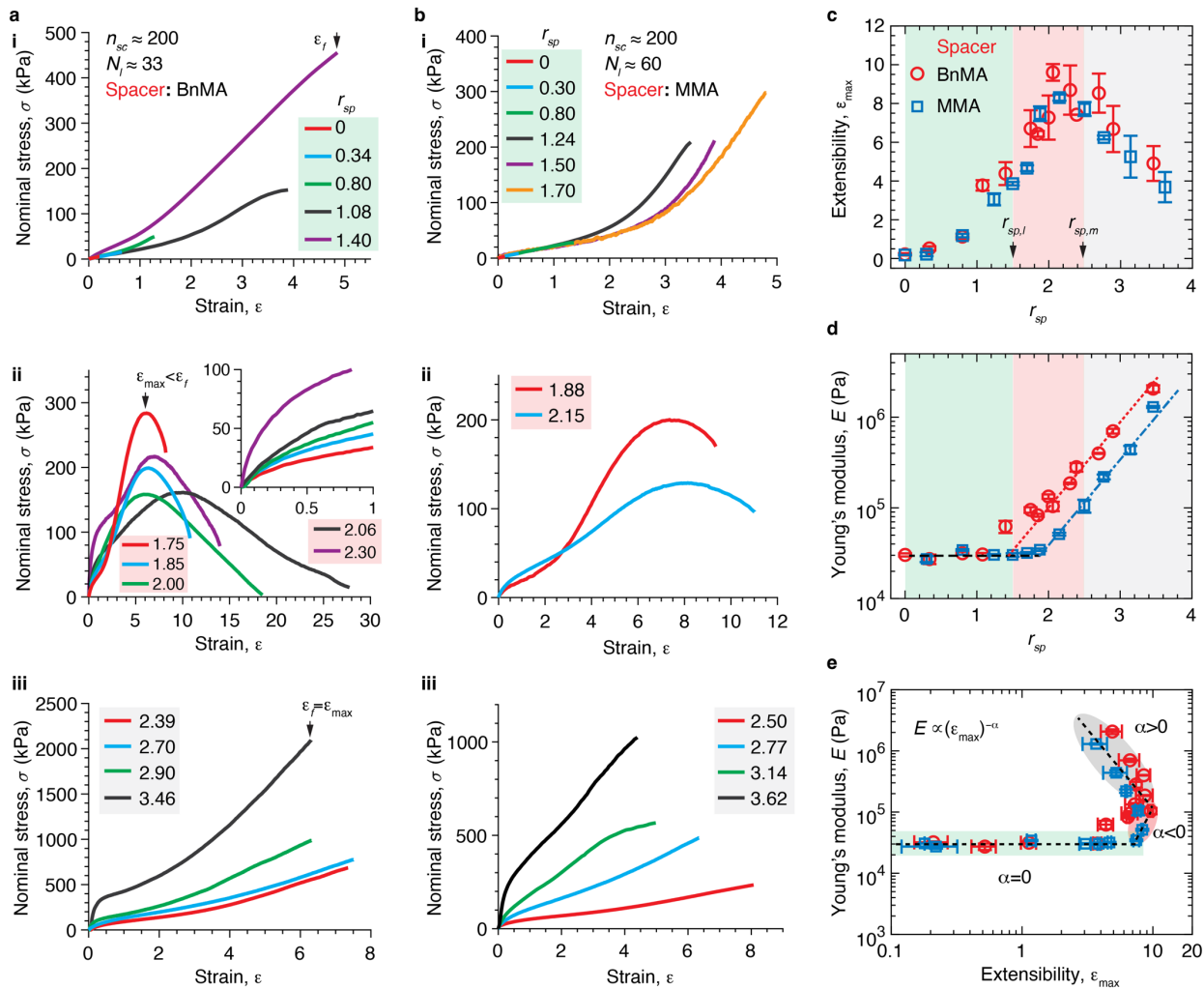

**Figure 3. Using fBB polymers as network strands provides a universal strategy for decoupling network stiffness and extensibility. a, b,** Nominal stress-strain curves of networks with **(a)** BnMA and **(b)** MMA as spacers. $\varepsilon_f$, tensile fracture strain; $\varepsilon_{max}$, critical strain at which the nominal stress is maximum. At either **(i)** low or **(iii)** high spacer ratios, $\varepsilon_f = \varepsilon_{max}$. By contrast, at intermediate spacer ratios, $\varepsilon_{max} < \varepsilon_f$ and the networks exhibit plastic deformation under large deformations, at which the nominal stress decreases dramatically with strain. All measurements are performed at RT and a fixed tensile strain rate of 0.02/sec. **c, d,** Dependencies of **(c)** network extensibility ($\varepsilon_{max}$) and **(d)** Young's modulus ($E$) on the spacer ratio. **e,** At relatively small spacer ratios, the network stiffness and extensibility are truly decoupled, where $\varepsilon_{max}$ can be increased from ~20% to 800% while keeping $E$ constant at ~30 kPa ($\alpha=0$). There exists a small window (intermediate spacer ratios) in which stiffness and extensibility are positively correlated ($\alpha<0$). At high spacer ratios, the folded bottlebrush itself has an elevated $T_g$ and becomes stiff (**Extended Data Fig. 6**), such that the network stiffness and extensibility resume the classical negative correlation ($\alpha>0$). Error bar, standard deviation for $n$=3-5.

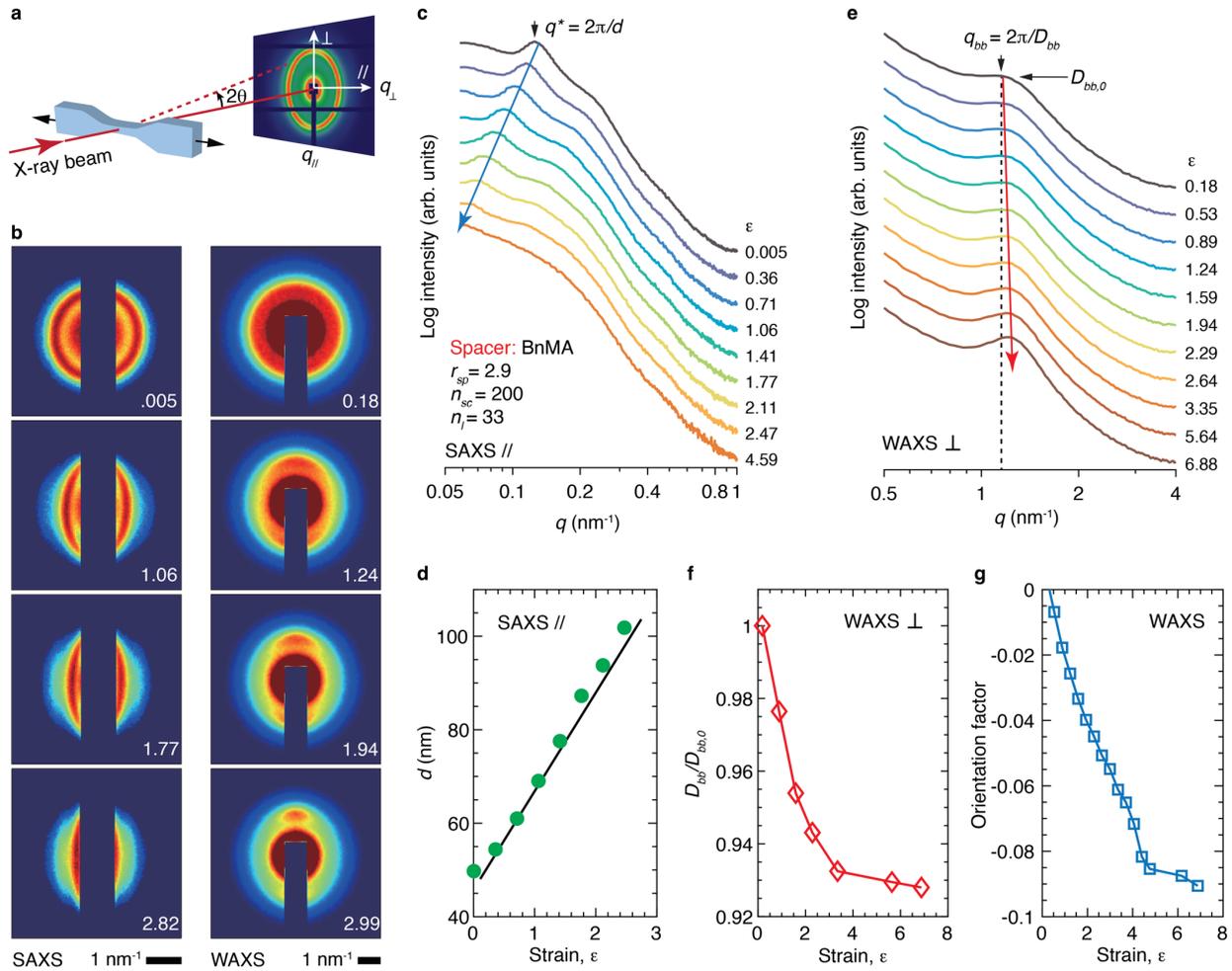

**Figure 4. Unfolding of a collapsed bottlebrush network strand under extension. a,** Illustration of *in situ* SAXS/WAXS measurements for a fBB elastomer with BnMA as the spacer monomer ([200, 33, 2.9]) under uniaxial tensile test at a strain rate of 0.01/sec. **b,** Two-dimensional (2D) (i) SAXS and (ii) WAXS patterns at various strains. **c, d,** Along the elongation direction, the inter-domain distance ($d$) increases linearly with strain ($\varepsilon$). **e, f,** Perpendicular to the elongation direction, the inter-backbone distance ($D_{bb}$) decreases with the increase of strain because of the strain-triggered unfolding of a collapsed bottlebrush backbone. **g,** For WAXS, the orientation factor is negative and decreases with the increase of strain (eq. 2, **Methods**). This behavior indicates that the bottlebrush backbones become more aligned along the stretching direction, and therefore, the orientation of inter-backbone spacing becomes more ordered perpendicular to the stretching direction (**Extended Data Fig. 8b-d**).

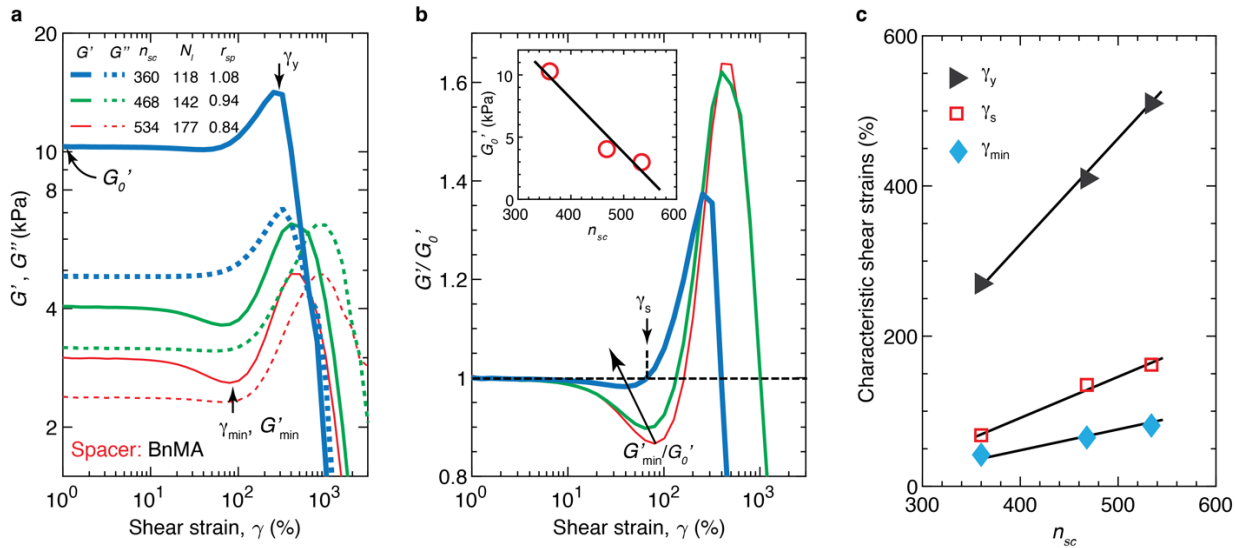

**Extended Data Fig. 1 | Strain-softening of fBB polymer networks.**
**a**, LAOS measurements of fBB polymer networks are performed at 20 °C and 1 rad/sec. All networks have nearly the same spacer ratio ($r_{sp}$) but different numbers of side chains ($n_{sc}$). The DP of each linear end block is about 30% of the number of side chains ($N_l \approx 0.3 n_{sc}$); this ensures that the volume fraction of the end linear blocks is nearly the same at ~10%. All networks exhibit a remarkable strain-softening followed by classical strain-stiffening attributed to stretching the network strand to its length limit. To describe the dynamic mechanical properties of a network under LAOS measurement, we introduce three parameters: yield strain ($\gamma_y$), the strain ($\gamma_{min}$) at which the shear storage modulus is minimum ($G'_{min}$), and shear storage modulus at the lowest strain 1% ($G'_0$). **b**, Remarkably, as $n_{sc}$ decreases from 534 to 360, the strain-softening behavior nearly vanishes, as evidenced by the increase of the ratio $G'_{min}/G'_0$ from 86.6% to 98.2% (northwest pointing arrow). The width of the strain-softening regime, $\gamma_s$, defined as the critical strain above which the strain-softening regime ends ($G'/G'_0 > 1$), increases with the number of side chains. Additionally, the network shear modulus is inversely proportional to the number of side chains or the volume of the fBB polymer network strand (inset). **c**, All the characteristic shear strains, $\gamma_{min}$, $\gamma_s$, and $\gamma_y$, increase linearly with the number of side chains. These results suggest that for a fBB polymer network, a relatively large number of side chains is needed to exhibit pronounced strain-softening.

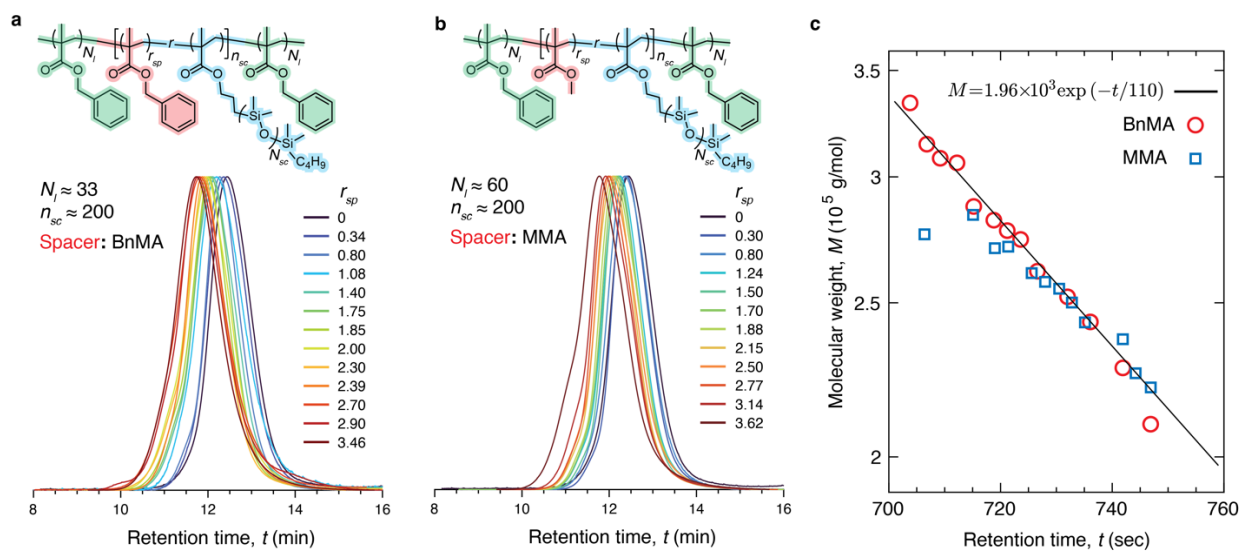

**Extended Data Fig. 2 | GPC profiles of linear-fBB-linear triblock copolymers.**
**a**, **b,** GPC profiles of polymers with **(a)** BnMA and **(b)** MMA as spacers. **c**, The logarithmic MW, $M$, of all triblock polymers decreases linearly with the increase of peak retention time, $t$. These results demonstrate controlled synthesis of linear-fBB-linear polymers with desired molecular architecture parameters ($[n_{sc}, N_l, r_{sp}]$).

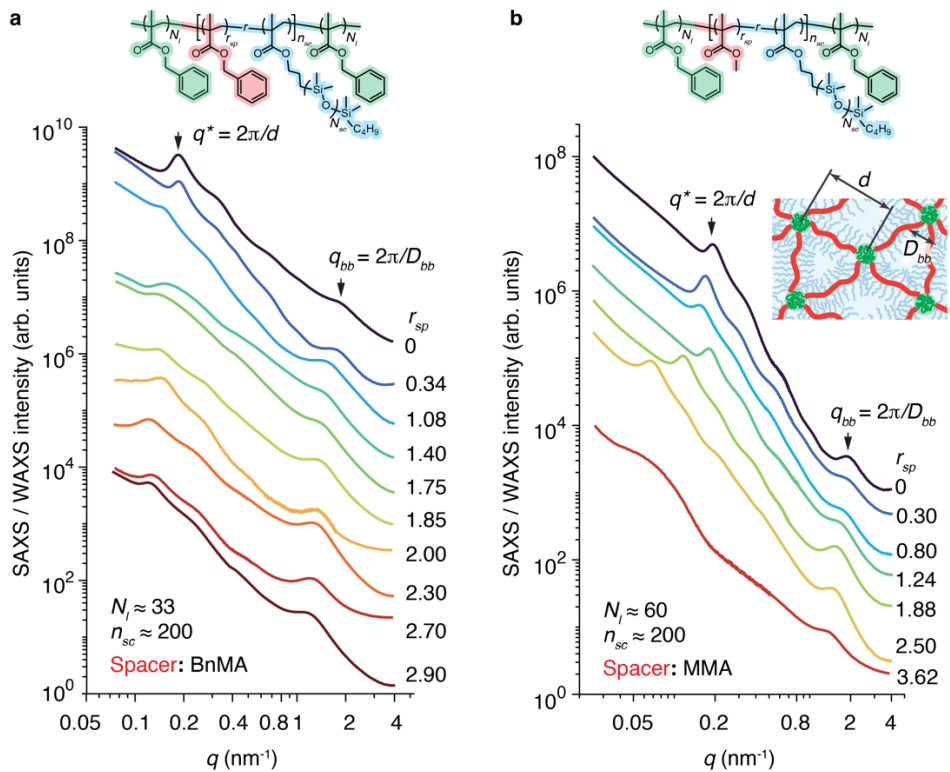

**Extended Data Fig. 3 | Microstructure of self-assembled fBB polymer networks.**
**a, b,** Radially averaged one-dimensional (1D) SAXS/WAXS intensity profiles as a function of the magnitude of the wavevector ($q$) for the self-assembled fBB polymer networks with **(a)** BnMA and **(b)** MMA as spacer monomers. All polymers form an end-crosslinked network, in which the fBB polymers are crosslinked by glassy nodules aggregated by linear end blocks (inset, **b**). For both series of polymers, the inter-backbone distance ($D_{bb}$) increases with the spacer ratio ($r_{sp}$). These results confirm that fBB polymer network strands remain folded in the self-assembled networks.

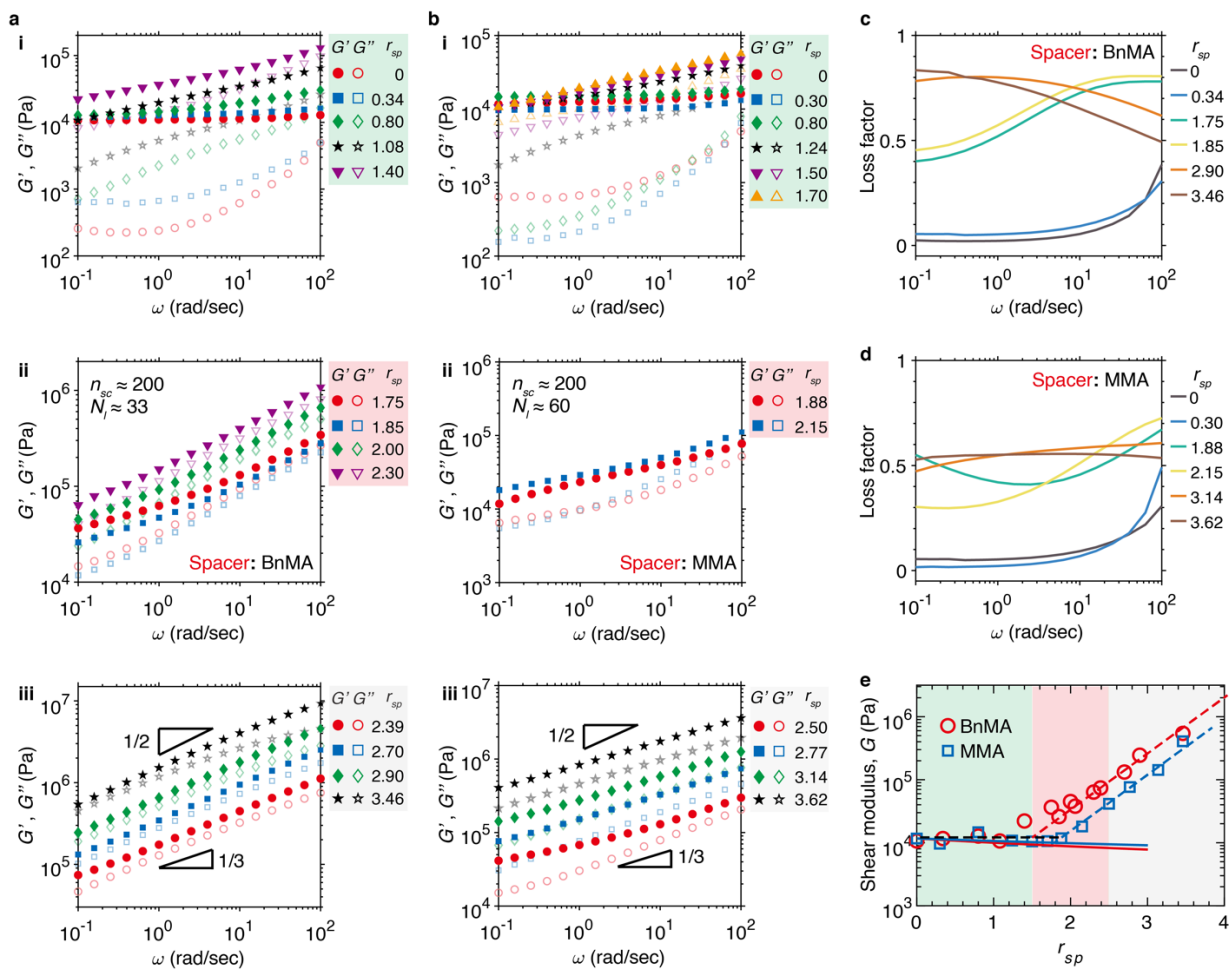

**Extended Data Fig. 4 | Dynamic mechanical properties of fBB polymer networks.**
**a, b,** Frequency dependencies of the storage (solid symbols, $G'$) and loss (open symbols, $G''$) moduli of fBB polymer networks with **(a)** BnMA and **(b)** MMA as spacer monomers. All measurements are performed at 20 °C with a fixed strain of 5%. At low spacer ratios (Regime I), $G'$ exhibits a weak dependence on frequency and reaches a plateau at low frequency. By contrast, as the spacer ratio becomes relatively high (Regimes II & III), $G'$ increases rapidly with frequency, indicating enhanced dissipative behavior in the polymer networks. **c, d,** Consistent with this understanding, the loss factor, $\tan\delta = G''/G'$, increases with the spacer ratio. Yet, all polymers are a solid network with $G'$ greater than $G''$ ($\tan\delta < 1$). **e,** The network shear modulus $G$ ($G'$ at the lowest frequency 0.1 rad/sec) remains nearly a constant at low spacer ratios (symbols in the light green region). The experimental measurements can be well-explained by the theory that the shear modulus of an unentangled network is $k_BT$ per volume of the network strand: $G = k_B T N_{Av} \rho / M$. Here, $T = 293$ K, $N_{Av}$ is Avogadro number, $\rho \approx 1$ g/cm$^3$ is polymer density, and $M = (n_{sc}M_{sc} + r_{sp}n_{sc}M_{sp})$ is the molar mass of the fBB polymer (with $M_{sc} \approx 1000$ g/mol and $M_{sp}$ being the side chain and spacer molecular weight, respectively). Black dashed line: $G \approx 11$ kPa for the control network ($r_{sp}=0$); red solid line: networks with BnMA spacer; blue solid line: networks with MMA spacer. Note that the predicted shear modulus is nearly independent of the spacer ratio, as the molecular weight of the fBB polymer is dominated by the side chains. At relatively high spacer ratios (light red and gray regions), *G* increases exponentially with the spacer ratio (dashed lines). The dependence of shear modulus *G* on the spacer ratio aligns well with the network Young's moduli *E*, following the classical relation *E*=3*G* (**Fig. 3d**).

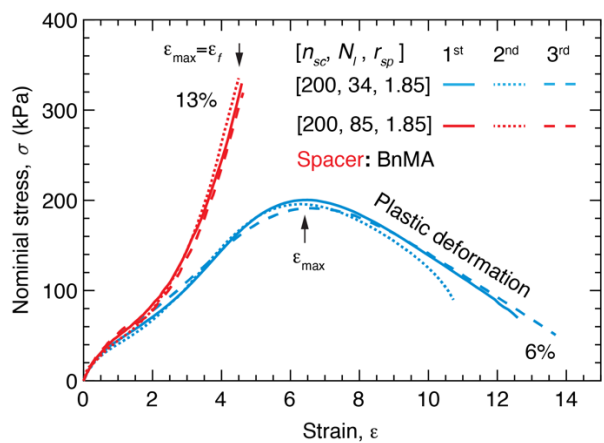

**Extended Data Fig. 5 | Plastic deformation disappears for stronger glassy nodules.**

At intermediate spacer ratios, fBB polymer networks exhibit plastic deformation under large deformation, at which the nominal stress decreases dramatically with the increase of tensile strain (**Fig. 3a,iii & b,ii**). We propose that the plastic deformation is because of the end-chain pullout from the glassy nodules. During this process, the network continues to stretch without breaking, but the network strands cannot efficiently sustain stress, resulting in an apparent decrease in nominal stress. We test this hypothesis by synthesizing two networks with the same fBB polymer network strand but different end block volume fractions. As the end block volume fraction increases, the glassy nodules increase in size and strength, preventing the end-chain pullout from the glassy nodules. Consistent with this understanding, the plastic deformation disappears as the end block volume fraction increases from 6% to 13%, a value close to the upper limit for sphere morphology. Moreover, the network Young's moduli remain the same, as shown by the overlap of strain-stress curves at low strains. These results demonstrate that the plastic deformation occurs only for relatively weak glassy nodules, as illustrated in **Extended Data Fig. 8e**. Thus, we use $\varepsilon_{max}$, the strain at which the nominal stress is maximum, to describe the network extensibility attributed to the fBB polymer network strands.

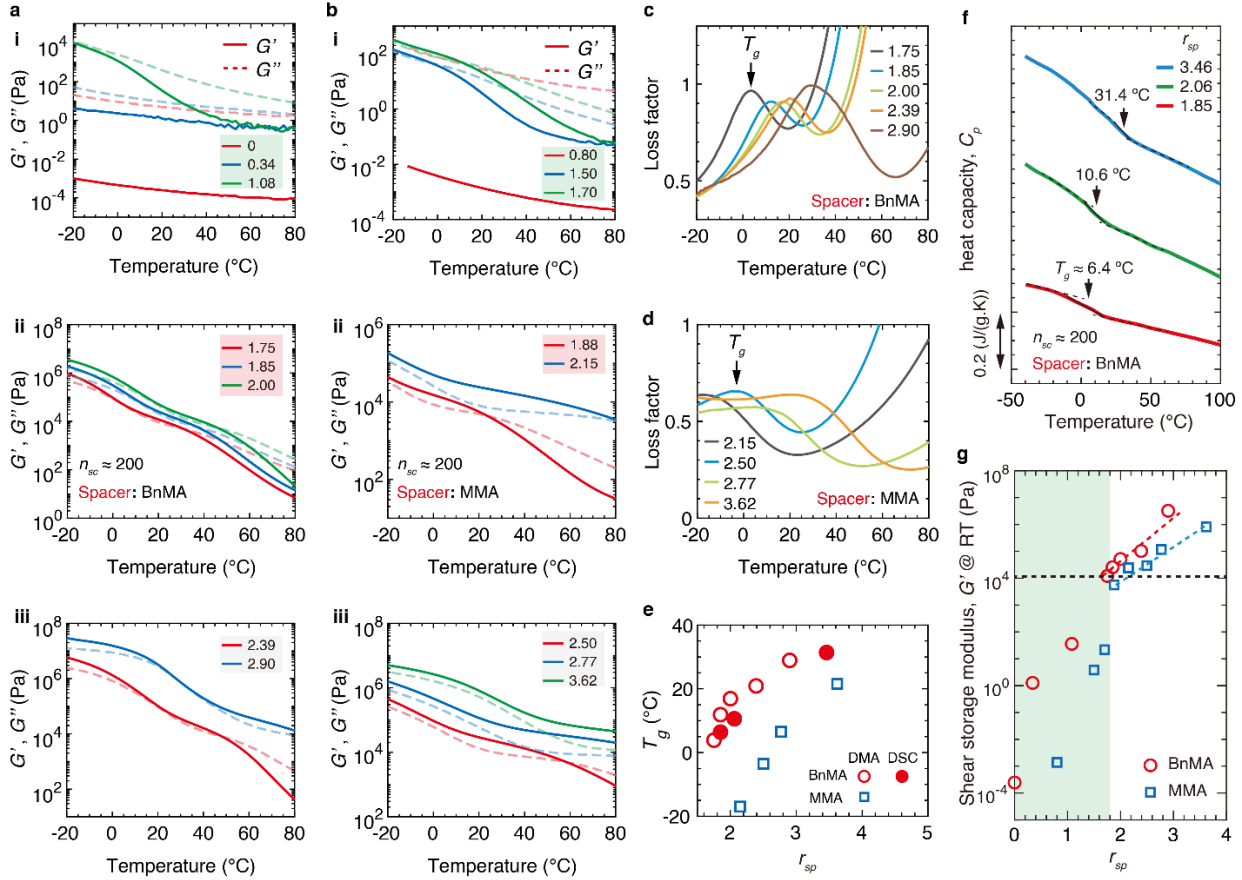

**Extended Data Fig. 6 | Melts of fBB polymers are viscoelastic solids at high spacer ratios.**
**a, b,** Dependence of shear moduli on temperature for the melts of fBB polymers with **(a)** BnMA and **(b)** MMA as spacer monomers. All measurements are performed at 1 rad/sec with a fixed oscillatory shear strain of 5%. **c, d,** Dynamic mechanical analysis (DMA) reveals that at relatively high spacer ratios, the loss factor of fBB polymer melts exhibits a pronounced peak at the glass transition temperature $T_g$ associated with the collapsed bottlebrush backbone. **e,** The $T_g$ of fBB polymers, measured by DMA, increases with the spacer ratio (empty symbols). **f,** In parallel, we use differential scanning calorimetry (DSC) to measure the $T_g$ of fBB polymers. The curves for the heat capacity $C_p$ of fBB polymers are shifted vertically for clarity. The values of $T_g$ measured by DSC (filled symbols in **(e)**) are consistent with those measured by DMA (empty symbols in **(e)**). At low spacer ratios, the volume fraction of the backbone within the fBB polymer is relatively small (<20%). Consequently, the $T_g$ of fBB polymers is much below RT, such that the fBB polymers remain elastic at RT. However, at relatively high spacer ratios, the $T_g$ of fBB polymers becomes close to and even higher than RT, such that the fBB polymers become a viscoelastic solid at RT. **g,** The shear storage modulus $G'$ (measured at 1 rad/sec and 20 °C) of fBB polymers increases with the spacer ratio. At relatively high spacer ratios ($r_{sp}$>2), the fBB polymers themselves become stiff, with $G'$ increasing exponentially with the spacer ratio (dashed red and blue lines for BnMA and MMA spacers, respectively). Moreover, the stiffness of fBB polymers becomes greater than the elastic contribution of the fBB polymers as a network strand (~11 kPa, horizontal dashed line). As a result, the stiffness of fBB polymer networks is dominated by the fBB polymer and increases exponentially with the spacer ratio (**Fig. 2d**).

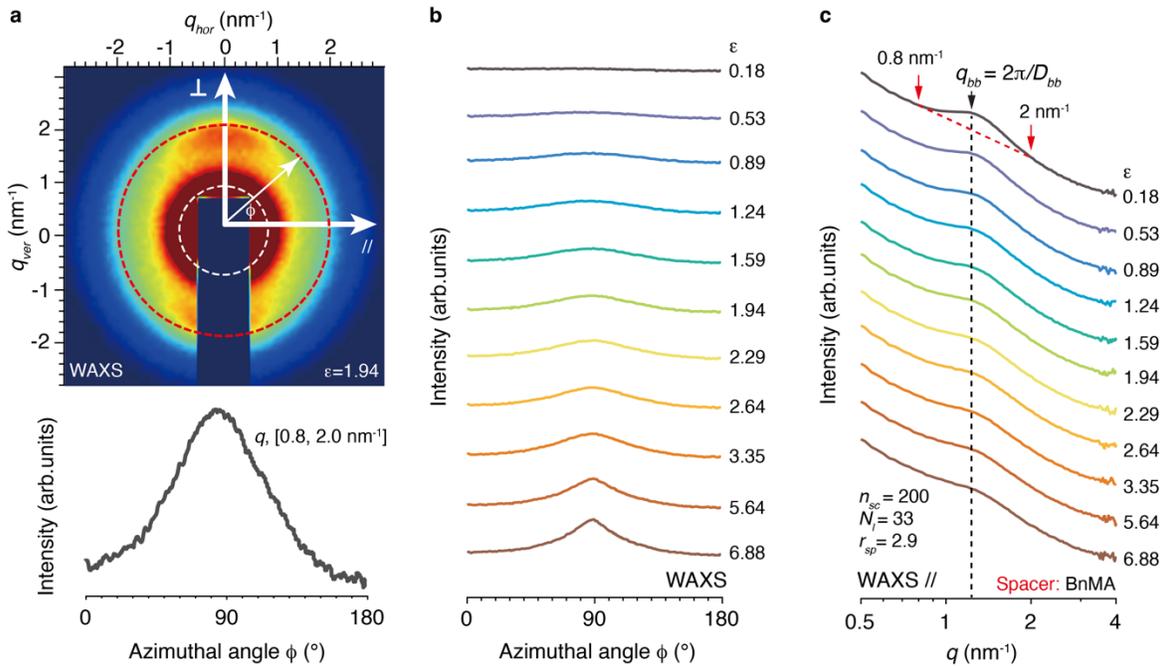

**Extended Data Fig. 7 | In situ SAXS/WAXS scattering of a fBB polymer network under uniaxial tension.**

**a,** A 2D WAXS pattern of a fBB polymer network (spacer: BnMA; [200, 33, 2.9]) at a tensile strain of 1.94. To determine the azimuthal spread of the peak intensity associated with the inter-backbone spacing, we choose the region with $q$ values from 0.8 nm$^{-1}$ to 2 nm$^{-1}$ (the ring between two dashed circles). This range is sufficient to cover the width of the characteristic scattering peak located at 1.2 nm$^{-1}$, as illustrated in **(c)**. An example of the scattering intensity as a function of the azimuthal angle is shown by the lower panel. **b,** Without deformation, the scattering intensity exhibits no preference for the azimuthal angle, indicating no preference for the orientation of the inter-backbone spacing. As the strain increases, the scattering intensity becomes more pronounced at 90°, indicating that the orientation of inter-backbone spacing becomes more ordered perpendicular to the stretching direction. **c,** Along the stretching direction, the 1D scattering intensity profiles exhibit a characteristic peak associated with the inter-backbone distance, as illustrated in **Extended Data Fig. 8c**. Remarkably, as the strain increases, the peak location does not change, yet the peak intensity decreases. This behavior is consistent with the understanding that the fBB polymer network strands are compressed perpendicular to the stretching direction, such that along the stretching direction, the inter-backbone distance does not change, yet the orientation of inter-backbone spacing becomes less ordered (**Extended Data Fig. 8c**).

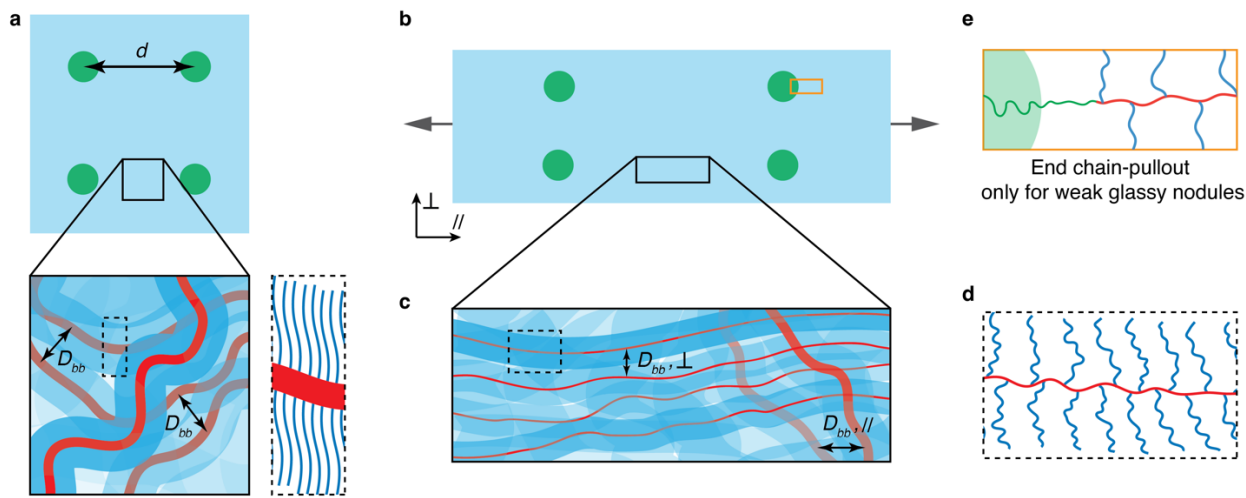

**Extended Data Fig. 8 | Illustration of molecular mechanisms for stretching fBB polymer networks.**

**a**, In a network self-assembled by linear-fBB-linear triblock copolymers, the end blocks aggregate into spherical glassy domains (green spheres) that crosslink the fBB polymer network strands. Without deformation, the backbone of fBB network strands remains folded (thick red line, lower panel). Moreover, the glassy nodules and the orientation of the collapsed bottlebrush backbone are randomly distributed. Consequently, the 2D SAXS/WAXS intensity profiles are independent of the azimuthal angle and exhibit symmetric, circular patterns (**Fig. 4b**). **b**, Upon elongation, the inter-domain distance ($d$) increases along the elongation direction but decreases perpendicular to the elongation direction (**Fig. 4c, d**). **c**, Along the elongation direction, the backbone of the fBB polymer network strands becomes more aligned, resulting in more ordered inter-backbone spacing perpendicular to the elongation direction. Consequently, as the strain increases, the characteristic WAXS scattering peaks exhibit increased intensity (**Fig. 4e**, **Extended Data Fig. 7b**) and enhanced orientation function (**Fig. 4g**). **d,** Simultaneously, along the elongation direction, the collapsed backbone unfolds, resulting in reduced side chain grafting density. Thus, both the side chain size and the bottlebrush backbone diameter decrease, resulting in reduced inter-backbone distance ($D_{bb}$) perpendicular to the stretching direction (**Fig. 4f**). By contrast, the fBB polymer network strands are not stretched but compressed, resulting in negligible change in inter-backbone distance along the stretching direction (**Extended Data Fig. 7c**). **e,** In some cases, when the glassy nodules are relatively weak, it is possible to pull the linear end blocks out from the glassy nodules. This process enables extremely stretchable networks. However, as the end-chain is being pullout, the network strands cannot efficiently sustain stress. Thus, the network exhibits plastic deformation with the nominal stress decreasing dramatically with the increase of tensile strain (**Figs. 3a&b, ii**). However, this plastic deformation disappears when the glassy nodules become relatively strong (**Extended Data Fig. 5**) or the crosslinkers become permanent covalent bonds (**Extended Data Fig. 9**).

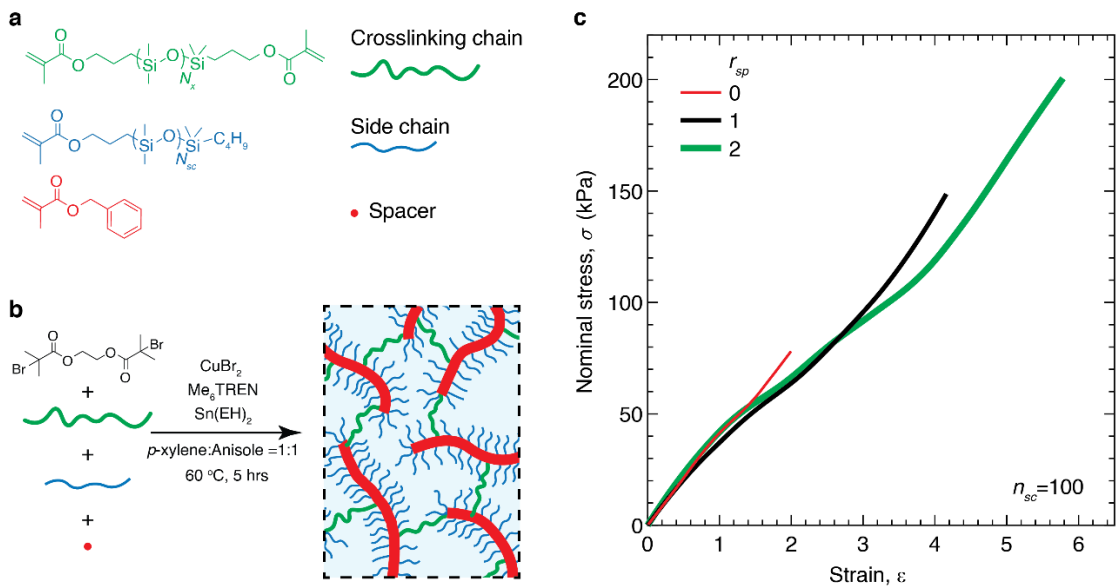

**Extended Data Fig. 9 | Randomly crosslinked fBB polymer networks.**

**a,** We create randomly crosslinked fBB polymer networks by copolymerizing three precursor monomers: (i) a mono-functional linear PDMS polymer (MCR-M11, ~1000 g/mol) as the side chain, (ii) a di-functional linear PDMS polymer (DMS-R18, ~5000 g/mol) as the crosslinking chain, and (iii) BnMA as the spacer monomer. **b,** We denote a randomly crosslinked fBB polymer network using two parameters, [$n_{sc}$, $r_{sp}$], where $n_{sc}$ is the average number of side chains per crosslinking chain. Using the same method for synthesizing fBB polymers, we fix the side chain/crosslinking chain molar ratio at 100:1 ($n_s$=100) and only change the spacer ratio ($r_{sp}$). For a polymer network, we use a mixed solvent, DMF and THF at a volume ratio of 1:1, to wash away any unreacted monomers and catalyst. The polymer networks are dried under vacuum before mechanical test. **c,** Compared to the control network ($r_{sp}$=0), which fractures at a tensile strain $\varepsilon_f$=200% (thin red line), the network with a low spacer ratio ($r_{sp}$=1.0) has nearly the same Young's modulus of 46 kPa but a significantly higher tensile breaking strain $\varepsilon_f$=416% (intermediate thick black line). At an intermediate spacer ratio ($r_{sp}$=2.0), the network has a higher Young's modulus of 58 kPa and a large extensibility of $\varepsilon_f$=578% (thick green line). As expected, unlike the self-assembled fBB polymer networks that exhibit plastic deformation at intermediate spacer ratios ($\varepsilon_f > \varepsilon_{max}$), the randomly crosslinked fBB polymer networks do not exhibit plastic deformation ($\varepsilon_f = \varepsilon_{max}$). These results demonstrate that our strategy of using fBB polymers as network strands to decouple the stiffness and extensibility of unentangled single-network elastomers applies to randomly crosslinked networks. All tensile tests are performed at RT and a fixed strain rate of 0.02/sec.

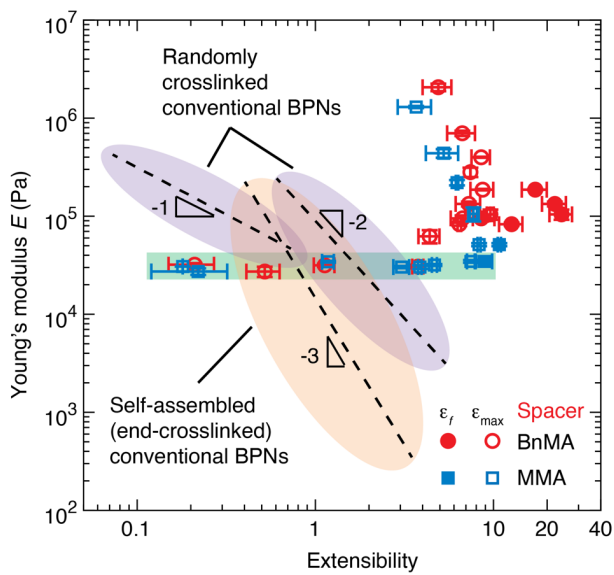

**Extended Data Fig. 10 | Comparison between fBB polymer networks and conventional bottlebrush polymer networks.**

Our fBB polymer networks allow for truly decoupled stiffness and extensibility (empty circles and squares, light green region). By contrast, the stiffness and extensibility of conventional bottlebrush polymer networks are negatively correlated (**Supplementary Table S1**). Because the backbone is strained by the steric repulsion among overlapping side chains, conventional bottlebrush polymers networks (BNPs) are often brittle with low extensibility unless using bottlebrush network strands of extremely high molecular weight (when the networks become very soft). By contrast, our fBB polymer networks can exhibit remarkable extensibility while maintaining constant stiffness. Filled symbols represent the tensile fracture strain $\varepsilon_f$ for fBB polymer networks with intermediate spacer ratios.

**Extended Data Table 1 | Molecular architecture parameters, microstructure, and mechanical properties of fBB polymer networks.**

$n_{sc}$, number of side chains per bottlebrush polymer; $N_l$, DP of each end linear blocks; $r_{sp}$, spacer/side chain number ratio; $f$, volume fraction of the end blocks; PDI, polydispersity index; $G$, shear modulus ($G'$ at 0.1 rad/sec and 20 °C); $E$, Young's modulus measured by tensile tests; $\epsilon_{max}$, the strain at which the nominal tensile stress is maximum ($\sigma_{max}$); $\epsilon_f$, tensile fracture strain; $d$, inter-domain distance of glassy nodules; $D_{bb}$, inter-backbone distance for fBB polymers in both the melt and the self-assembled networks. S1 and S3: samples with BnMA as the spacer; S2: samples with MMA as the spacer. The side chains are linear PDMS of 1000 g/mol and the linear end blocks are PBnMA. NA: not applicable or not measured.

| Sample | fBB middle block | | | Linear-fBB-linear triblock | | | Mechanical properties | | | | | | | | | Microstructure | | fBB melts |
|---|---|---|---|---|---|---|---|---|---|---|---|---|---|---|---|---|---|---|
| | | | | | | | Shear modulus $G$ (kPa) | Young's modulus $E$ (kPa) | | $\epsilon_{max}$ | | $\sigma_{max}$ (kPa) | | $\epsilon_f$ | | fBB networks | | |
| | $n_{sc}$ | $r_{sp}$ | PDI | $N_l$ | $f$ (%) | PDI | | Mean | STDEV | Mean | STDEV | Mean | STDEV | Mean | STDEV | $d$ (nm) | $D_{bb}$ (nm) | $D_{bb}$ (nm) |
| S1-1 | 197 | 0.00 | 1.24 | 36 | 6.1 | 1.27 | 10.7 | 30.3 | 1.2 | NA | NA | 8.2 | 1.9 | 0.21 | 0.06 | 32.22 | 3.54 | 3.59 |
| S1-2 | 203 | 0.34 | 1.18 | 35 | 5.7 | 1.22 | 11.6 | 27.3 | 3.3 | NA | NA | 12.6 | 2.6 | 0.52 | 0.11 | 32.39 | 3.90 | 3.95 |
| S1-3 | 202 | 0.80 | 1.25 | 36 | 5.9 | 1.28 | 12.8 | 31.6 | 1.1 | NA | NA | 35.3 | 10.1 | 1.13 | 0.15 | 32.90 | 4.06 | 3.95 |
| S1-4 | 195 | 1.08 | 1.23 | 40 | 6.7 | 1.27 | 10.7 | 30.6 | 1.1 | NA | NA | 139.1 | 15.3 | 3.77 | 0.27 | 40.28 | 4.21 | 4.03 |
| S1-5 | 200 | 1.40 | 1.38 | 35 | 5.8 | 1.46 | 22.1 | 62.4 | 9.6 | NA | NA | 498.8 | 85.7 | 4.38 | 0.59 | 39.77 | 4.39 | 4.13 |
| S1-6 | 200 | 1.75 | 1.36 | 35 | 5.8 | 1.45 | 36.5 | 95.0 | 6.7 | 6.70 | 0.95 | 292.0 | 18.0 | 8.53 | 1.59 | 40.80 | 4.63 | 4.59 |
| S1-7 | 200 | 1.85 | 1.24 | 35 | 5.8 | 1.27 | 26.0 | 88.5 | 7.5 | 6.43 | 0.15 | 195.9 | 4.3 | 12.63 | 1.89 | 41.61 | 4.77 | 4.52 |
| S1-8 | 200 | 2.00 | 1.48 | 32 | 5.3 | 1.50 | 45.0 | 133.4 | 7.4 | 7.27 | 1.14 | 169.3 | 0.6 | 22.09 | 3.39 | 41.68 | 4.87 | 4.80 |
| S1-9 | 198 | 2.06 | 1.22 | 34 | 5.7 | 1.25 | 38.0 | 104.5 | 12.5 | 9.60 | 0.43 | 162.7 | 2.1 | 24.09 | 3.50 | NA | NA | 4.69 |
| S1-10 | 196 | 2.30 | 1.25 | 34 | 5.8 | 1.27 | 63.6 | 186.6 | 3.2 | 8.69 | 1.27 | 221.7 | 2.3 | 17.15 | 2.74 | 49.47 | 4.77 | 5.11 |
| S1-11 | 207 | 2.39 | 1.35 | 37 | 5.9 | 1.41 | 74.0 | 281.4 | 30.6 | NA | NA | 352.1 | 75.9 | 7.42 | NA | 40.54 | 5.14 | 5.19 |
| S1-12 | 200 | 2.70 | 1.24 | 35 | 5.8 | 1.28 | 132.3 | 399.0 | 4.9 | NA | NA | 694.9 | 73.0 | 8.53 | 1.01 | 47.60 | 5.12 | NA |
| S1-13 | 200 | 2.90 | 1.31 | 35 | 5.8 | 1.36 | 244.7 | 701.3 | 42.6 | NA | NA | 968.3 | 10.0 | 6.68 | 1.19 | 49.47 | 5.39 | NA |
| S1-14 | 200 | 3.46 | 1.48 | 34 | 5.7 | 1.64 | 543.0 | 2067.2 | 181.6 | NA | NA | 1114.7 | 48.5 | 4.90 | 0.90 | 49.09 | 5.51 | 5.71 |
| S2-1 | 200 | 0.00 | 1.20 | 60 | 10.6 | 1.24 | 11.6 | 30.7 | NA | NA | NA | 3.1 | NA | 0.18 | NA | 33.07 | 3.40 | 3.59 |
| S2-2 | 200 | 0.30 | 1.21 | 56 | 10.0 | 1.24 | 9.7 | 27.3 | 2.1 | NA | NA | 6.8 | 1.9 | 0.28 | 0.09 | 36.96 | 3.47 | 3.54 |
| S2-3 | 200 | 0.80 | 1.21 | 60 | 10.6 | 1.27 | 14.7 | 34.3 | 0.7 | NA | NA | 34.4 | 4.7 | 1.24 | 0.11 | 41.89 | 3.70 | 3.72 |

| ID | | | | | | | | | | | | | | | | | | |
|---|---|---|---|---|---|---|---|---|---|---|---|---|---|---|---|---|---|---|
| S2-4 | 198 | 1.24 | 1.31 | 58 | 10.2 | 1.34 | 10.8 | 30.3 | 0.7 | NA | NA | 178.0 | 64.3 | 3.07 | 0.34 | 34.91 | 3.95 | 3.93 |
| S2-5 | 200 | 1.50 | 1.24 | 57 | 10.0 | 1.25 | 10.5 | 30.3 | 1.3 | NA | NA | 210.7 | 5.7 | 3.86 | 0.09 | NA | NA | 4.13 |
| S2-6 | 200 | 1.70 | 1.23 | 60 | 10.6 | 1.29 | 10.6 | 31.8 | 0.7 | NA | NA | 303.1 | 7.1 | 4.66 | 0.15 | NA | NA | 4.21 |
| S2-7 | 200 | 1.88 | 1.38 | 57 | 10.0 | 1.40 | 11.8 | 34.6 | 1.1 | 7.47 | 0.38 | 200.2 | 19.1 | 8.99 | 0.89 | 54.64 | 4.05 | 4.05 |
| S2-8 | 200 | 2.15 | 1.41 | 51 | 8.2 | 1.44 | 18.1 | 51.3 | 1.8 | 8.30 | 0.16 | 129.7 | 5.0 | 10.78 | 0.23 | NA | NA | 4.30 |
| S2-9 | 200 | 2.50 | 1.41 | 60 | 10.6 | 1.50 | 41.4 | 105.7 | 17.3 | NA | NA | 216.1 | 37.3 | 7.32 | 0.80 | NA | NA | 4.36 |
| S2-10 | 200 | 2.77 | 1.43 | 43 | 7.0 | 1.46 | 76.5 | 219.9 | 13.7 | NA | NA | 482.8 | 16.1 | 6.24 | 0.10 | NA | NA | 4.49 |
| S2-11 | 200 | 3.14 | 1.58 | 60 | 10.6 | 1.88 | 142.5 | 438.8 | 44.1 | NA | NA | 550.8 | 106.4 | 5.07 | 0.94 | NA | NA | NA |
| S2-12 | 192 | 3.62 | 1.66 | 50 | 8.4 | 1.90 | 406.9 | 1298.3 | 33.2 | NA | NA | 780.6 | 209.9 | 3.68 | 0.78 | NA | NA | 4.65 |
| S1-7v | 200 | 1.85 | 1.24 | 85 | 13.0 | 1.29 | 38.1 | 91.3 | 1.0 | NA | NA | 327.9 | 8.3 | 4.63 | 0.10 | 44.9 | 4.80 | 4.52 |
| S3-1 | 550 | 0 | 1.30 | 176 | 10.1 | 1.58 | 2.7 | NA | NA | NA | NA | NA | NA | NA | NA | 70.3 | 3.54 | 3.59 |
| S3-2 | 534 | 0.84 | 1.44 | 177 | 10.4 | 1.52 | 3.0 | NA | NA | NA | NA | NA | NA | NA | NA | 95.9 | 4.06 | 3.95 |
| S3-3 | 468 | 0.94 | 1.44 | 142 | 9.7 | 1.65 | 4.0 | NA | NA | NA | NA | NA | NA | NA | NA | 98.1 | NA | NA |
| S3-4 | 360 | 1.04 | 1.40 | 118 | 10.4 | 1.67 | 10.3 | NA | NA | NA | NA | NA | NA | NA | NA | 71.4 | NA | NA |

# Supplementary Information

# A universal strategy for decoupling stiffness and extensibility of polymer networks

Baiqiang Huang[1], Shifeng Nian[1], Li-Heng Cai[1,2,3,*]

**Affiliations:**

[1]Soft Biomatter Laboratory, Department of Materials Science and Engineering, University of Virginia, Charlottesville, VA 22904, USA

[2]Department of Chemical Engineering, University of Virginia, Charlottesville, VA 22904, USA

[3]Department of Biomedical Engineering, University of Virginia, Charlottesville, VA 22904, USA

*Corresponding author. Email: liheng.cai@virginia.edu

**The PDF file includes:**

   Supplementary Notes
   Supplementary Figures
   Supplementary Tables
   Captions for Supplementary Videos
   Supplementary Data
   Supplementary References

**Other Supplementary Materials for this manuscript include the following:**

   Supplementary Videos 1 to 7

# Table of Contents

**Supplementary Notes**

**Supplementary Note 1. Inherent stiffness-extensibility trade-off of conventional unentangled single-network elastomers**

For an unentangled single-network elastomer, the stiffness is about $k_B T$ per volume $V$ of a network strand:

$$G \approx k_B T / V \quad (S1)$$

Here, $G$ is the network shear modulus, which is 1/3 of the network bulk modulus (Young's modulus $E$). The network extensibility, or tensile strain at break $\epsilon_{max}$, is determined by the network strand size,

$$\epsilon_{max} = \frac{L_{max}}{R_0} - 1 \quad (S2)$$

where $L_{max}$ and $R_0$ are the contour length and unperturbed end-to-end distance of the network strand, respectively. The relations for network stiffness [eq. (S1)] and extensibility [eq. (S2)] apply to unentangled single-network elastomers regardless of the type of network strands. The correlation between network stiffness and extensibility follows a power law relation:

$$G \propto (\epsilon_{max})^{-\alpha} \quad (S3)$$

For conventional single-network elastomers, $\alpha$ must be positive. However, the value of $\alpha$ depends on the type of network strands. Below we derive the value of $\alpha$ for network strands made of classical flexible linear polymers and emerging bottlebrush polymers.

**1.1. Flexible linear polymers**

For a network strand made of a flexible linear polymer with $N$ Kuhn monomers of size $b$, $V \approx Nb^3$, $R_0 \approx bN^{1/2}$, and $L_{max} = Nb$. Substituting these relations to eqs. (S1) and (S2) and considering $N \gg 1$, one obtains the stiffness-extensibility relation:

$$G \approx \frac{k_B T}{b^3} (\epsilon_{max})^{-2}, \text{ flexible linear network strand} \quad (S4)$$

**1.2. Conventional bottlebrush polymers**

A bottlebrush polymer consists of a long, linear backbone grafted with many relatively short linear side chains. The degree of polymerization (DP) of the side chain is $N_{sc}$, and the average DP of the backbone section between two neighboring grafting points is $N_g$. The number of side chains per bottlebrush polymer, $n_{sc}$, is much larger than the DP of the side chain, $n_{sc} \gg N_{sc}$, such that the effects of extra space near the two ends of a grafted polymer on the polymer conformation can be ignored. We use $l$ and $v$, respectively, to denote the length of a main-chain bond and the volume of a chemical monomers. Throughout the theory, we denote the side chains and the bottlebrush backbone with '*sc*' and '*bb*' as subscripts or superscripts, respectively. Additionally, we provide simplified expressions that disregard the difference in polymer physics parameters between the side chains and the bottlebrush backbone polymer. This simplification aids in distilling the essential physical pictures for the bottlebrush polymers.

In a conventional bottlebrush polymer, the side chains and the backbone are assumed to be compatible. An example is a poly(dimethyl siloxane) (PDMS) bottlebrush in which the bottlebrush backbone and the side chains are both linear PDMS[1]. The molecular structure of a conventional bottlebrush polymer is largely determined by how to pack the densely grafted side chains into the limited space surrounding the bottlebrush backbone. In brief, at relatively low grafting density when the side chains are far apart from each other, the side chain adopts an unperturbed Gaussian conformation. The mean-square end-to-end distance of an unperturbed side chain is:

$$R_{sc,0}^2 \approx b_{sc} L_{max}^{sc} \approx b_{sc} l_{sc} N_{sc} \approx bl N_{sc} \tag{S5}$$

As the grafting density ($1/N_g$) increases, the side chains from the same bottlebrush polymer can overlap. Yet, they will not completely fill the volume pervaded by one side chain of unperturbed size, $V_P \approx R_{sc,0}^3$, until a crossover grafting density, $1/N_g^*$:

$$R_{sc,0}^3 \approx N_{sc} v_{sc} \left(\frac{g}{N_g^*}\right) \tag{S6}$$

Here, $g$ is the number of monomers in a backbone section passing through the pervaded volume, and $g/N_g^*$ corresponds to the number of side chains within $V_P$. At $1/N_g^*$, the side chains are not crowded and there is no steric repulsion among the side chains. Thus, the backbone polymer is not extended and adopts an unperturbed Gaussian conformation. Because the size of the backbone section with $g$ monomers is about that of the side chain, $(g l_{bb} b_{bb})^{1/2} \approx R_{sc,0}$, eq. (S6) can be rewritten as:

$$N_g^* \approx N_{sc}^{\frac{1}{2}} \frac{v_{sc}}{(b_{sc} l_{sc})^{\frac{1}{2}} (l_{bb} b_{bb})} \tag{S7}$$

As the grafting density further increases ($1/N_g > 1/N_g^*$), the space near the bottlebrush backbone is not enough to accommodate all the side chains from the same bottlebrush polymer if the backbone remains an unperturbed Gaussian conformation. To avoid the crowding of side chains, the backbone polymer must extend. This process ensures a constant distance in space between two neighboring grafting sites. In doing so, the side chains remain an unperturbed Gaussian conformation of size $R_{sc,0}$ [eq. (S5)]. By contrast, the backbone polymer continues to extend as the grafting density increases. However, the extension of the backbone cannot continue forever; instead, it will stop at certain grafting density $1/N_g^{**}$, at which the spacer segment between two neighboring grafting sites is stretched to its maximum contour length.

At very high grafting density ($1/N_g > 1/N_g^{**}$), there is no other way for the side chains to avoid crowding except by radially extending away from the backbone polymer. The side chain size $R_{sc}$ is determined by volume conservation: a side chain of volume $v_{sc}N_{sc}$ fills the cylindrical space $R_{sc}^2 N_g l_{bb}$ surrounding the backbone polymer: $R_{sc}^2 N_g l_{bb} \approx v_{sc} N_{sc}$. This gives:

$$R_{sc} \approx \left(\frac{v_{sc} N_{sc}}{N_g l_{bb}}\right)^{\frac{1}{2}} \approx R_{sc,0} \left(\frac{N_g^{**}}{N_g}\right)^{1/2}, \text{ for a densely grafted bottlebrush } (1 < N_g < N_g^{**}) \quad (S8)$$

And $1/N_g^{**}$ is the crossover grafting density above which the side chains start to extend:

$$N_g^{**} \approx \frac{v_{sc}}{b_{sc} l_{sc} l_{bb}} \quad (S9)$$

Assuming the backbone and the side chains are compatible with the Flory Huggins parameter $\chi = 0$, the value of $N_g^{**}$ for PDMS bottlebrush polymers with 1000 g/mol side chains and a methacrylate-based backbone is approximately 1.5.[2]

### 1.2.1. Inter-backbone distance

In the melt, a bottlebrush molecule ($N_g < N_g^*$) is effectively a 'fat' linear polymer with a cross-section about twice the side chain size (**Fig. S1a**). The diameter of a bottlebrush polymer approximately equals the inter-backbone distance $D_{bb}$ between neighboring bottlebrush polymers (**Fig. S1b**):

$$D_{bb} \approx 2R_{sc} \propto \begin{cases} (N_{sc})^{1/2}(N_g)^{-1/2}, & \text{for } 1 < N_g < N_g^{**} \\ (N_{sc})^{1/2}, & \text{for } N_g^{**} < N_g < N_g^* \end{cases} \quad (S10)$$

Thus, for the melt of conventional bottlebrush polymers, $D_{bb}$ decreases with decrease of grafting density ($1/N_g$) by a power of 1/2 (black dashed line, **Fig. 2g**). This understanding has been pointed out in pioneering experimental works[1,3], confirmed by simulations[4], and detailed in a seminal work by Rubinstein Lab[5].

### 1.2.2. Unperturbed bottlebrush network strands

The network stiffness is inversely proportional to the volume of a network strand. The contribution to the volume of the bottlebrush polymer is predominately from the side chains, and therefore: $V = (n_{sc}M_{sc} + N_g n_{sc} m_{bb})/(\rho N_{Av}) \approx n_{sc} M_{sc}/(\rho N_{Av})$, where the molar mass of a side chain $M_{sc}$ is much larger than that of backbone or spacer monomers ($m_{bb}$), $\rho$ is the density of the polymer, and $N_{Av}$ is the Avogadro number. As a result, the number of side chains per bottlebrush polymer is related to the network shear modulus by:

$$G \approx \frac{k_B T \rho N_{Av}}{n_{sc} M_{sc}} \propto (n_{sc})^{-1} \quad (S11)$$

To understand the network extensibility, one needs to determine the relation between the size and the molecular architecture parameters of a bottlebrush network strand. The contour length of the bottlebrush backbone is:

$$L_{max}^{bb} = N_g n_{sc} l_{bb} \tag{S12}$$

The bottlebrush polymer is effectively a 'fat' semiflexible polymer, the end-to-end distance of which can be determined using the worm-like-chain model[6]:

$$R_{bb,0}^2 = 2l_p L_{max}^{bb} - 2l_p^2 \left(1 - \exp\left(-\frac{L_{max}^{bb}}{l_p}\right)\right) \tag{S13}$$

in which $l_p \approx D_{bb}$ is the persistence length of the bottlebrush polymer.

The end-to-end distance of the bottlebrush polymer can be simplified into two cases depending on flexibility of a bottlebrush, which is defined as[7]:

$$\kappa \equiv L_{max}^{bb}/(2l_p) \tag{S14}$$

For a relatively stiff bottlebrush with $\kappa < 1$, eq. (S13) can be approximated as:

$$R_{0,bb} \approx L_{max}^{bb}\left(1 - \frac{1}{6}\frac{L_{max}^{bb}}{l_p}\right), \text{ for } \kappa < 1 \tag{S15}$$

For a flexible bottlebrush with $\kappa \gg 1$, eq. (S13) can be approximated as:

$$R_{0,bb} \approx \left(2l_p L_{max}^{bb}\right)^{1/2}, \text{ for } \kappa \gg 1 \tag{S16}$$

Thus, the network extensibility [eq. (S2)]:

$$\epsilon_{max} = \frac{L_{max}^{bb}}{R_{bb,0}} - 1 \propto \begin{cases} L_{max}^{bb}/l_p \sim n_{sc}, & \text{for } \kappa < 1 \\ \left(L_{max}/l_p\right)^{1/2} \sim n_{sc}^{\frac{1}{2}}, & \text{for } \kappa \gg 1 \end{cases} \tag{S17}$$

Recalling the network stiffness [eq. (S11)], the stiffness-extensibility tradeoff of conventional bottlebrush polymer networks is:

$$G \propto \begin{cases} (\epsilon_{max})^{-1}, & \kappa < 1 \\ (\epsilon_{max})^{-2}, & \kappa \gg 1 \end{cases} \tag{S18}$$

These relationships were demonstrated both experimentally and theoretically in our previous work[8].

### 1.2.3. Pre-strained bottlebrush network strands

In a network self-assembled by linear-bottlebrush-linear (LBBL) triblock copolymers, the bottlebrush backbone is pre-strained to balance interfacial repulsion between incompatible bottlebrush and end block domains.

To understand the stiffness-extensibility trade-off of networks with pre-strained bottlebrush polymer network strands, we consider a sphere microstructure self-assembled by a LBBL triblock copolymer. The volume fraction of end blocks is $f$, the radius of a spherical domain is $r$, and the distance between the centers of two neighboring spherical domains is $d$.

The interfacial free energy density is the product of interfacial tension $\gamma$ and the interfacial area $A$ per unit volume,

$$F_{int} \approx \gamma A \tag{S19}$$

Here, the interfacial area $A$ per unit volume is the product of the number density of spherical domains, $1/d^3$, and the surface area of a spherical domain, $4\pi r^2$: $A \approx 4\pi r^2/d^3$. Considering that the density of different blocks is nearly the same, the domain radius and distance is related by the volume fraction, $f \approx \left(\frac{2r}{d}\right)^3$. Thus, the interfacial free energy density can be rewritten as:

$$F_{int} \approx \gamma f/r \tag{S20}$$

which increases linearly with $f$ but is inverse to the radius of the spherical domain.

The entropic free energy is associated with stretching the bottlebrush backbone. For a semiflexible bottlebrush polymer ($\kappa \approx 1$), at unperturbed state the bottlebrush backbone is already stretched close to its contour length. As a result, the corresponding bottlebrush polymer networks are very brittle with low fracture strain. Therefore, we consider a flexible bottlebrush ($\kappa \gg 1$), whose entropic free energy obeys the same form as Gaussian coil. The end-to-end distance of the bottlebrush polymer is:

$$R_{bb} \approx d - 2r \approx 2r\left(f^{-\frac{1}{3}} - 1\right) = d\left(1 - f^{\frac{1}{3}}\right) \approx rf^{-\frac{2}{3}} \tag{S21}$$

The entropic free energy associated with stretching one bottlebrush polymer is:

$$\frac{F_{ent}^0}{k_B T} \approx \frac{R_{bb}^2}{R_{bb,0}^2} \approx \frac{r^2}{R_{bb,0}^2} f^{-\frac{2}{3}}, \qquad \text{for } f \ll 1 \tag{S22}$$

The volume of a bottlebrush polymer is $V_c = pR_{bb,0}^2 \approx 2pl_p L_{max}^{bb}$ [eq. (S16)], where $p$ is the packing length of the bottlebrush polymer. Thus, the entropic free energy density due to the stretching of the bottlebrush backbone is:

$$F_{ent} = \frac{F_{ent}^0}{V_c} \approx k_B T f^{-2/3} \frac{1}{p l_p^2} \frac{r^2}{\left(L_{max}^{bb}\right)^2} \tag{S23}$$

Using eqs. (S20) and (S23), the density of total free energy is:

$$\frac{F_{tot}}{k_B T} \approx \frac{\gamma}{k_B T} \frac{f}{r} + \frac{f^{-2/3}}{p l_p^2} \frac{r^2}{L_{max}^2} \tag{S24}$$

Minimizing the density of total free energy by setting $\left.\frac{\partial}{\partial r}\left(\frac{F_{tot}}{k_B T}\right)\right|_{r=r^*} = 0$ gives the radius of the spherical domain in equilibrium:

$$r^* \approx h^{-2/3} f^{5/9} p^{1/3} l_p^{4/3} \kappa^{2/3} \tag{S25}$$

where $h \equiv (k_B T/\gamma)^{1/2}$ is related to the interfacial thickness and has the scale of a monomer length. The effective Kuhn monomer of a bottlebrush is nearly spherical, such that the packing length is comparable to the persistence length ($p \approx l_p$). Thus, the equilibrium size of the bottlebrush polymer in the self-assembled network is:

$$R_{bb,e} \approx f^{-\frac{1}{3}} r^* \approx h^{-2/3} f^{2/9} l_p^{5/3} \kappa^{2/3} \tag{S26}$$

The above scaling theory was documented in our previous work[9]. A similar scaling theory was also documented in the supplementary information of a previous publication[10].

Recalling the definition of $\kappa$ [eq. (S14)], the network extensibility can be rewritten as:

$$\epsilon_{max} = \frac{L_{max}^{bb}}{R_{bb,e}} - 1 \propto h^{\frac{2}{3}} f^{-\frac{2}{9}} (L_{max}^{bb})^{\frac{1}{3}}/l_p \sim (n_{sc})^{1/3}, \quad \text{for } \kappa \gg 1 \tag{S27}$$

Recalling that the network shear modulus is inversely proportional to the number side chains per bottlebrush polymer [eq. (S11)], one obtains the stiffness-extensibility relation for conventional bottlebrush polymer networks self-assembled by LBBL triblock polymers:

$$G \propto (\epsilon_{max})^{-3}, \text{ pre-strained conventional bottlebrush network strand} \tag{S28}$$

**Supplementary Note 2. Inter-backbone distance of foldable bottlebrush polymers in melts**

The theory for the inter-backbone distance of foldable bottlebrush polymers in melts has been documented in our previous publication[2]. Here, we briefly review the theory and describe essential physical pictures associated with foldable bottlebrush polymers.

We consider the strongly segregated case characterized by sharp interface between distinct domains. To achieve this, it is necessary for all the grafting sites to reside on the surface rather than the interior of the cylindrical core formed by the folded backbone polymer. To determine the diameter of the bottlebrush polymer, we consider the free energy associated with its folded structure. The steric repulsion among the strongly overlapping side chains tends to elongate the cylindrical core. Conversely, the backbone polymer tends to collapse into a cylinder of a larger diameter to minimize the interfacial area between the side chains and the backbone polymer (**Fig. 1c**). Thus, the total free energy $F_{tot}$ of a bottlebrush polymer can be expressed as:

$$F_{tot} = F_{sc} + F_{bb} + F_{int} \approx F_{sc} + F_{int} \tag{S29}$$

Here, $F_{int}$ represents the interfacial free energy between the incompatible backbone and side chains. $F_{sc}$ and $F_{bb}$ correspond to the entropic free energies resulting from the stretching of the side chains and the backbone polymer, respectively. However, even in a densely grafted conventional bottlebrush polymer, the stretching of the bottlebrush backbone is comparable to that of one side chain, as detailed in our previous publication[2] and by others[5]. Moreover, there are many side chains in each bottlebrush polymer. Thus, compared to $F_{sc}$, $F_{bb}$ is much smaller and can be neglected.

The interfacial free energy between the side chains and the bottlebrush backbone is:

$$\frac{F_{int}}{k_B T} \approx r_c L_c \left(\frac{\chi}{a_0}\right) \tag{S30}$$

Here, $r_c L_c$ is the total surface area of the cylindrical core, and $\chi/a_0$ is the interfacial free energy per contact area $a_0$ of a Kuhn segment. The length $L_c$ of the cylinder is given by the volume conservation of the backbone polymer: $n_{sc} N_g v_{bb} \approx r_c^2 L_c$. Thus, eq. (S30) can be re-written as:

$$\frac{F_{int}}{k_B T} \approx \left(\frac{\chi}{a_0}\right) \frac{n_{sc} N_g v_{bb}}{r_c} \tag{S31}$$

To calculate the entropic free energy of side chains, we consider a section of the cylindrical core with the length of $b_{bb}$. The number of side chains grafted to this cylindrical section is $b_{bb} r_c^2 / (N_g v_{bb})$, in which $b_{bb} r_c^2$ is the volume of the section of the cylindrical core, and $N_g v_{bb}$ is the volume of backbone section between two neighboring grafting sites. Surrounding the cylindrical core, these side chains fill the space with a volume of $[(R_{sc} + r_c)^2 - r_c^2] b_{bb}$:

$$[(R_{sc} + r_c)^2 - r_c^2] b_{bb} \approx \frac{b_{bb} r_c^2}{N_g v_{bb}} N_{sc} v_{sc} \tag{S32}$$

Assuming that the side chain size is much larger than the diameter of the cylindrical core, $R_{sc} \gg r_c$, the effects of cylindrical core on the volume available to the side chains can be neglected:

$$(R_{sc} + r_c)^2 - r_c^2 \approx R_{sc}^2, \text{ for } R_{sc} \gg r_c \tag{S33}$$

Re-writing eq. (S32) gives the side chain size:

$$R_{sc} \approx r_c \left(\frac{N_{sc}v_{sc}}{N_g v_{bb}}\right)^{\frac{1}{2}}, \text{ for } R_{sc} \gg r_c \tag{S34}$$

The entropic free energy due to the stretching of $n_{sc}$ side chains of the bottlebrush polymer is:

$$\frac{F_{ent}}{k_B T} \approx n_{sc} \left(\frac{R_{sc}}{R_{sc,0}}\right)^2 \approx n_{sc} r_c^2 \frac{v_{sc}}{N_g v_{bb} b_{sc} l_{sc}} \tag{S35}$$

Using eqs. (S29), (S31), and (S35), one obtains the total free energy of the bottlebrush polymer:

$$\frac{F_{tot}}{k_B T} = \frac{F_{ent}}{k_B T} + \frac{F_{int}}{k_B T} \approx \left(\frac{\chi}{a_0}\right) \frac{n_{sc} N_g v_{bb}}{r_c} + n_{sc} r_c^2 \frac{v_{sc}}{N_g v_{bb} b_{sc} l_{sc}} \tag{S36}$$

Minimizing the total free energy gives the equilibrium cross-section size of the cylindrical core:

$$r_{c,e} \approx \left(\frac{\chi}{a_0}\right)^{\frac{1}{3}} \left(\frac{v_{bb}^2 b_{sc} l_{sc}}{v_{sc}}\right)^{\frac{1}{3}} N_g^{\frac{2}{3}} \equiv r_0 N_g^{2/3} \tag{S37}$$

in which $r_0 \equiv \left(\frac{\chi}{a_0}\right)^{\frac{1}{3}} \left(\frac{v_{bb}^2 b_{sc} l_{sc}}{v_{sc}}\right)^{\frac{1}{3}}$ is a length scale determined by the Flory-Huggins interaction parameter $\chi$. Substituting eq. (S37) into eq. (S34) gives the equilibrium size of the side chain:

$$R_{sc,e} \approx r_{c,e} \left(\frac{N_{sc} v_{sc}}{N_g v_{bb}}\right)^{\frac{1}{2}} \approx r_0 \left(\frac{v_{sc}}{v_{bb}}\right)^{\frac{1}{2}} N_{sc}^{\frac{1}{2}} N_g^{\frac{1}{6}}, \text{ for } 1 < N_g < N_{k,bb} \tag{S38}$$

Here, $N_{k,bb}$ is the number of chemical monomers per Kuhn segment of the bottlebrush backbone. Consequently, the inter-backbone distance becomes [eqs. (S37) and (S38)]:

$$D_{bb} \approx R_{sc,e} + r_{c,e} \approx r_0 \left[\left(\frac{v_{sc}}{v_{bb}}\right)^{\frac{1}{2}} N_{sc}^{\frac{1}{2}} N_g^{\frac{1}{6}} + N_g^{\frac{2}{3}}\right] \propto N_{sc}^{\frac{1}{2}} N_g^{\frac{1}{6}} + N_g^{\frac{2}{3}}, \text{ for } 1 < N_g < N_{k,bb} \tag{S39}$$

Note that our theory is based on the minimization of free energy for individual foldable bottlebrush polymers, which requires that the polymers are in a thermodynamic equilibrium. This would not be an issue for foldable bottlebrush polymers with relatively low spacer ratios ($N_g < 3$), at which the glass transition temperature $T_g$ of the foldable bottlebrush polymers is below room temperature (RT) (**Extended Data Fig. 6**). At relatively high spacer ratios ($N_g > 3$), the $T_g$ of the foldable bottlebrush polymers is higher than RT, such that the conformation of a foldable bottlebrush polymer may be trapped in a metastable state at RT. To this end, we use a combination of solvent and thermal annealing to prepare both foldable bottlebrush polymer melts and networks to ensure that they are in an equilibrated state.

Our theory [eq. (S39)] predicts that the inter-backbone distance $D_{bb}$ increases with the decrease of grafting density ($1/N_g$). At very high grafting density (small $N_g$) or with long side chains (large $N_{sc}$), the $D_{bb}$ scales with $N_g$ by a power of 1/6: $D_{bb} \sim N_g^{1/6}$. However, the window for this regime is very small, because the diameter of the cylindrical core $r_c$ becomes noticeable at relatively large $N_g$; this makes the approximation in eq. (S33) inappropriate. Nevertheless, $D_{bb}$ is expected to increase with $N_g$ but with an exponent intermediate between 1/6 and 2/3. These predications were verified by experiments in our previous work for the melt of foldable bottlebrush polymers (filled squares, **Fig. 2g**)[2].

**Supplementary Figures**

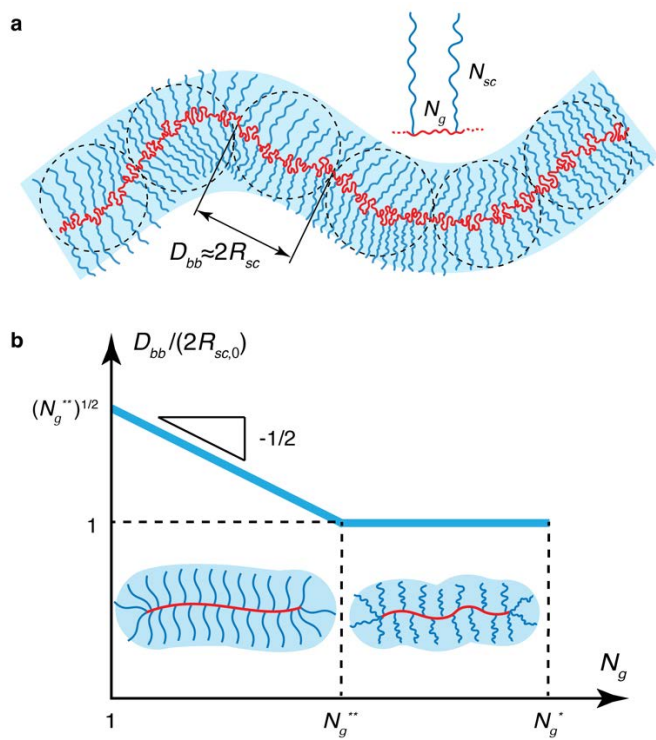

**Fig. S1. Molecular structure of conventional bottlebrush polymers in melts.**
**a,** A bottlebrush molecule is effectively a 'fat' linear polymer with a diameter about twice the end-to-end distance of a side chain, $R_{sc}$. **b,** In the melt of bottlebrush polymers, the average distance $D_{bb}$ between the backbones of two neighboring bottlebrush molecules is about the bottlebrush diameter ($D_{bb} \approx 2R_{sc}$) [eq. (S10)]. The prevailing understanding is that $D_{bb}$ decreases with the increase of $N_g$ by a power of -1/2 up to $N_g^{**}$ [eq. (S9)], above which the side chain adopts an unperturbed Gaussian conformation with the size of $R_{sc,0}$ [eq. (S5)] (blue line). For $N_g < N_g^*$ [eq. (S7)], the side chains from the same bottlebrush completely fill the cylindrical space surrounding the bottlebrush backbone.

**Supplementary Tables**

Table S1. Mechanical properties of conventional bottlebrush polymer networks.

| Network topology | Chemical specie of the bottlebrush side chain | Young's modulus $E$ (kPa) | Extensibility ($\epsilon_{max} = \epsilon_f$) | References |
|---|---|---|---|---|
| End-crosslinked (Self-assembled by linear-bottlebrush-linear triblock copolymers) | Poly(dimethyl siloxane) (PDMS) | 4.5 | 1.7 | Ref. [11] |
| | | 5.4 | 1.5 | |
| | | 5.7 | 1.0 | |
| | | 6.6 | 0.8 | |
| | | 3.0 | 2.7 | |
| | | 3.6 | 2.8 | |
| | | 4.5 | 2.1 | |
| | | 5.1 | 1.5 | |
| | | 3.3 | 2.8 | |
| | | 3.6 | 3.4 | |
| | | 4.8 | 2.3 | |
| | | 6.3 | 1.1 | |
| | | 3.3 | 3.5 | |
| | | 3.9 | 2.7 | |
| | | 3.9 | 2.2 | |
| | | 4.2 | 1.9 | |
| | | 2.1 | 2.1 | |
| | | 1.5 | 1.8 | |
| | | 1.8 | 1.4 | |
| | | 1.8 | 0.8 | |
| | | 1.5 | 1.6 | |
| | | 2.1 | 1.0 | |
| | | 2.4 | 1.5 | |
| | | 1.5 | 1.1 | |
| | | 2.4 | 0.9 | |
| | | 3.3 | 2.5 | Ref. [10] |
| | | 8.4 | 1.9 | |
| | | 30.0 | 0.9 | |
| | | 18.6 | 1.1 | |
| | | 40.5 | 0.5 | |
| | | 2.4 | 1.6 | |
| | | 3.9 | 1.5 | |
| | | 8.7 | 0.6 | |
| | | 21.3 | 0.4 | |
| | | 2.7 | 3.5 | Ref. [12] |
| | | 13.2 | 2.5 | |
| | | 18.3 | 1.7 | |
| | | 0.6 | 3.5 | |
| | | 0.2 | 5.6 | |
| | Poly(butyl acrylate) | 18.3 | 0.5 | Ref. [13] |
| | | 24.8 | 0.6 | |
| | | 55.7 | 0.4 | |
| | | 13.3 | 0.5 | |
| | | 106.2 | 0.6 | |
| | | 130.0 | 0.7 | |
| | | 155.7 | 0.4 | |
| | | 11.2 | 0.6 | |
| | | 20.4 | 0.6 | |
| | | 22.3 | 0.5 | |
| | | 26.3 | 0.5 | |
| Randomly crosslinked | PDMS | 54.6 | 0.5 | Ref. [8] |
| | | 396.0 | 0.1 | |
| | | 252.3 | 0.2 | |
| | | 104.1 | 0.3 | |
| | | 61.2 | 0.5 | |
| | | 35.4 | 1.0 | |
| | | 26.8 | 1.1 | |
| | | 22.3 | 1.2 | |
| | | 20.3 | 1.4 | |
| | | 17.3 | 1.5 | |
| | | 36.0 | 0.8 | |
| | | 56.1 | 0.5 | |
| | | 135.3 | 0.2 | |
| | | 5.6 | 4.0 | |
| | | 266.7 | 0.1 | |
| | | 20.3 | 1.4 | |
| | | 26.0 | 1.0 | |
| | Poly(butyl acrylate) | 120.0 | 1.0 | Ref. [14] |
| | | 81.2 | 2.2 | |
| | | 45.1 | 3.8 | |
| | | 29.2 | 1.2 | |
| | | 16.1 | 1.8 | |
| | | 6.9 | 3.2 | |
| | | 30.3 | 1.0 | |
| | | 15.0 | 1.6 | |
| | | 6.7 | 2.8 | |
| | | 12.7 | 1.0 | |
| | | 5.5 | 2.0 | |
| | | 1.8 | 3.6 | |
| | Poly(4-methylcaprolactone) | 40.0 | 3.3 | Ref. [15] |
| | poly(ε-caprolactone-co-L-lactide) | 51.0 | 3.2 | Ref. [16] |
| | | 42.0 | 4.2 | |
| | Poly(butyl acrylate) | 42.4 | 1.4 | Ref. [17] |
| | | 20.7 | 2.0 | |
| | | 8.7 | 3.5 | |
| | | 87.5 | 0.4 | |
| | | 38.2 | 1.5 | |
| | | 18.6 | 2.9 | |
| | | 123.2 | 1.0 | |
| | | 69.0 | 1.4 | |
| | | 37.8 | 2.6 | |
| | | 33.9 | 3.5 | |
| | | 185.8 | 1.0 | |
| | | 112.4 | 2.4 | |
| | | 44.2 | 1.6 | |
| | | 18.5 | 2.0 | |
| | | 7.5 | 3.2 | |
| | | 76.9 | 1.0 | |
| | | 32.2 | 2.0 | |
| | | 9.9 | 3.0 | |
| | | 189.3 | 0.7 | |
| | | 99.2 | 1.0 | |
| | | 39.0 | 2.2 | |
| | | 18.4 | 1.0 | |
| | | 6.6 | 2.6 | |
| | | 2.0 | 3.9 | |
| | | 7.4 | 1.8 | |
| | | 187.2 | 0.6 | |
| | | 99.9 | 1.2 | |
| | | 45.9 | 2.2 | |
| | Polyisobutylene | 15.7 | 1.0 | Ref. [17] |
| | | 6.7 | 1.5 | |
| | | 4.9 | 3.0 | |
| | | 4.0 | 4.0 | |
| | | 13.5 | 2.7 | |
| | | 29.6 | 3.0 | |
| | | 77.6 | 2.4 | |
| | | 136.0 | 1.2 | |
| | | 69.3 | 2.2 | |
| | | 59.3 | 2.4 | |
| | | 43.7 | 1.5 | |
| | | 24.2 | 2.2 | |
| | | 9.9 | 2.5 | |
| | | 3.7 | 3.0 | |

| | PDMS | 8.2 | 4.2 | Ref. [17] |
|---|---|---|---|---|
| | | 3.1 | 5.6 | |
| | | 25.5 | 5.0 | |
| | | 150.0 | 3.5 | |
| | | 44.2 | 5.5 | |
| | | 5.6 | 1.6 | |
| | | 2.0 | 1.4 | |
| | | 1.7 | 1.3 | |
| | | 0.8 | 1.9 | |

**Captions for Supplementary Videos**

Video 1. Uniaxial tensile test of a self-assembled (end-crosslinked) foldable bottlebrush polymer network with a low spacer ratio (BnMA spacer, $r_{sp}$=0.34, sample S1-2).

Video 2. Uniaxial tensile test of a self-assembled (end-crosslinked) foldable bottlebrush polymer network with a spacer ratio close to the upper limit of Regime I (BnMA spacer, $r_{sp}$=1.40, sample S1-5).

Video 3. Uniaxial tensile test of a self-assembled (end-crosslinked) foldable bottlebrush polymer network with an intermediate spacer ratio (BnMA spacer, $r_{sp}$=2.00, sample S1-8).

Video 4. Uniaxial tensile test of a self-assembled (end-crosslinked) foldable bottlebrush polymer network with a high spacer ratio (BnMA spacer, $r_{sp}$=2.06, sample S1-9).

Video 5. Uniaxial tensile test of a self-assembled (end-crosslinked) foldable bottlebrush polymer network with a high spacer ratio (BnMA spacer, $r_{sp}$=2.70, sample S1-12).

Video 6. Uniaxial tensile test of a randomly crosslinked bottlebrush polymer network without spacer monomers ($r_{sp}$=0, $n_{sc}$=100).

Video 7. Uniaxial tensile test of a randomly crosslinked foldable bottlebrush polymer network with an intermediate spacer ratio (BnMA spacer, $r_{sp}$=2.00, $n_{sc}$=100).

**Supplementary Data | ¹H NMR spectra for all polymers**

**Data Set 1: Foldable bottlebrush polymers and networks with BnMA spacer**

1.1. Foldable bottlebrush middle block

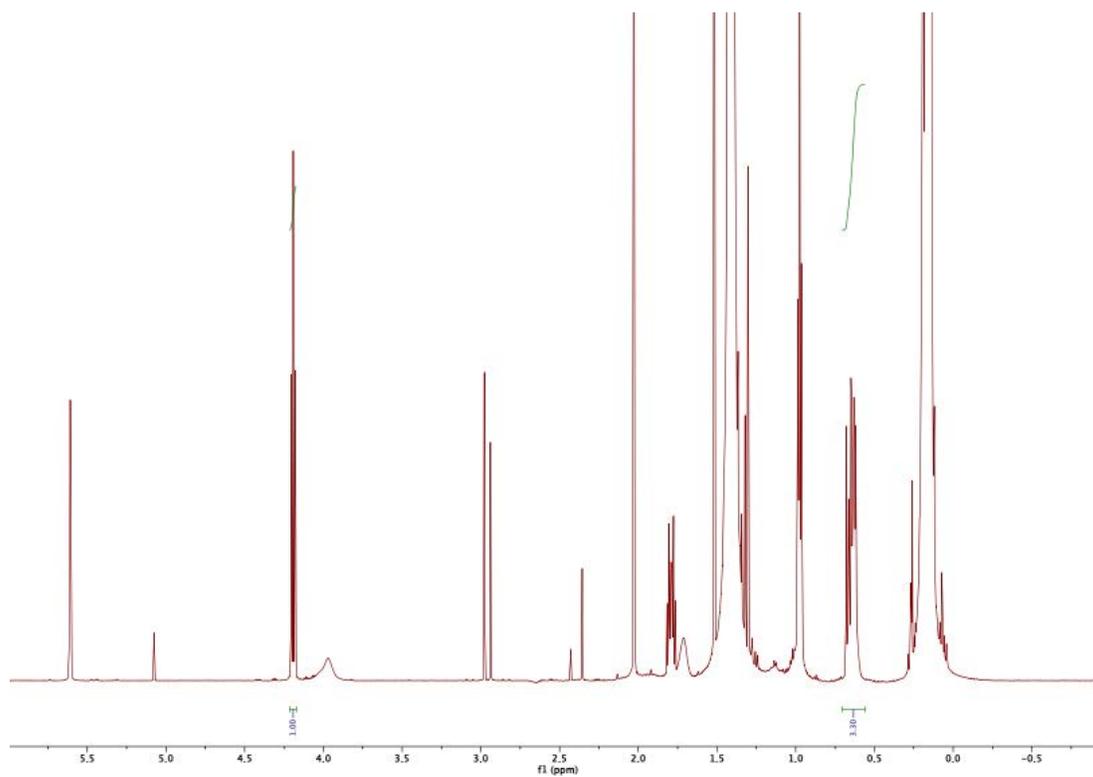

**Fig. S2.** ¹H NMR data of raw mixture of ARGET ATRP of $(PDMS^1)_{197}$. Conversion = $(1 - 1.000 \times 2 / 3.30) \times 100\% = 39.4\%$. The number of PDMS side chains is 500×39.4%=197.

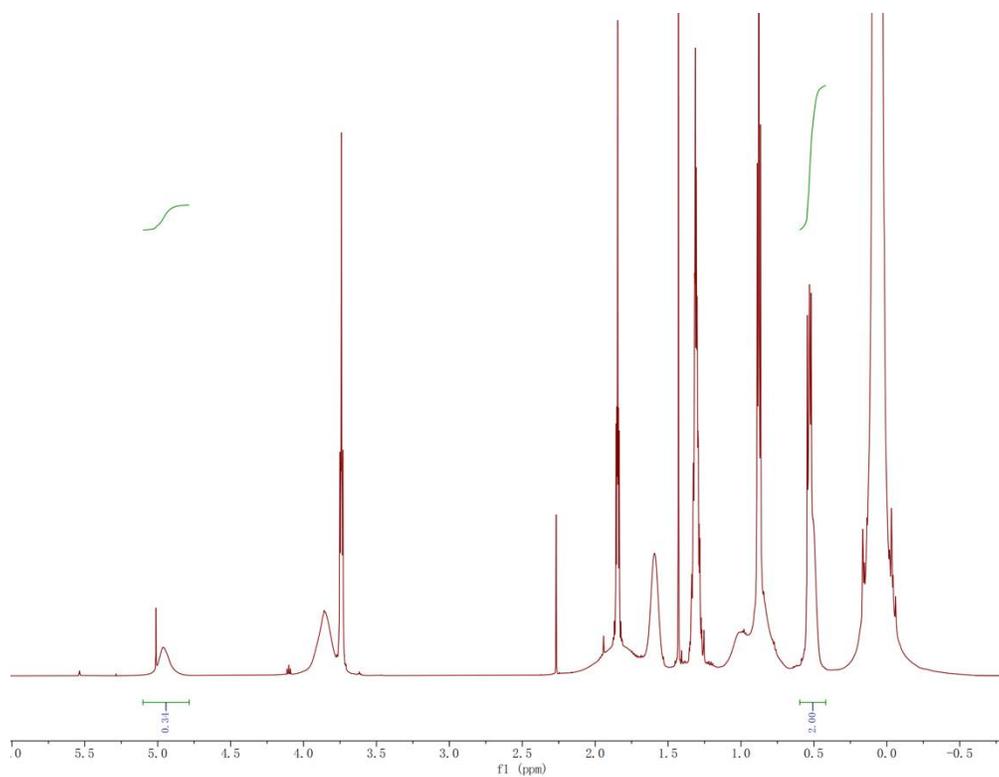

**Fig. S3.** $^1$H NMR of (BnMA$_{0.34}$-$r$-PDMS$^1$)$_{203}$. The number of PDMS side chains is 203, the spacer ratio is 0.34, and the number of BnMA monomers is 203×0.34=69.

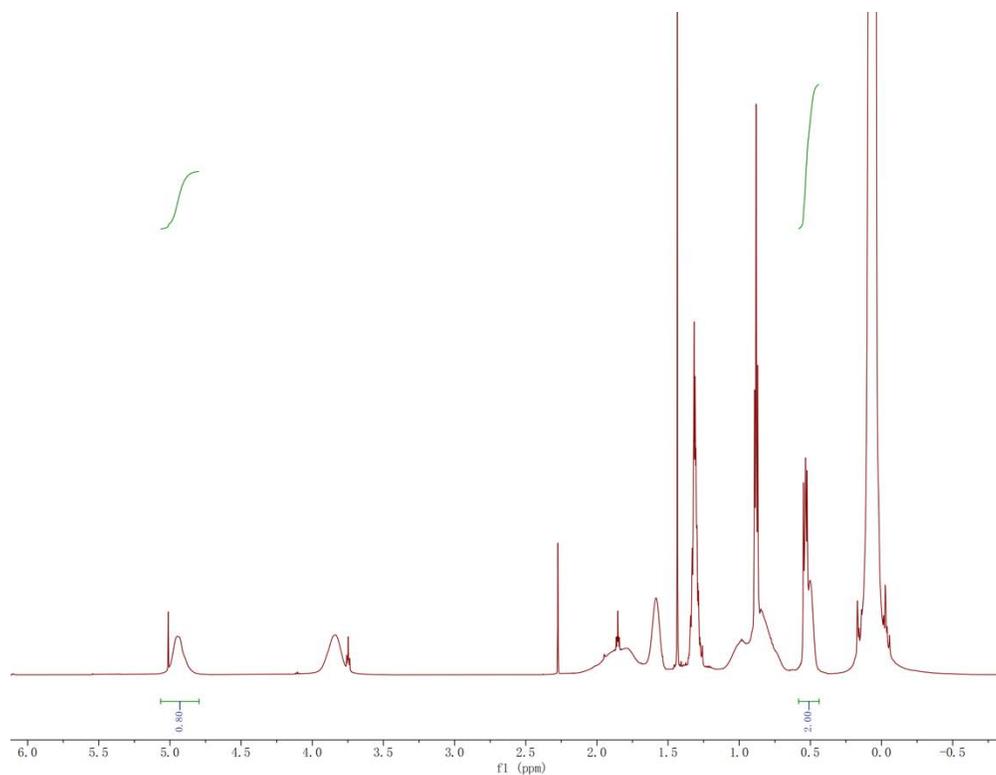

**Fig. S4.** $^1$H NMR of (BnMA$_{0.80}$-$r$-PDMS$^1$)$_{202}$. The number of PDMS side chains is 202, the spacer ratio is 0.80, and the number of BnMA monomers is 202×0.80=162.

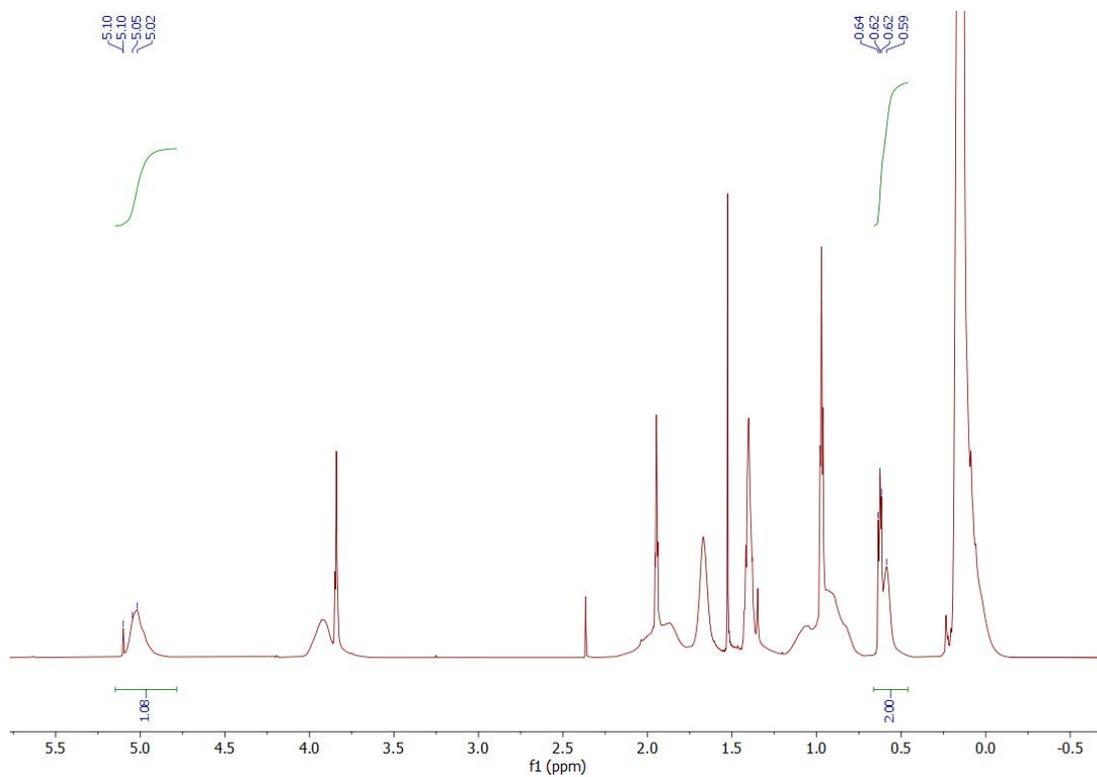

**Fig. S5.** $^1$H NMR of (BnMA$_{1.08}$-$r$-PDMS$^1$)$_{195}$. The number of PDMS side chains is 195, the spacer ratio is 1.08, and the number of BnMA monomers is 195×1.08=211.

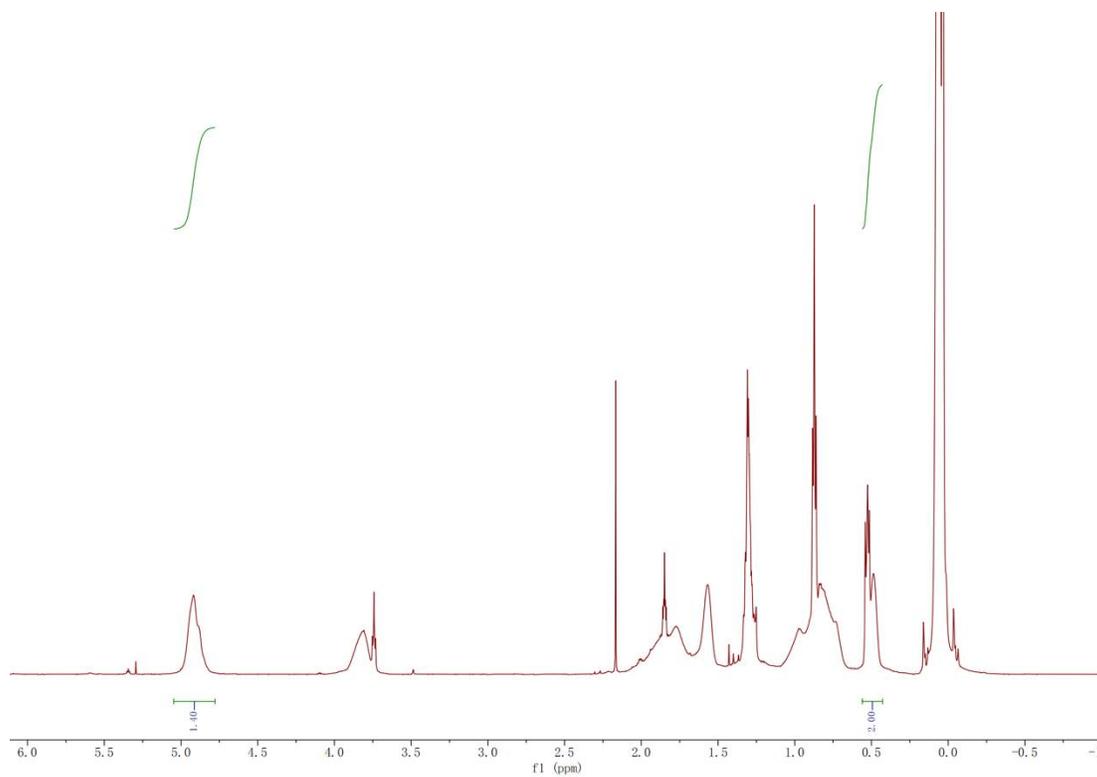

**Fig. S6.** $^1$H NMR of (BnMA$_{1.40}$-$r$-PDMS$^1$)$_{200}$. The number of PDMS side chains is 200, the spacer ratio is 1.40, and the number of BnMA monomers is 200×1.40=280.

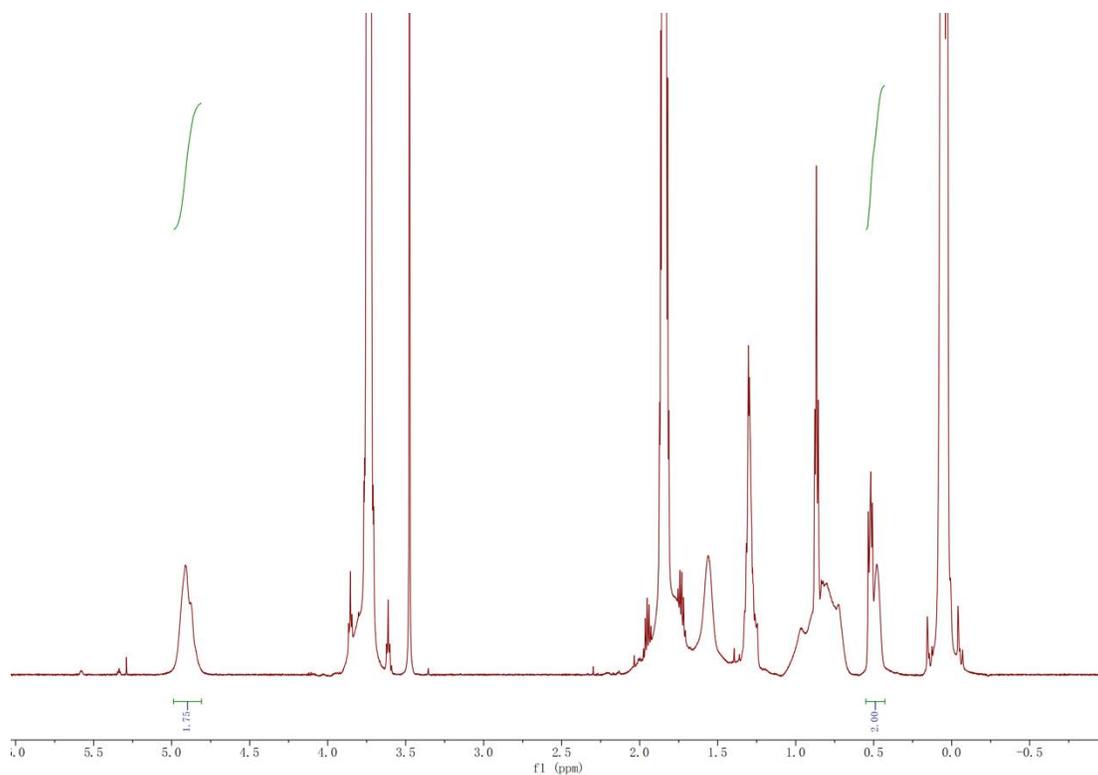

**Fig. S7.** $^1$H NMR of (BnMA$_{1.75}$-$r$-PDMS$^1$)$_{190}$. The number of PDMS side chains is 200, the spacer ratio is 1.75, and the number of BnMA monomers is 200×1.75=350.

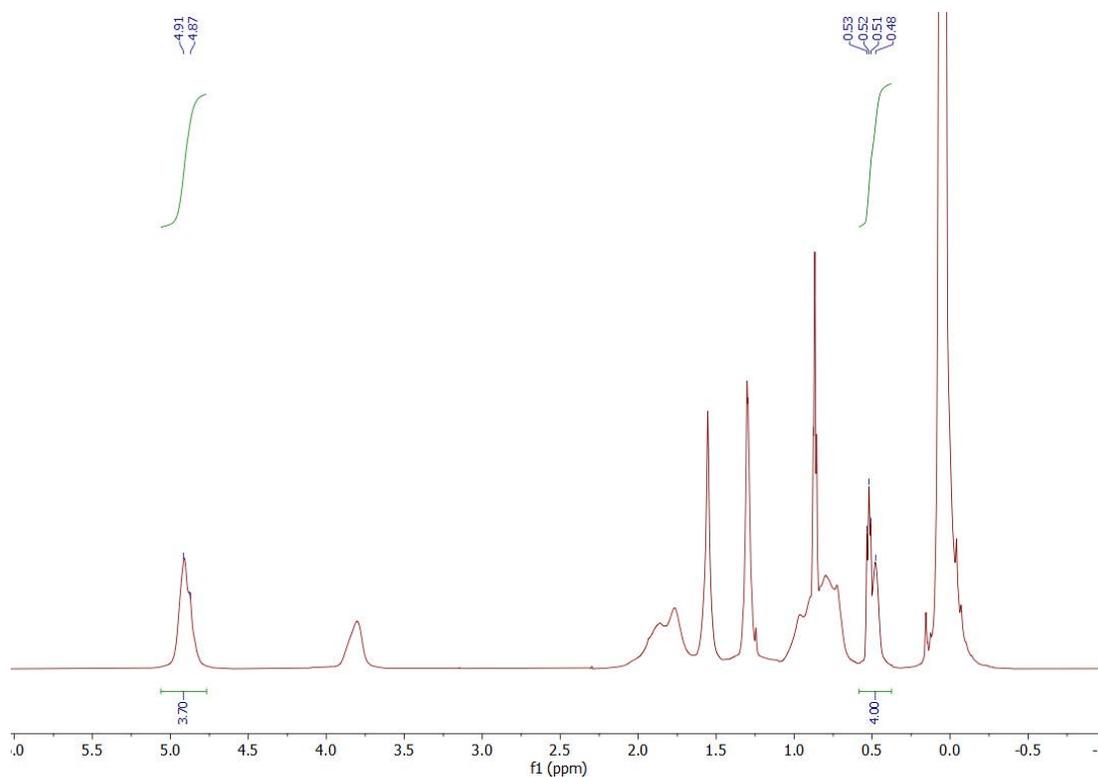

**Fig. S8.** $^1$H NMR of (BnMA$_{1.85}$-$r$-PDMS$^1$)$_{190}$. The number of PDMS side chains is 200, the spacer ratio is 3.70/2=1.85, and the number of BnMA monomers is 200×1.85=370.

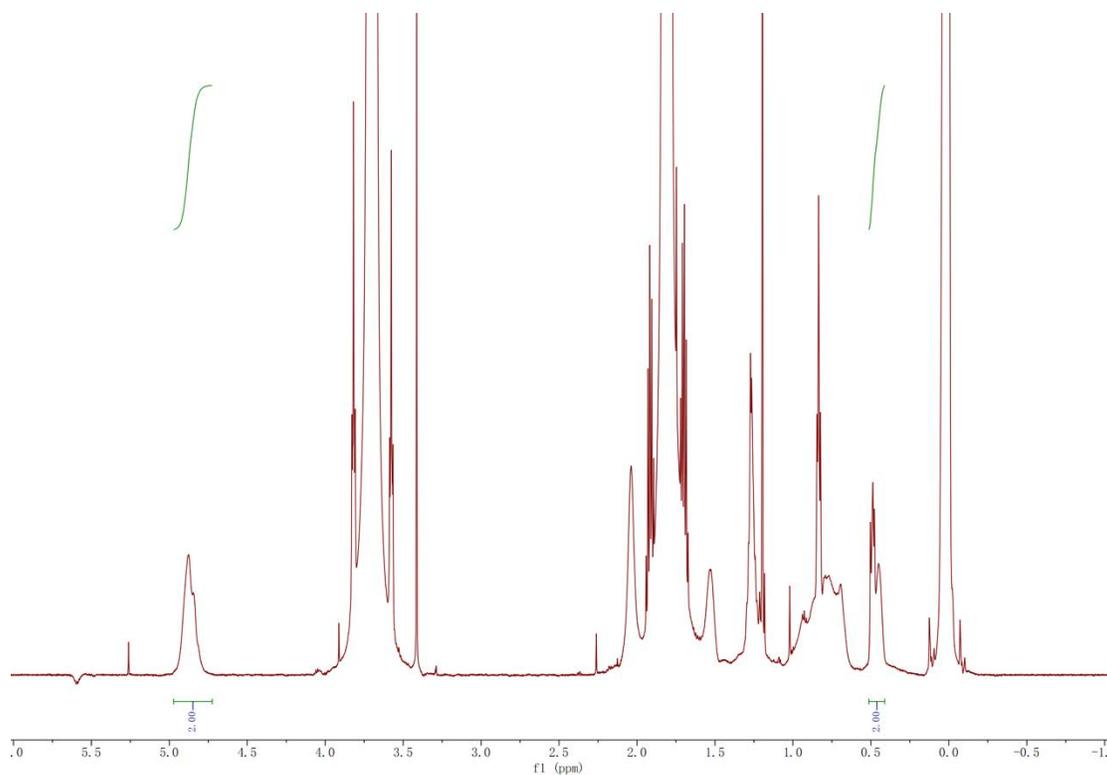

**Fig. S9.** $^1$H NMR of (BnMA$_{2.00}$-$r$-PDMS$^1$)$_{200}$. The number of PDMS side chains is 200, the spacer ratio is 2.00, and the number of BnMA monomers is 200×2.00=400.

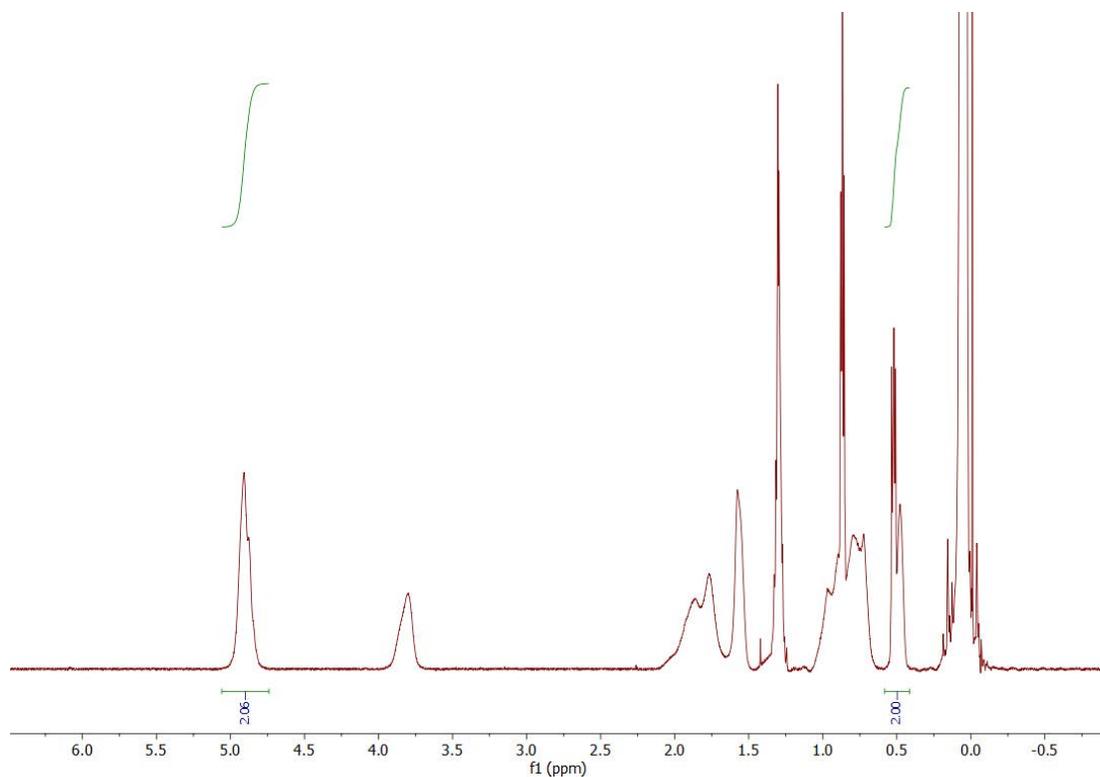

**Fig. S10.** $^1$H NMR of (BnMA$_{2.06}$-$r$-PDMS$^1$)$_{198}$. The number of PDMS side chains is 198, the spacer ratio is 2.06, and the number of BnMA monomers is 198×2.06=408.

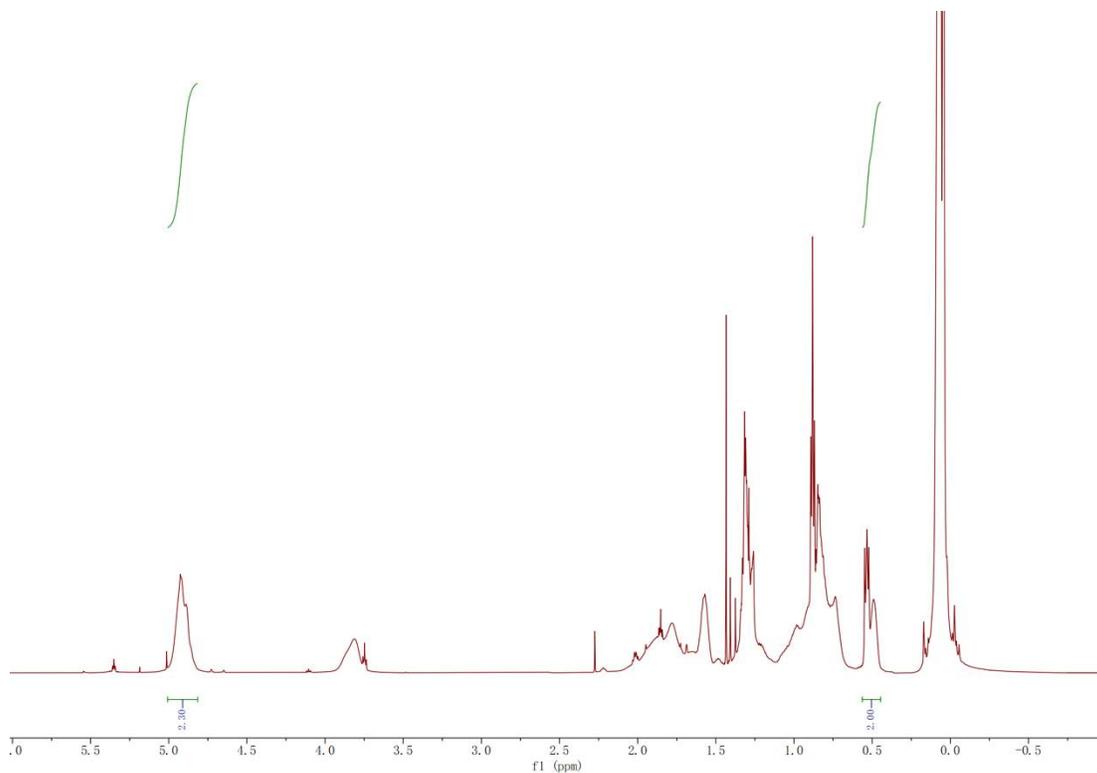

**Fig. S11.** $^1$H NMR of (BnMA$_{2.30}$-$r$-PDMS$^1$)$_{196}$. The number of PDMS side chains is 196, the spacer ratio is 2.30, and the number of BnMA monomers is 196×2.30=451.

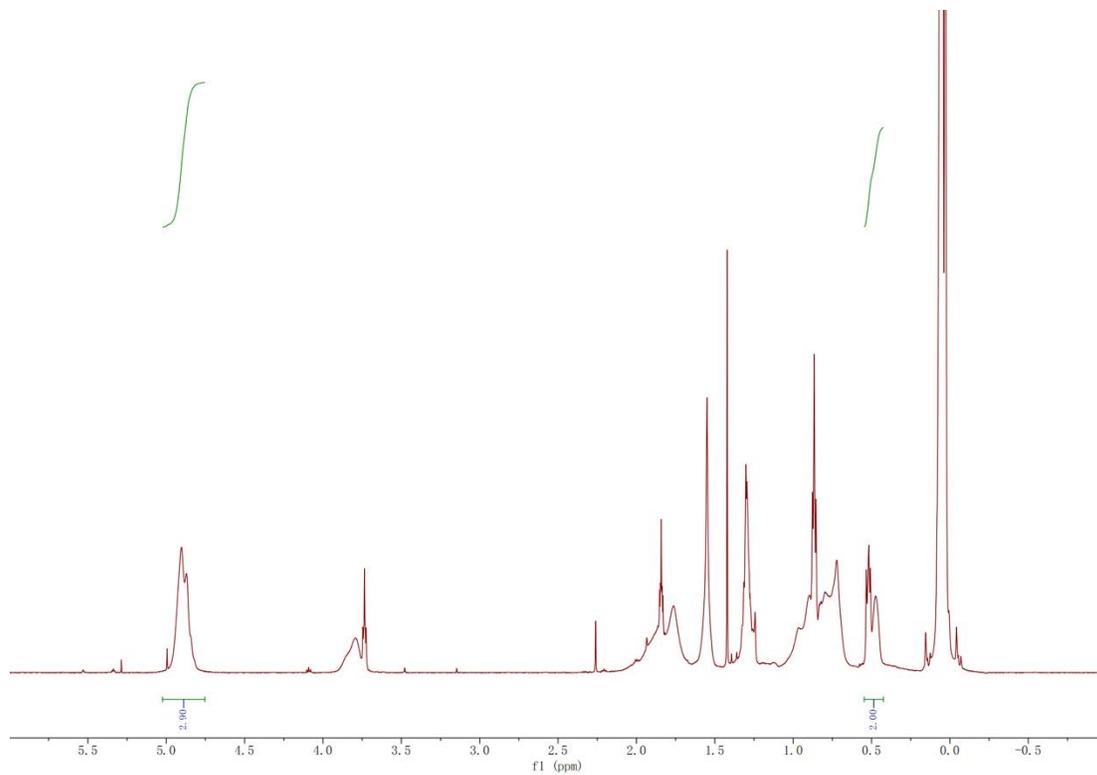

**Fig. S12.** $^1$H NMR of (BnMA$_{2.90}$-$r$-PDMS$^1$)$_{200}$. The number of PDMS side chains is 200, the spacer ratio is 2.90, and the number of BnMA monomers is 200×2.90=580.

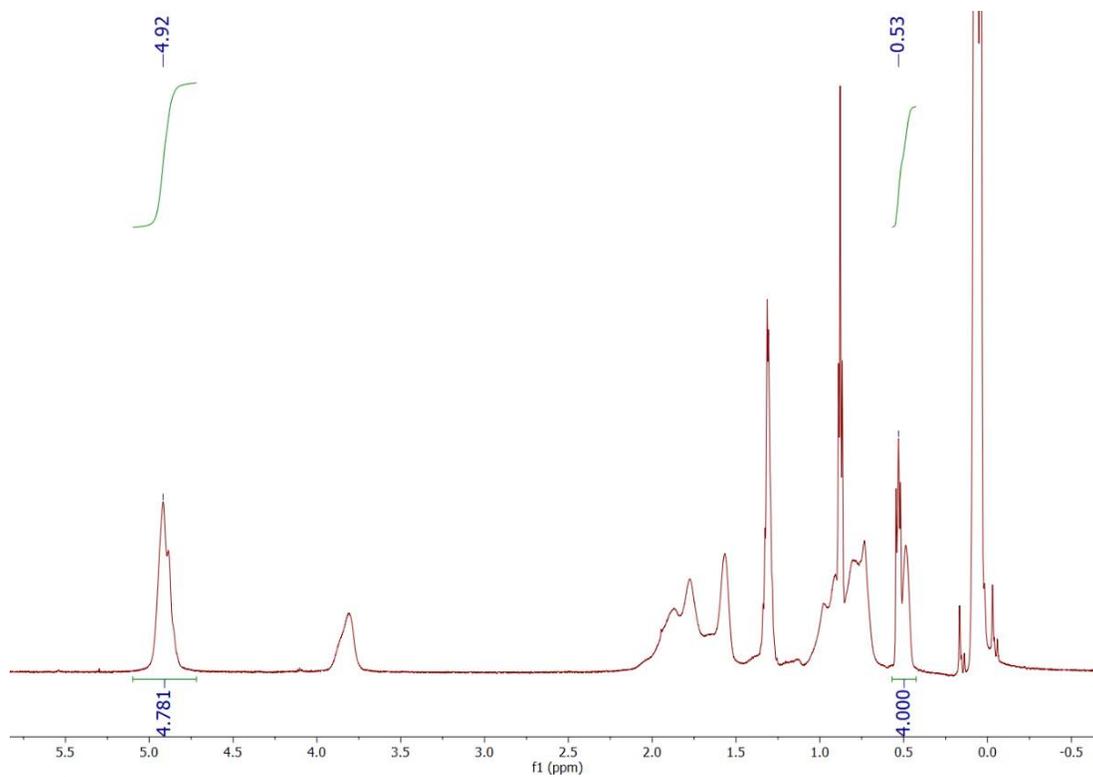

**Fig. S13.** $^1$H NMR of $(BnMA_{2.39}$-$r$-$PDMS^1)_{207}$. The number of PDMS side chains is 207, the spacer ratio is 4.781/2=2.39, and the number of BnMA monomers is 207×2.39=495.

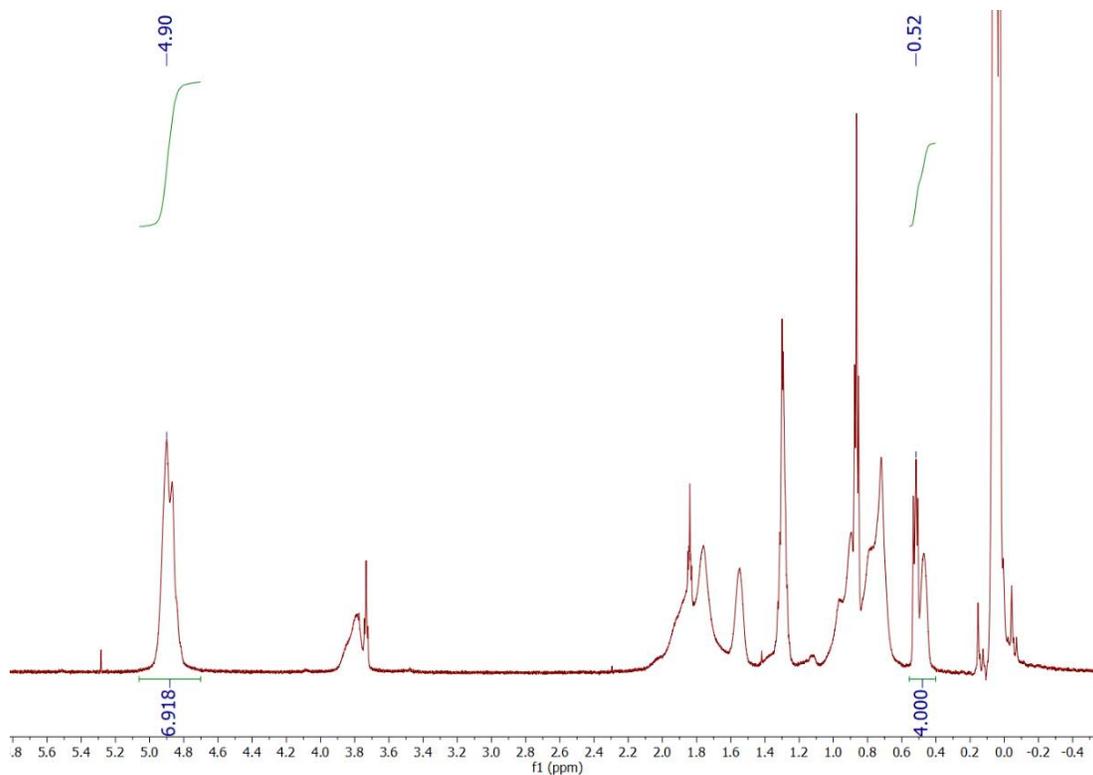

**Fig. S14.** $^1$H NMR of $(BnMA_{3.46}$-$r$-$PDMS^1)_{200}$. The number of PDMS side chains is 200, the spacer ratio is 6.918/2=3.459, and the number of BnMA monomers is 200×2.39=692.

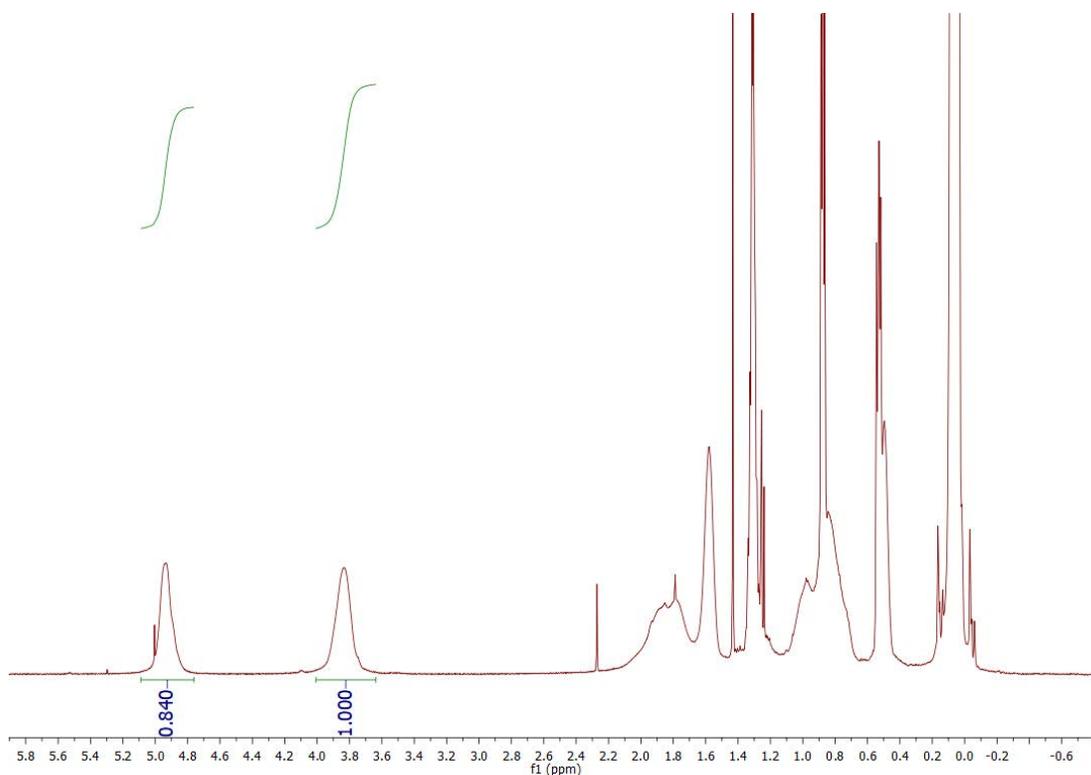

**Fig. S15.** $^1$H NMR of (BnMA$_{0.84}$-$r$-PDMS$^1$)$_{534}$. The number of PDMS side chains is 534, the spacer ratio is 0.84, and the number of BnMA monomers is 534×0.84=448.

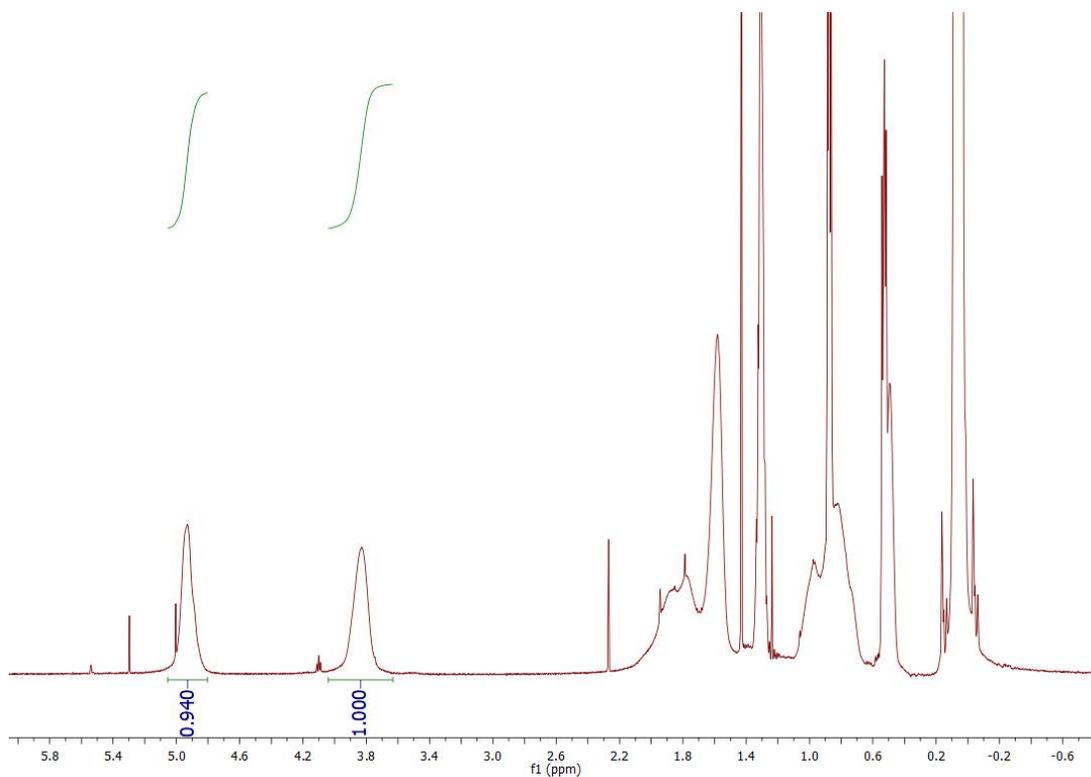

**Fig. S16.** $^1$H NMR of (BnMA$_{0.94}$-$r$-PDMS$^1$)$_{468}$. The number of PDMS side chains is 468, the spacer ratio is 0.94, and the number of BnMA monomers is 468×0.44=440.

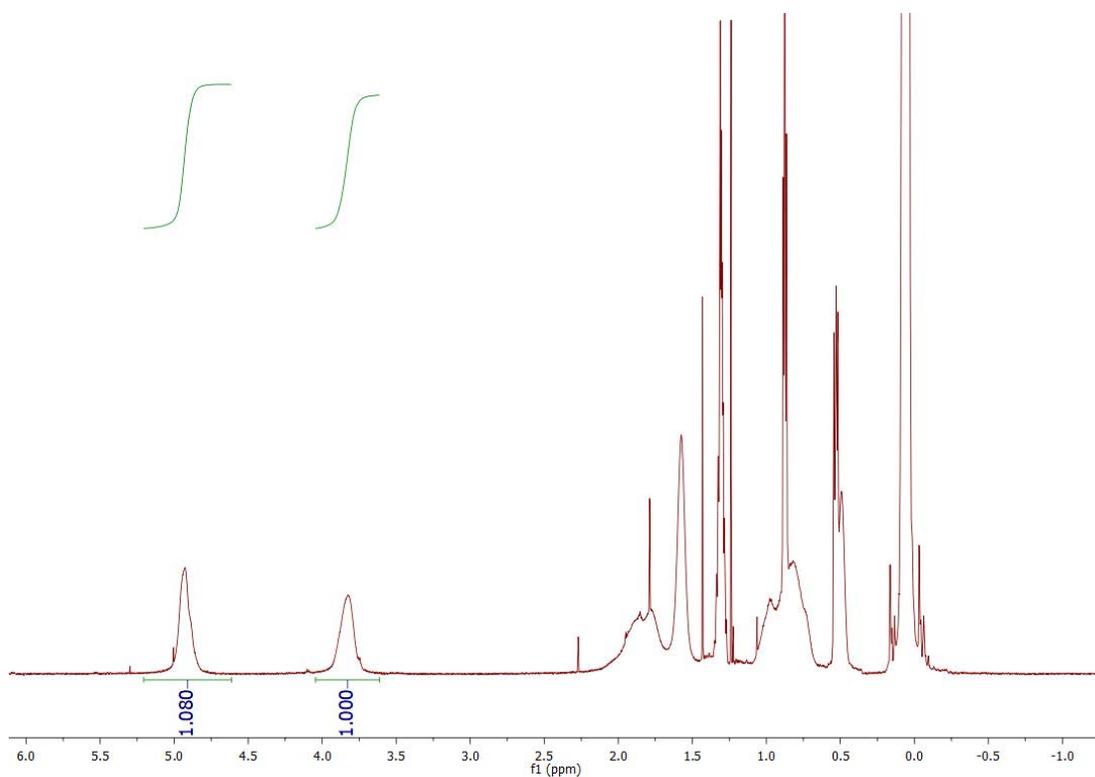

**Fig. S17.** $^1$H NMR of (BnMA$_{1.08}$-*r*-PDMS$^1$)$_{360}$. The number of PDMS side chains is 360, the spacer ratio is 1.08, and the number of BnMA monomers is 360×1.08=389.

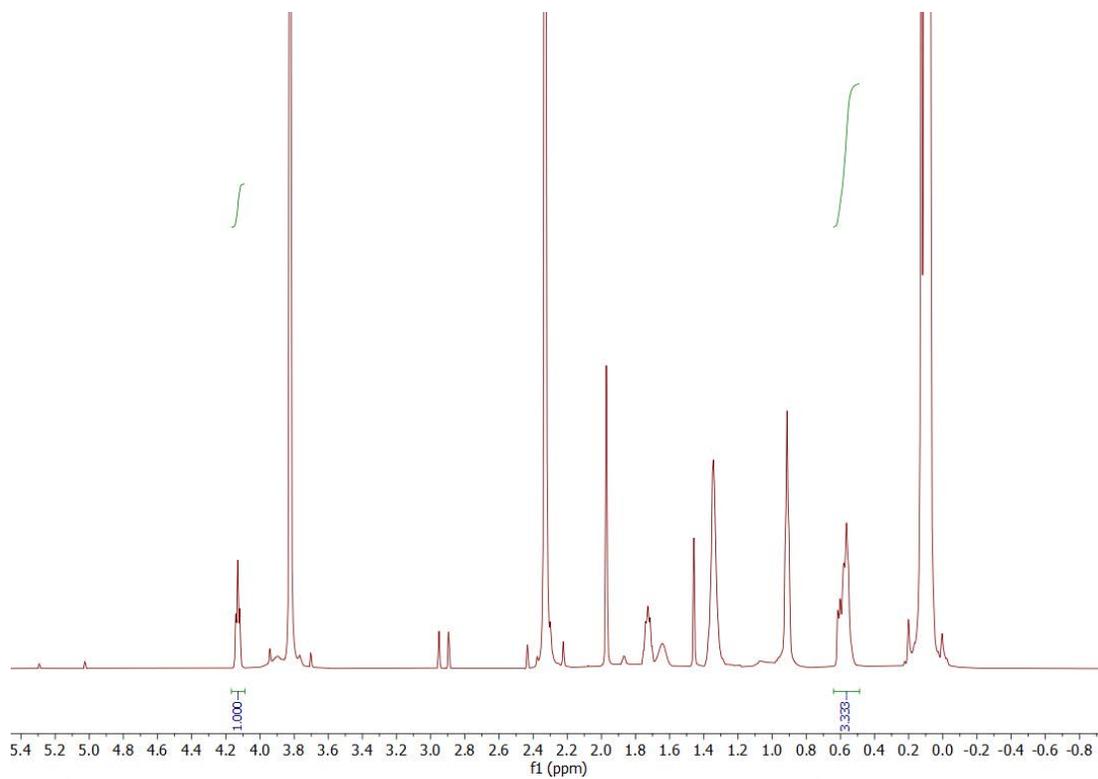

**Fig. S18.** $^1$H NMR data of raw mixture of ARGET ATRP of (PDMS$^1$)$_{550}$. Conversion = (1 − 1.000 × 2 / 3.333) × 100% = 40%. The number of PDMS side chains is 1375×40%=550.

1.2. Foldable bottlebrush polymer networks: [~200, ~33, 0-3.46]

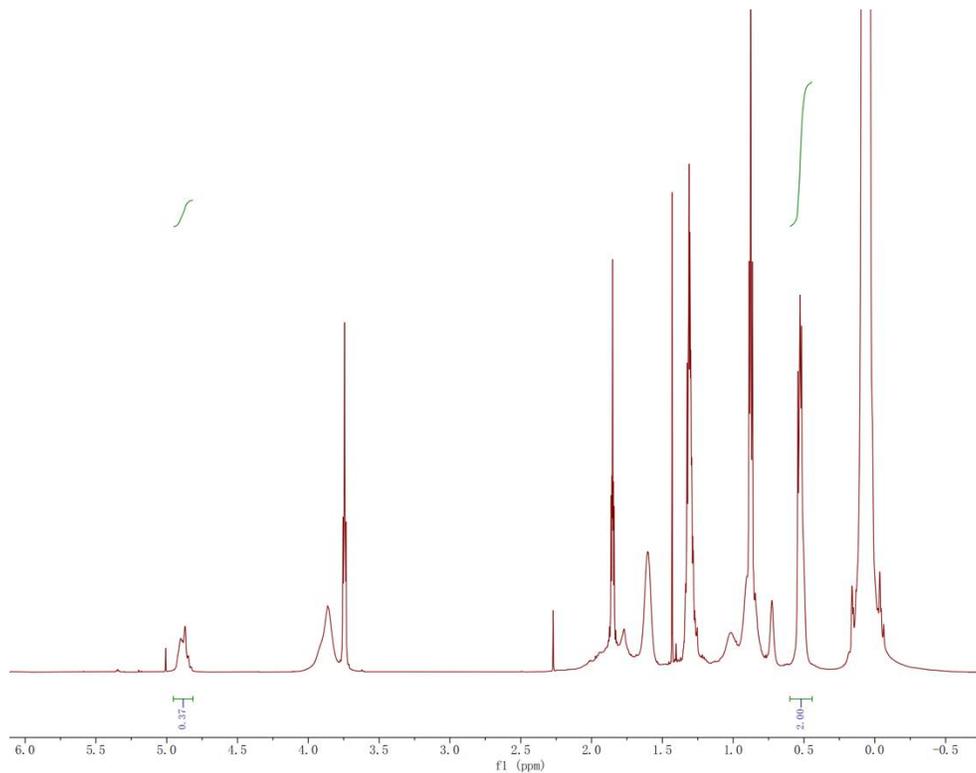

**Fig. S19.** $^1$H NMR of BnMA$_{36}$-*b*-(PDMS$^1$)$_{197}$-*b*-BnMA$_{36}$. The number of total BnMA monomers is 197×0.37=72.9. The number of BnMA monomers on each end block is 72.9/2=36.

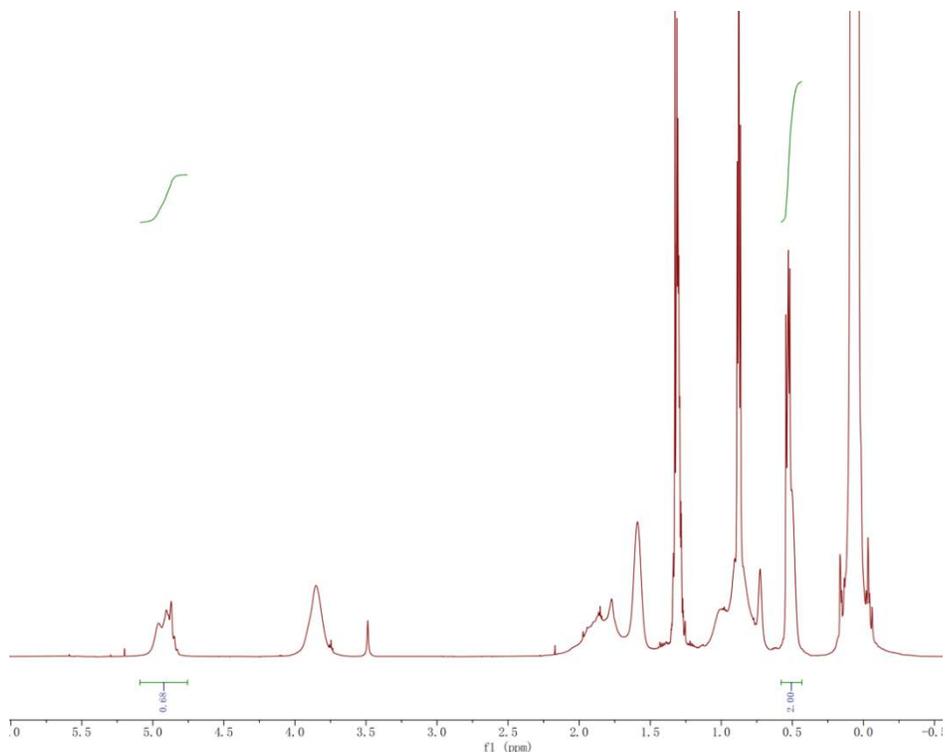

**Fig. S20.** $^1$H NMR of BnMA$_{35}$-$b$-(BnMA$_{0.34}$-$r$-PDMS$^1$)$_{203}$-$b$-BnMA$_{35}$. The number of total BnMA monomers is 203×0.68=138. The number of BnMA monomers on each end block is (138-69)/2=35.

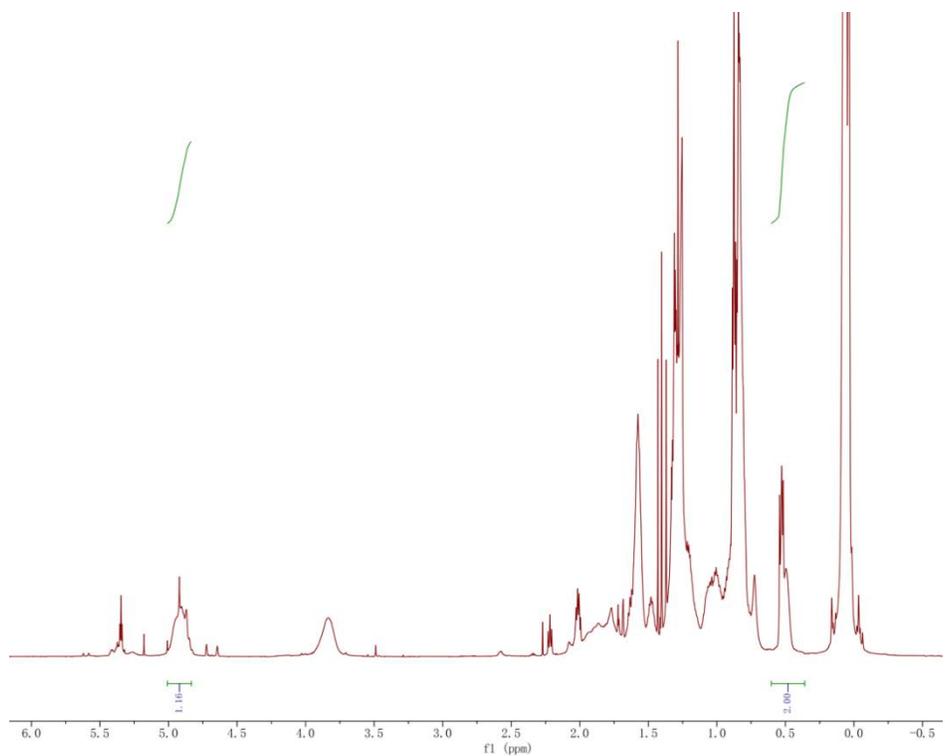

**Fig. S21.** $^1$H NMR of BnMA$_{36}$-$b$-(BnMA$_{0.80}$-$r$-PDMS$^1$)$_{202}$-$b$-BnMA$_{36}$. The number of total BnMA monomers is 202×1.16=234. The number of BnMA monomers on each end block is (234-162)/2=36.

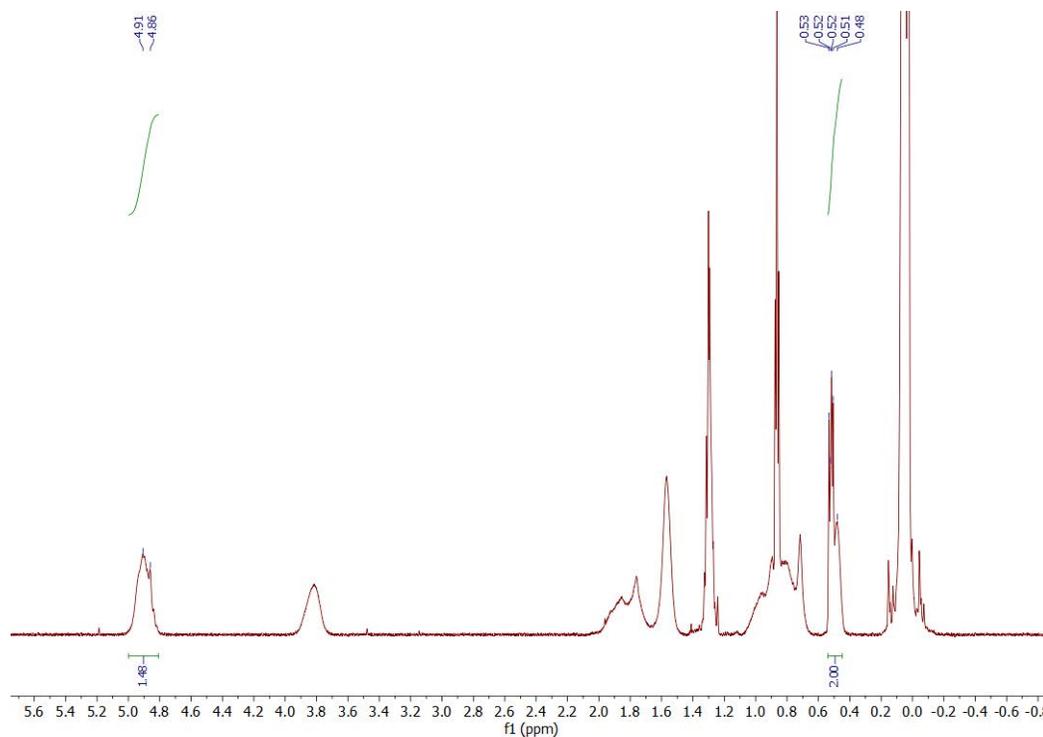

**Fig. S22.** $^1$H NMR of BnMA$_{40}$-*b*-(BnMA$_{1.08}$-*r*-PDMS$^1$)$_{195}$-*b*-BnMA$_{40}$. The number of total BnMA monomers is 195×1.48=289. The number of BnMA monomers on each end block is (289-210)/2=40.

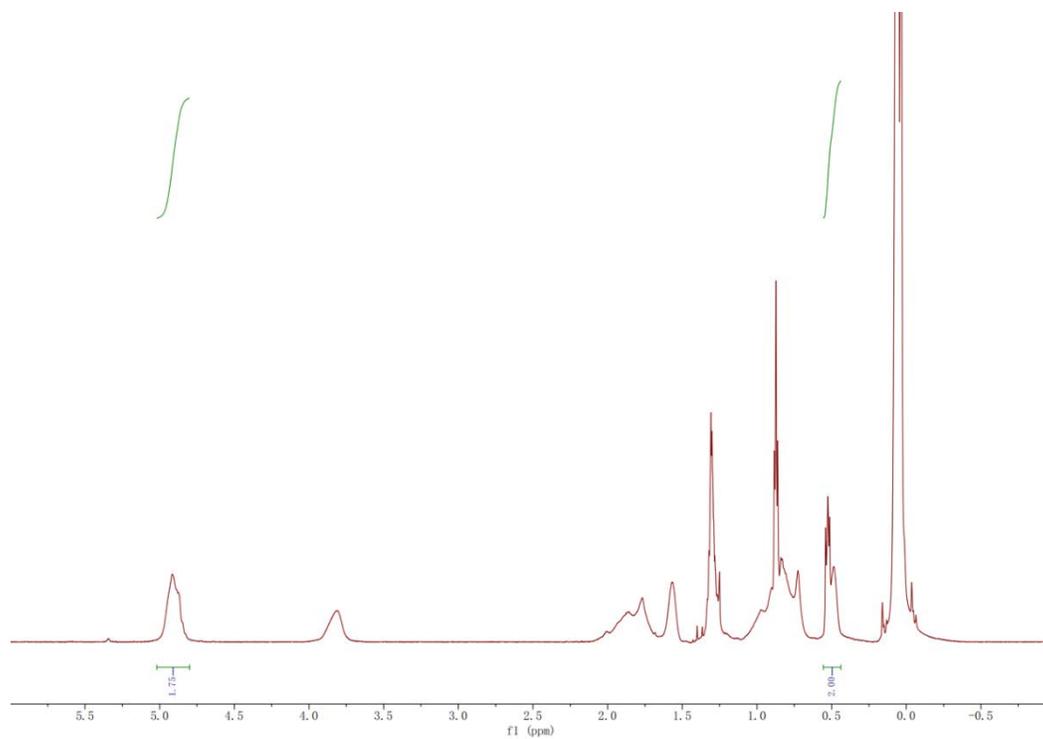

**Fig. S23.** $^1$H NMR of BnMA$_{35}$-*b*-(BnMA$_{1.40}$-*r*-PDMS$^1$)$_{200}$-*b*-BnMA$_{35}$. The number of total BnMA monomers is 200×1.75=350. The number of BnMA monomers on each end block is (350-280)/2=35.

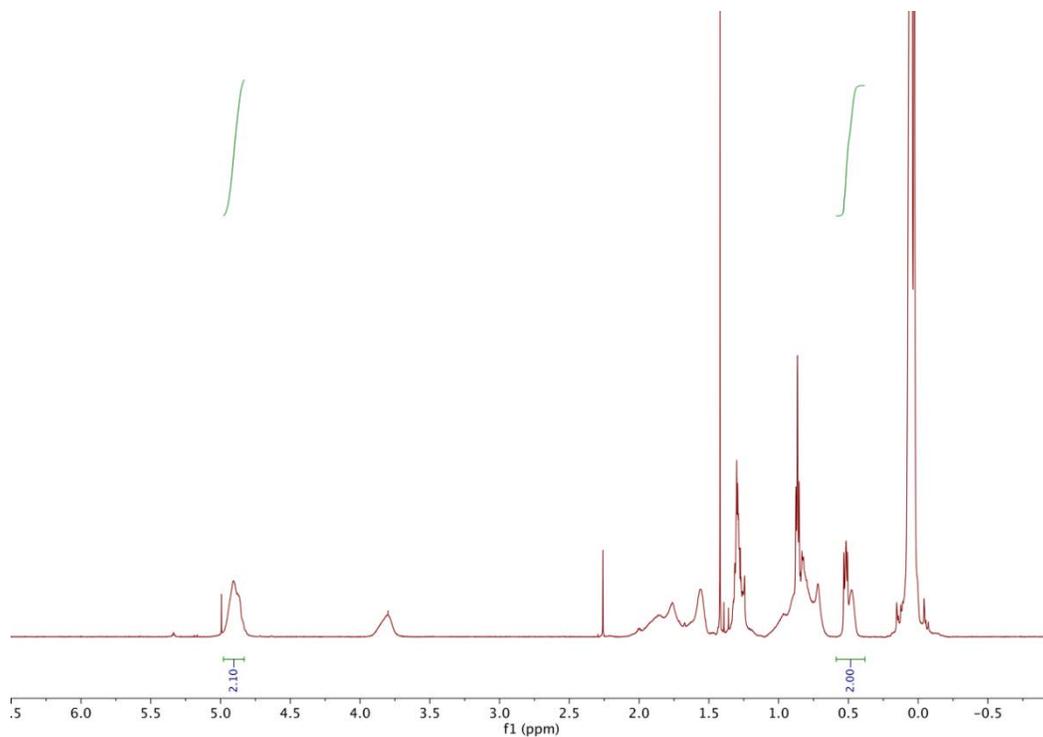

**Fig. S24.** $^1$H NMR of BnMA$_{35}$-*b*-(BnMA$_{1.75}$-*r*-PDMS$^1$)$_{200}$-*b*-BnMA$_{35}$. The number of total BnMA monomers is 200×2.10=420. The number of BnMA monomers on each end block is (420-350)/2=35.

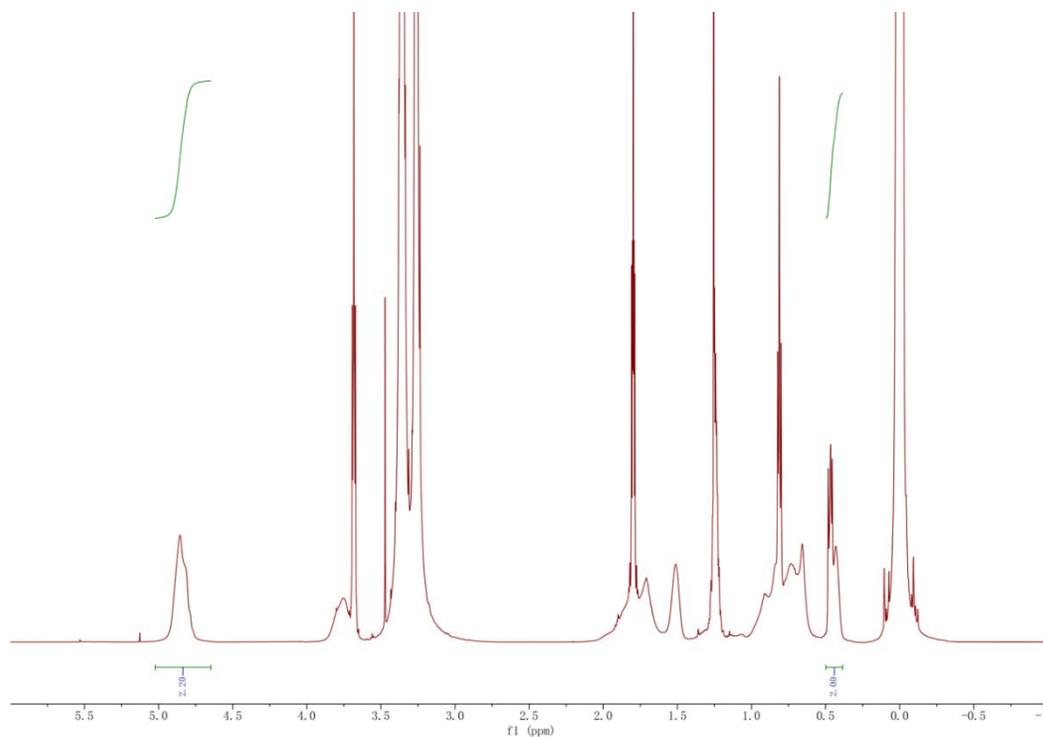

**Fig. S25.** $^1$H NMR of BnMA$_{35}$-*b*-(BnMA$_{1.85}$-*r*-PDMS$^1$)$_{200}$-*b*-BnMA$_{35}$. The number of total BnMA monomers is 200×2.20=440. The number of BnMA monomers on each end block is (440-370)/2=35.

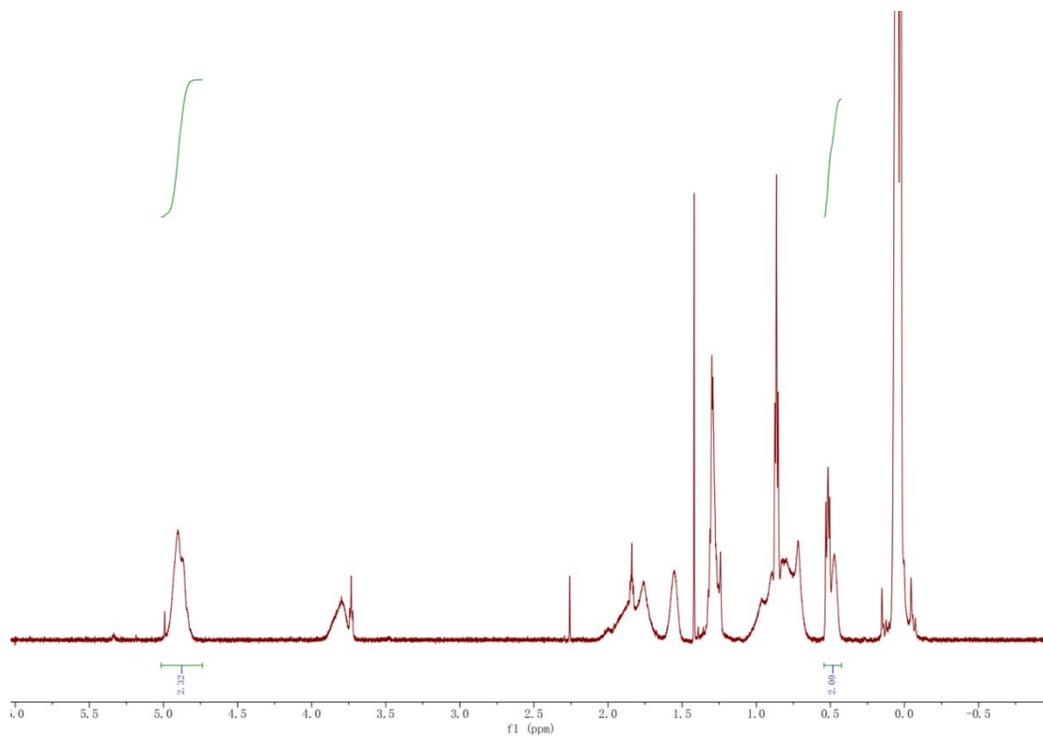

**Fig. S26.** $^1$H NMR of BnMA$_{32}$-*b*-(BnMA$_{2.00}$-*r*-PDMS$^1$)$_{200}$-*b*-BnMA$_{32}$. The number of total BnMA monomers is 200×2.32=464. The number of BnMA monomers on each end block is (464-400)/2=32.

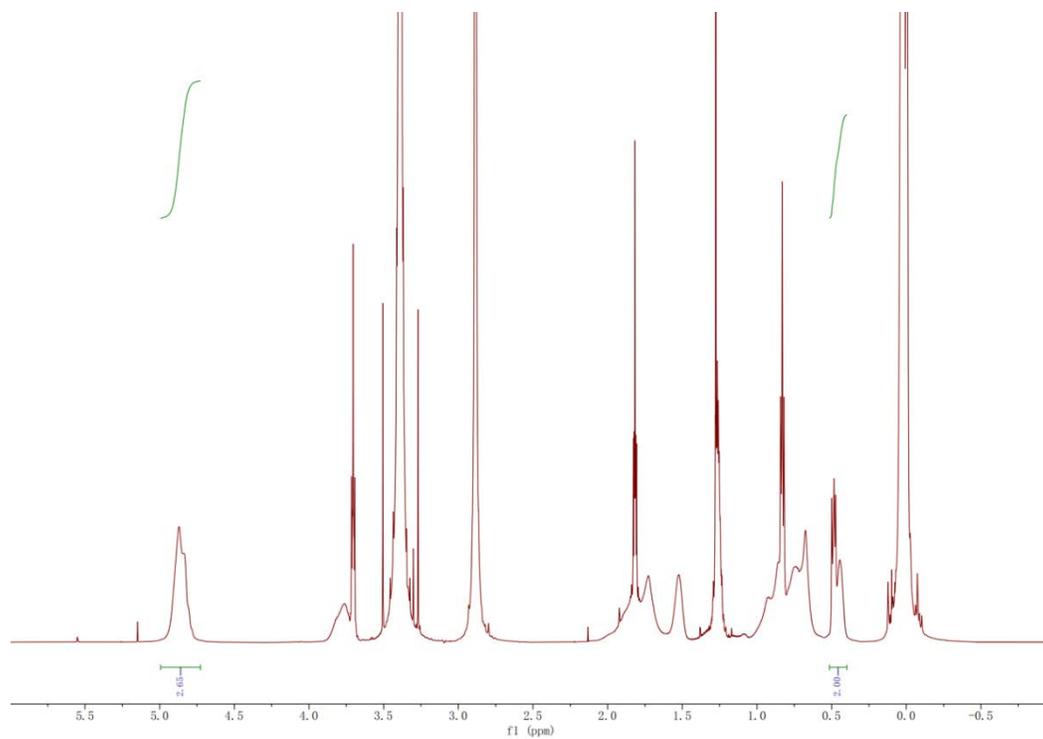

**Fig. S27.** $^1$H NMR of BnMA$_{34}$-*b*-(BnMA$_{2.30}$-*r*-PDMS$^1$)$_{196}$-*b*-BnMA$_{34}$. The number of total BnMA monomers is 196×2.65=519. The number of BnMA monomers on each end block is (519-451)/2=34.

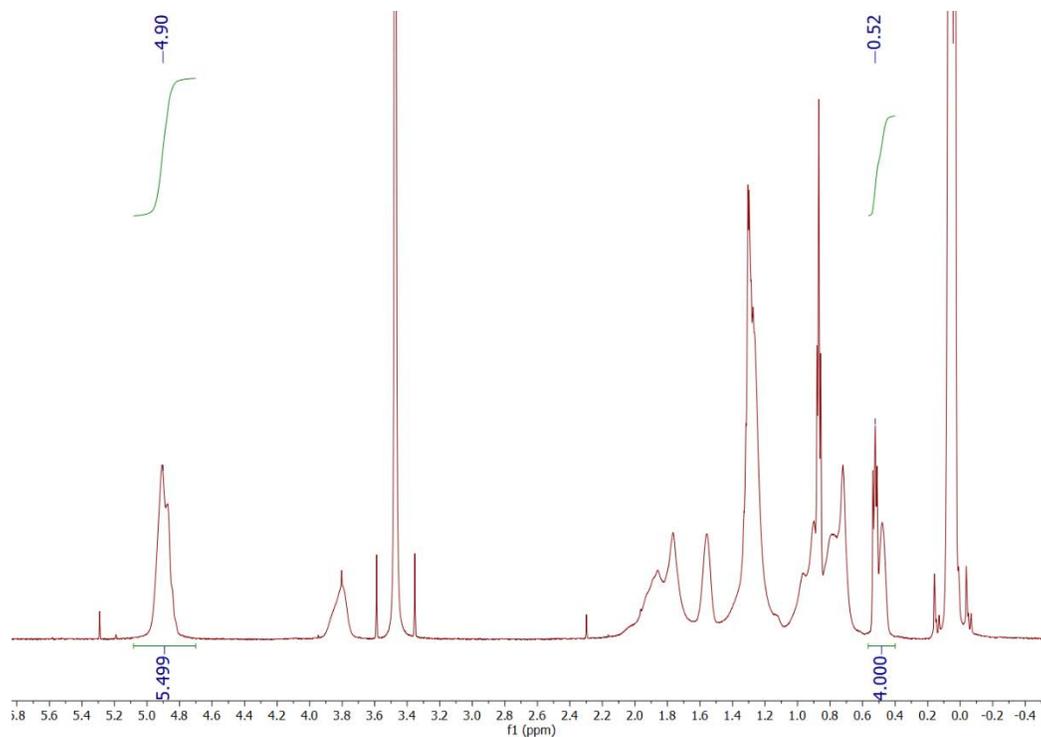

**Fig. S28.** $^1$H NMR of BnMA$_{37}$-*b*-(BnMA$_{2.39}$-*r*-PDMS$^1$)$_{207}$-*b*-BnMA$_{37}$. The number of total BnMA monomers is 207×5.499/2=569. The number of BnMA monomers on each end block is (569-495)/2=37.

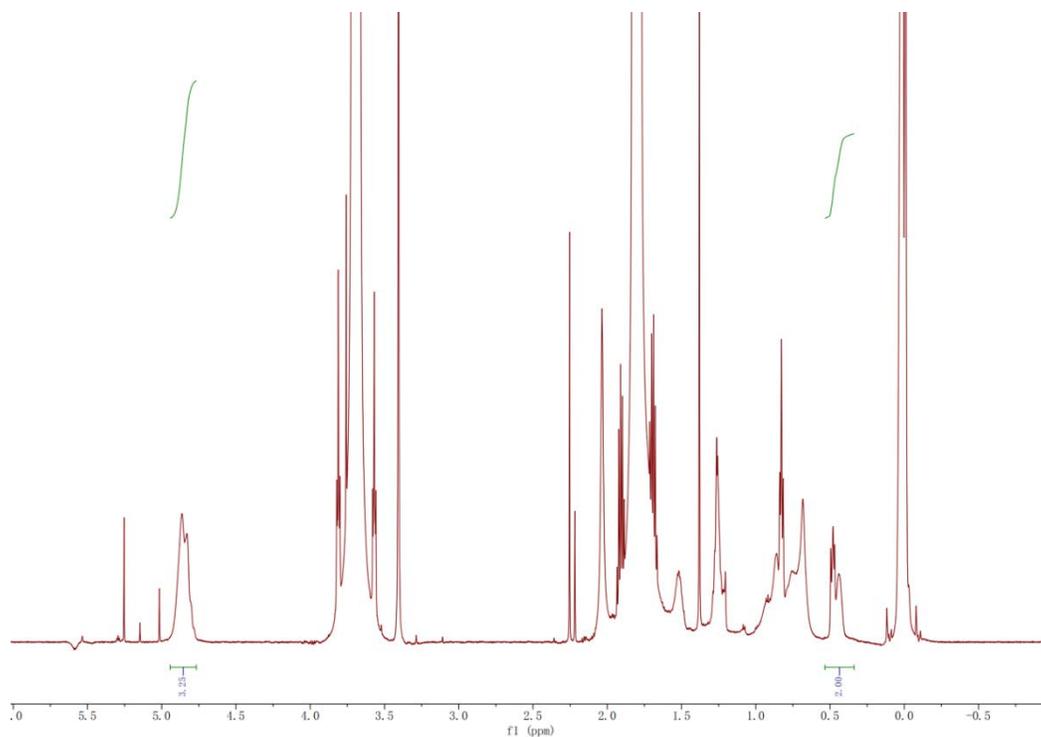

**Fig. S29.** $^1$H NMR of BnMA$_{35}$-*b*-(BnMA$_{2.90}$-*r*-PDMS$^1$)$_{200}$-*b*-BnMA$_{35}$. The number of total BnMA monomers is 200×3.25=650. The number of BnMA monomers on each end block is (650-540)/2=35.

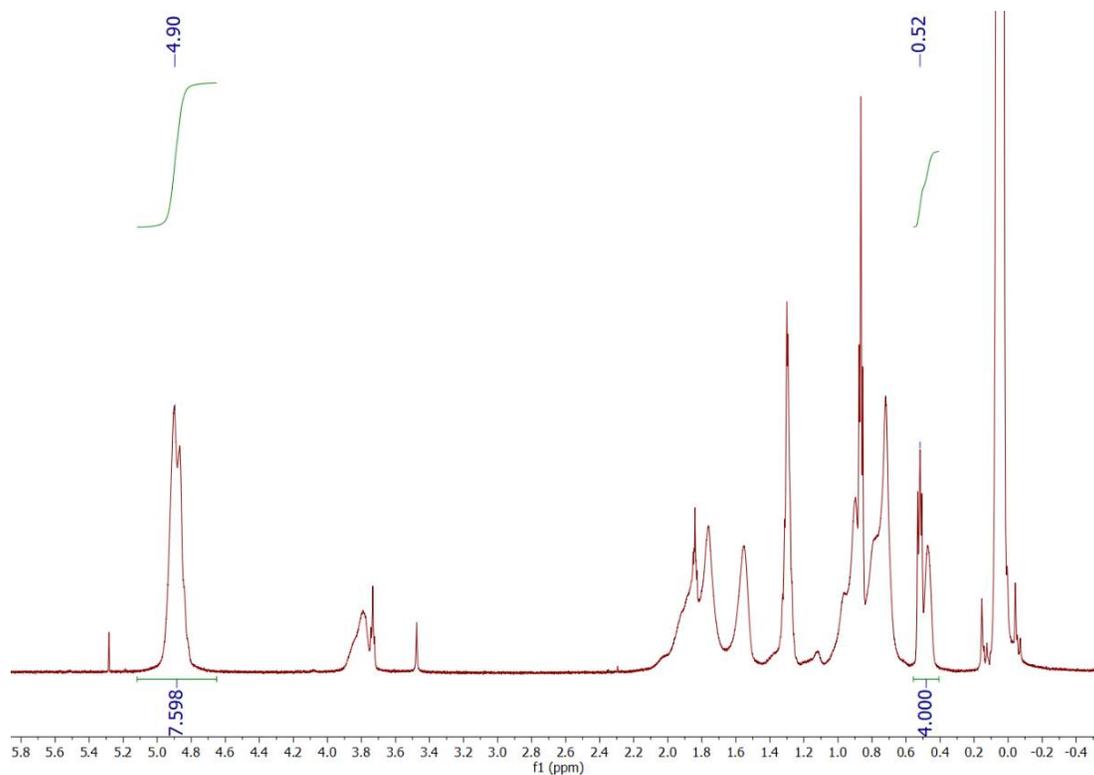

**Fig. S30.** $^1$H NMR of BnMA$_{34}$-$b$-(BnMA$_{3.46}$-$r$-PDMS$^1$)$_{200}$-$b$-BnMA$_{34}$. The number of total BnMA monomers is 200×7.598/2=760. The number of BnMA monomers on each end block is (760-692)/2=34.

1.3. Fix the spacer ratio and change the number of side chains

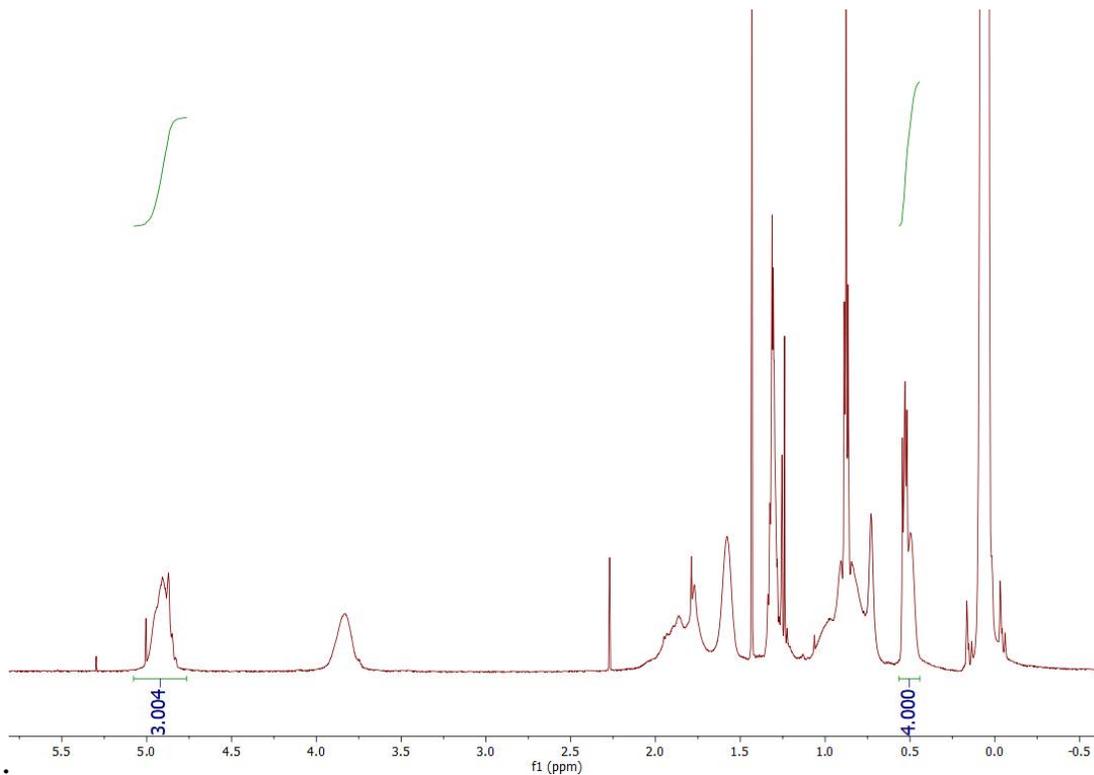

**Fig. S31.** $^1$H NMR of BnMA$_{177}$-$b$-(BnMA$_{0.84}$-$r$-PDMS$^1$)$_{534}$-$b$-BnMA$_{177}$. The number of total BnMA monomers is 534×3.004/2=801. The number of BnMA monomers on each end block is (801-448)/2=177.

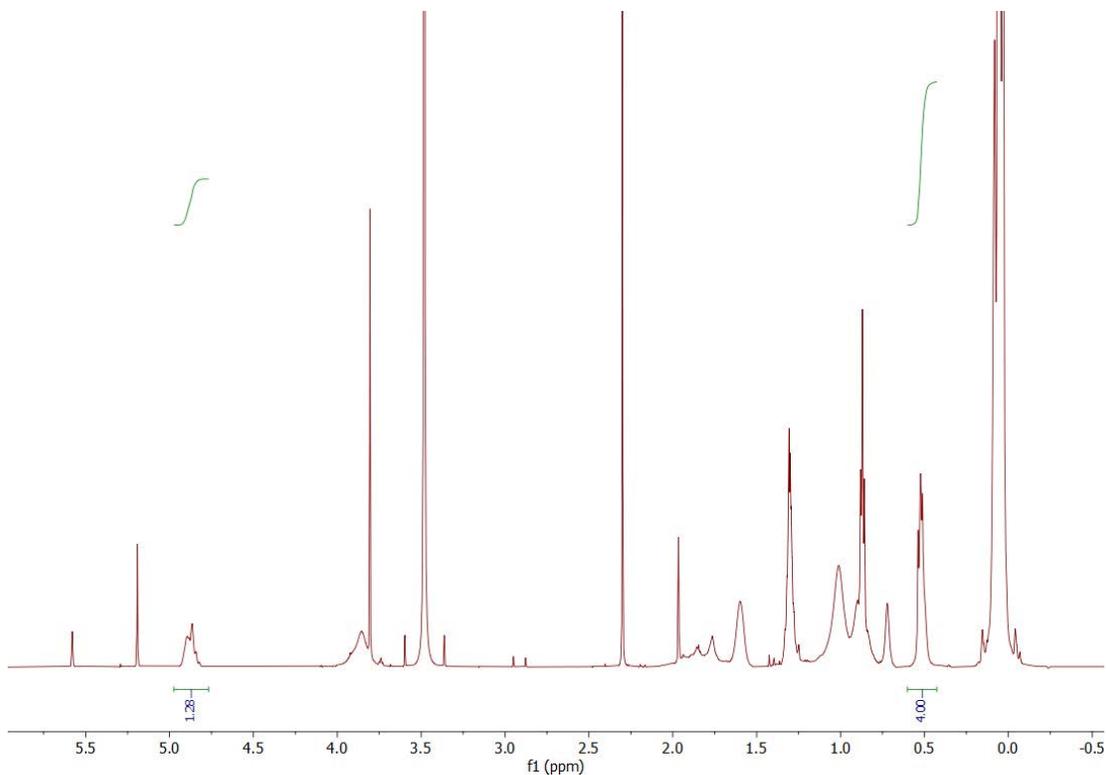

**Fig. S32.** $^1$H NMR of BnMA$_{176}$-$b$-(PDMS$^1$)$_{550}$-$b$-BnMA$_{176}$. The number of total BnMA monomers is 550×1.28/2=352. The number of BnMA monomers on each end block is 352/2=176.

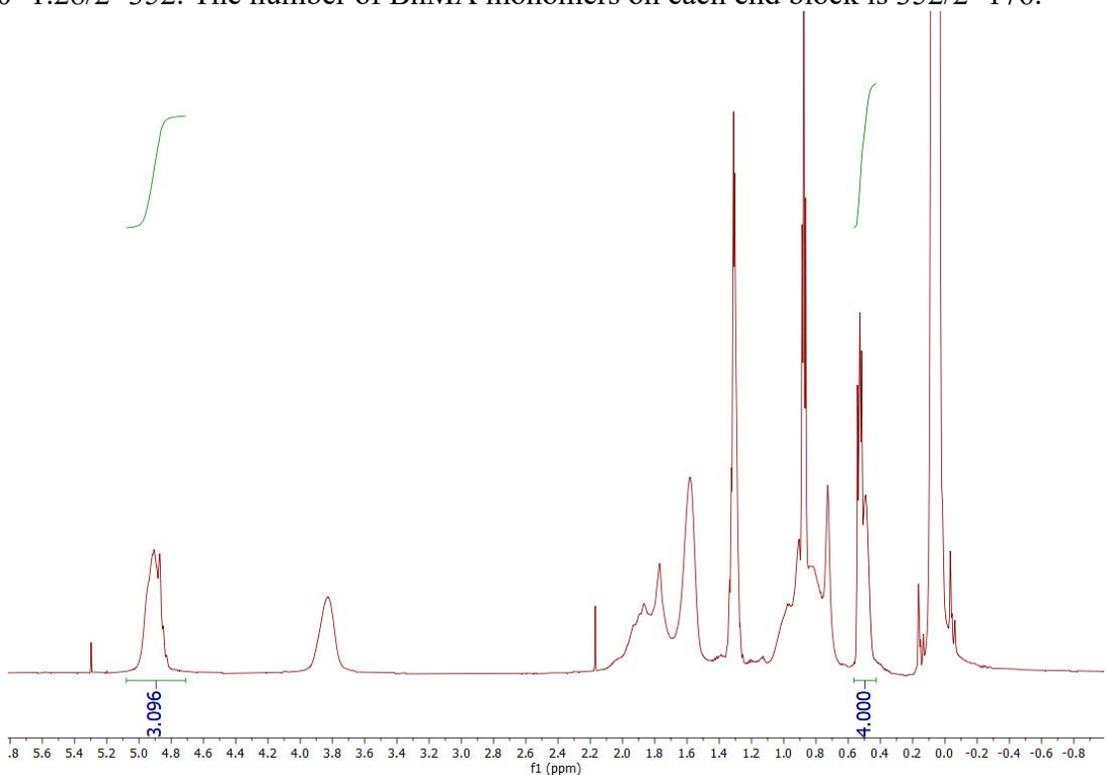

**Fig. S33.** $^1$H NMR of BnMA$_{142}$-$b$-(BnMA$_{0.94}$-$r$-PDMS$^1$)$_{468}$-$b$-BnMA$_{142}$. The number of total BnMA monomers is 468×3.096/2=724. The number of BnMA monomers on each end block is (724-440)/2=142.

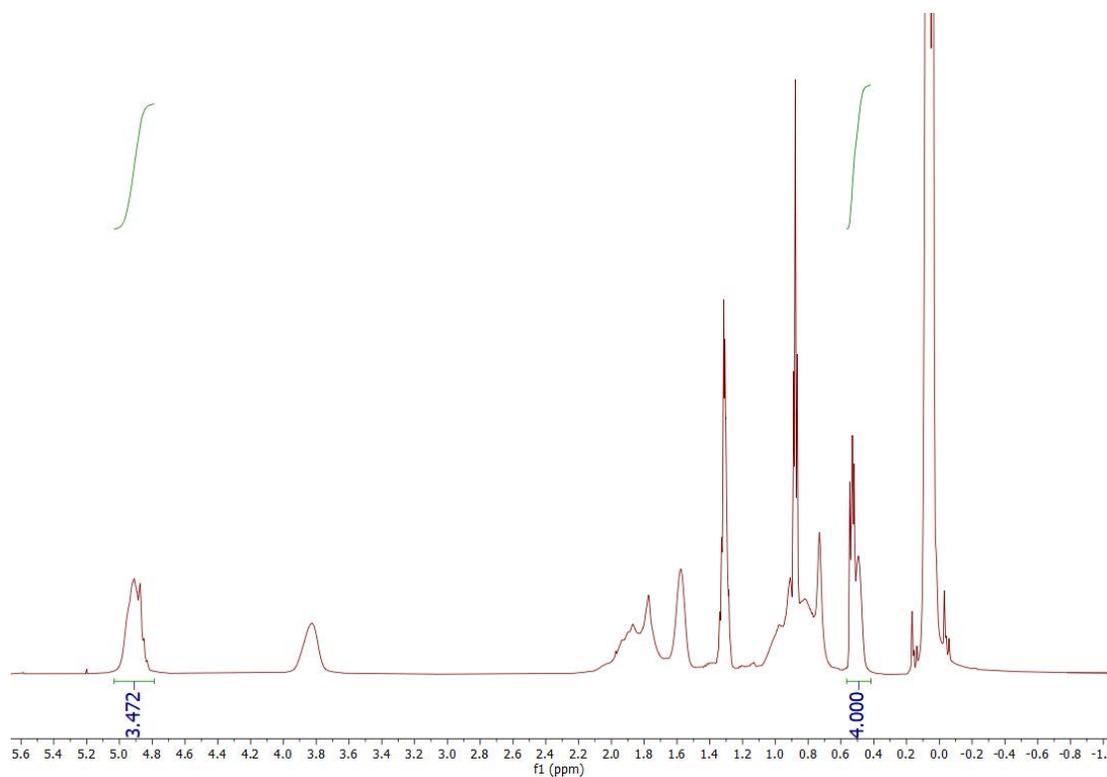

**Fig. S34.** $^1$H NMR of BnMA$_{118}$-$b$-(BnMA$_{1.08}$-$r$-PDMS$^1$)$_{360}$-$b$-BnMA$_{118}$. The number of total BnMA monomers is 360×3.372/2=625. The number of BnMA monomers on each end block is (625-389)/2=118.

1.4. Increase the end block volume fraction for a network with an intermediate spacer ratio

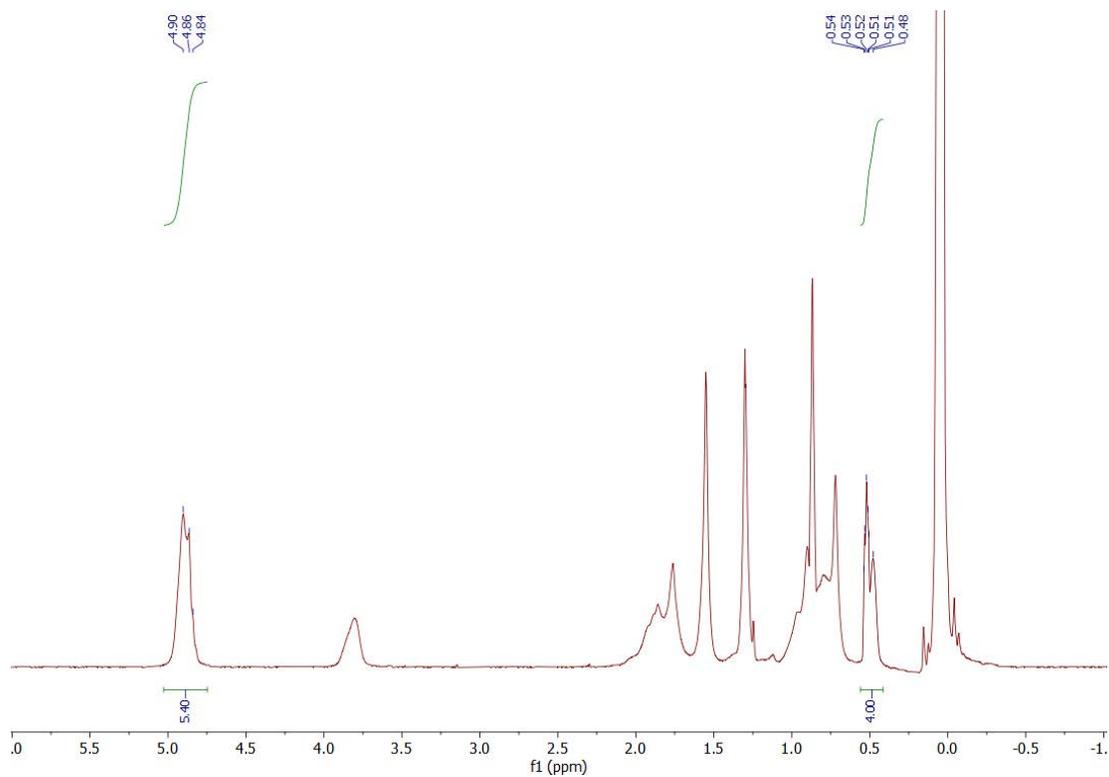

**Fig. S35.** $^1$H NMR of BnMA$_{85}$-$b$-(BnMA$_{1.85}$-$r$-PDMS$^1$)$_{200}$-$b$-BnMA$_{85}$. End block volume fraction, $f$=13%

**Data Set 2: Foldable bottlebrush polymers and networks with MMA spacer**

2.1. Foldable bottlebrush middle block

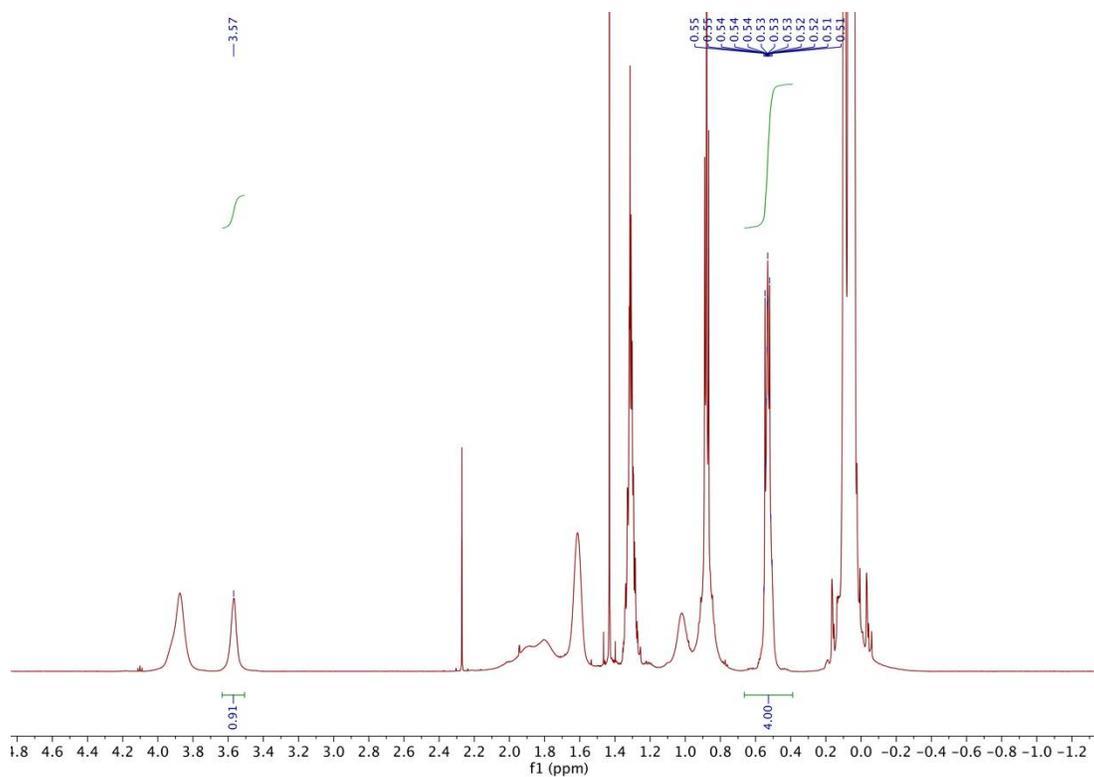

**Fig. S36.** $^1$H NMR of $(MMA_{0.30}\text{-}r\text{-}PDMS^1)_{200}$. The number of PDMS side chains is 200, the spacer ratio is 0.91/3=0.30, and the number of MMA monomers is 200×0.30=60.

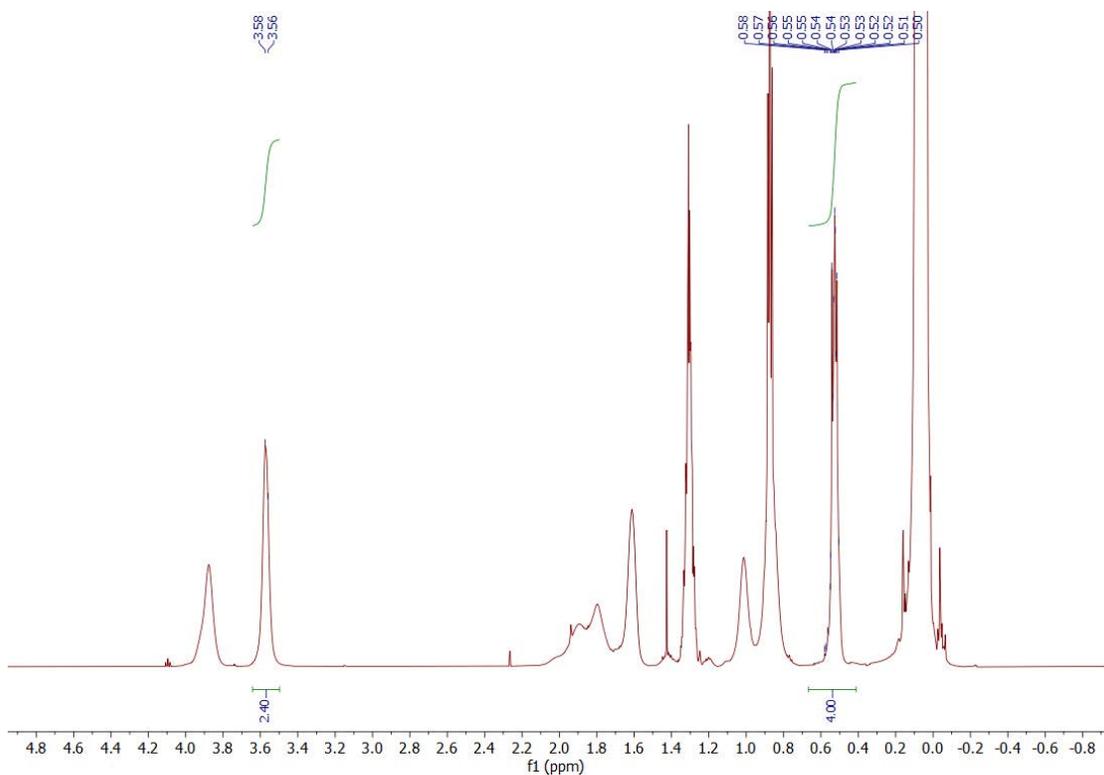

**Fig. S37.** $^1$H NMR of (MMA$_{0.80}$-$r$-PDMS$^1$)$_{200}$. The number of PDMS side chains is 200, the spacer ratio is 2.40/3=0.80, and the number of MMA monomers is 200×0.80=160.

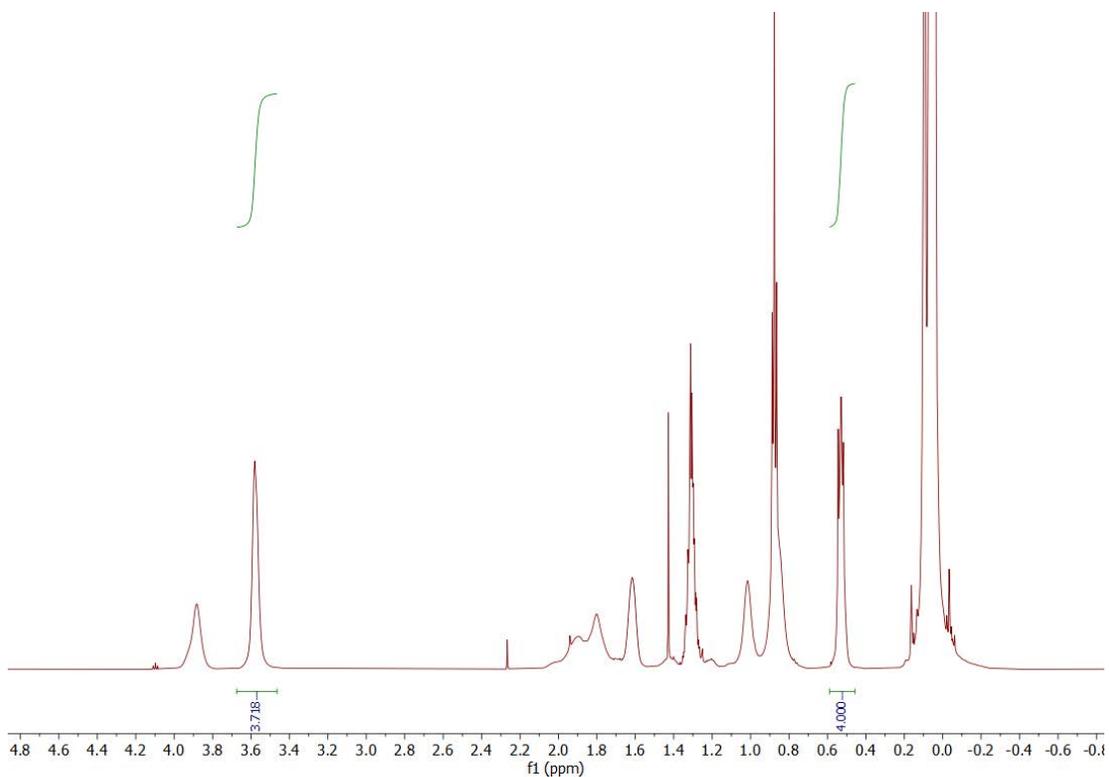

**Fig. S38.** $^1$H NMR of (MMA$_{1.24}$-$r$-PDMS$^1$)$_{198}$. The number of PDMS side chains is 198, the spacer ratio is 3.718/3=1.24, and the number of MMA monomers is 198×1.24=246.

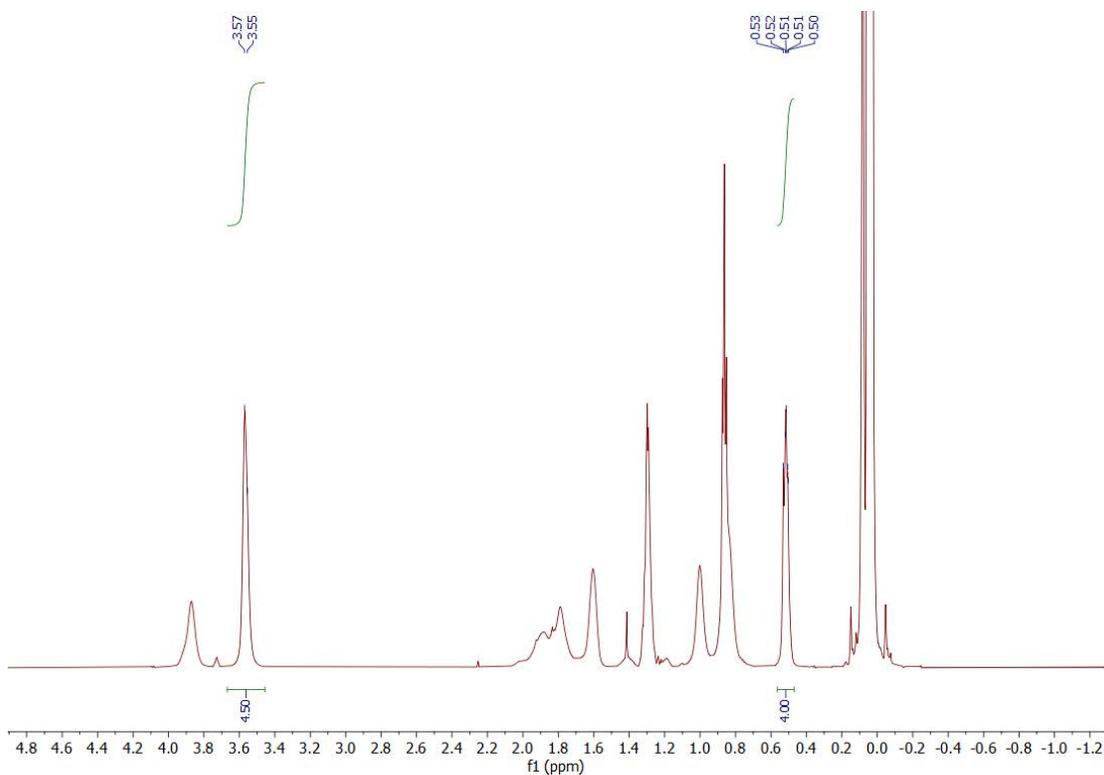

**Fig. S39.** $^1$H NMR of (MMA$_{1.50}$-*r*-PDMS$^1$)$_{200}$. The number of PDMS side chains is 200, the spacer ratio is 4.50/3=1.50, and the number of MMA monomers is 200×1.50=300.

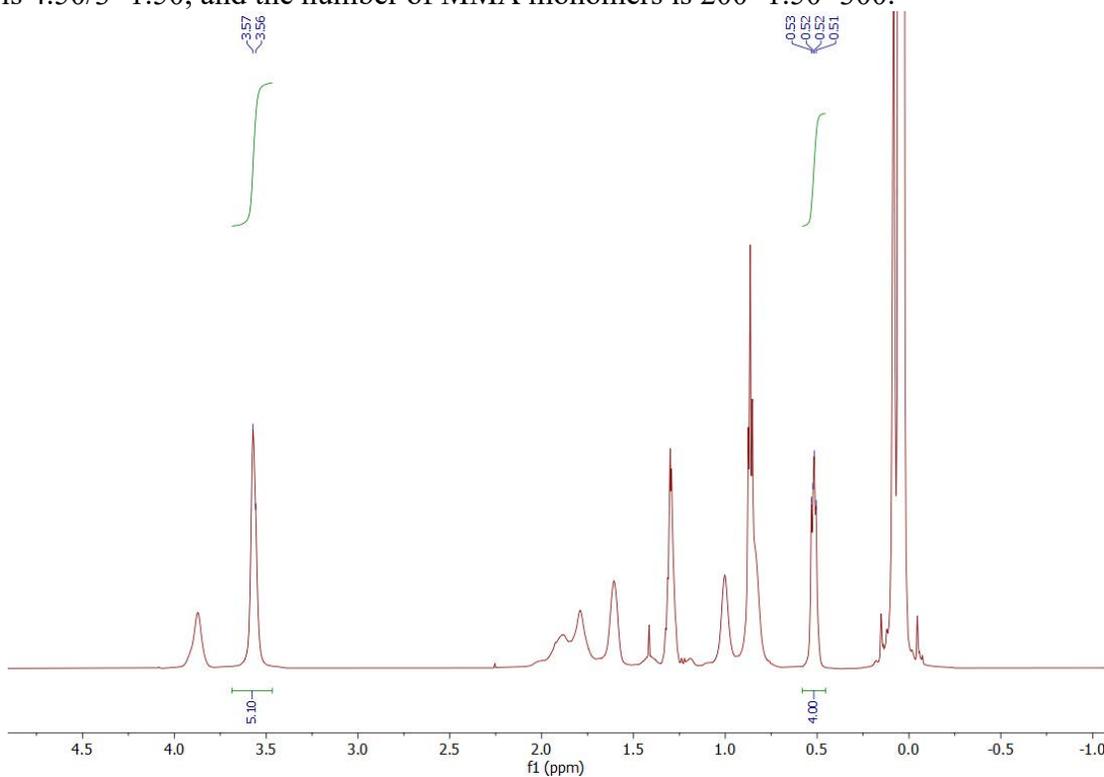

**Fig. S40.** $^1$H NMR of (MMA$_{1.70}$-*r*-PDMS$^1$)$_{200}$. The number of PDMS side chains is 200, the spacer ratio is 5.10/3=1.70, and the number of MMA monomers is 200×1.70=340.

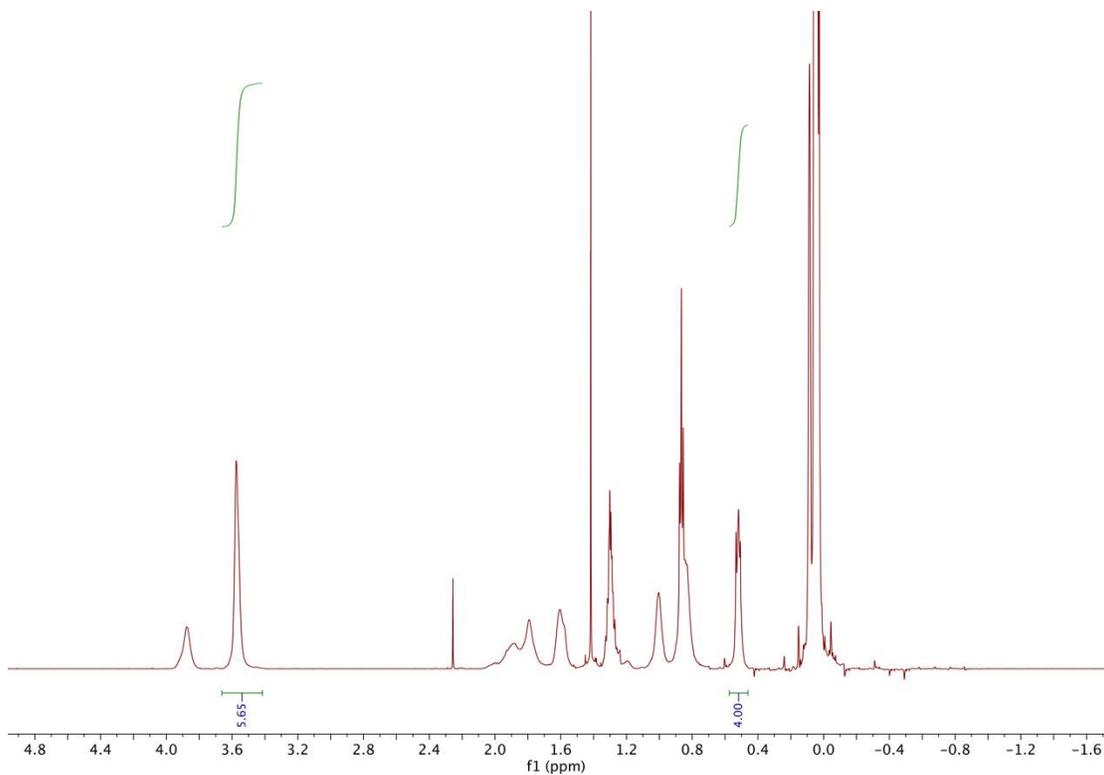

**Fig. S41.** $^1$H NMR of $(MMA_{1.88}\text{-}r\text{-}PDMS^1)_{200}$. The number of PDMS side chains is 200, the spacer ratio is 5.65/3=1.88, and the number of MMA monomers is 200×1.88=376.

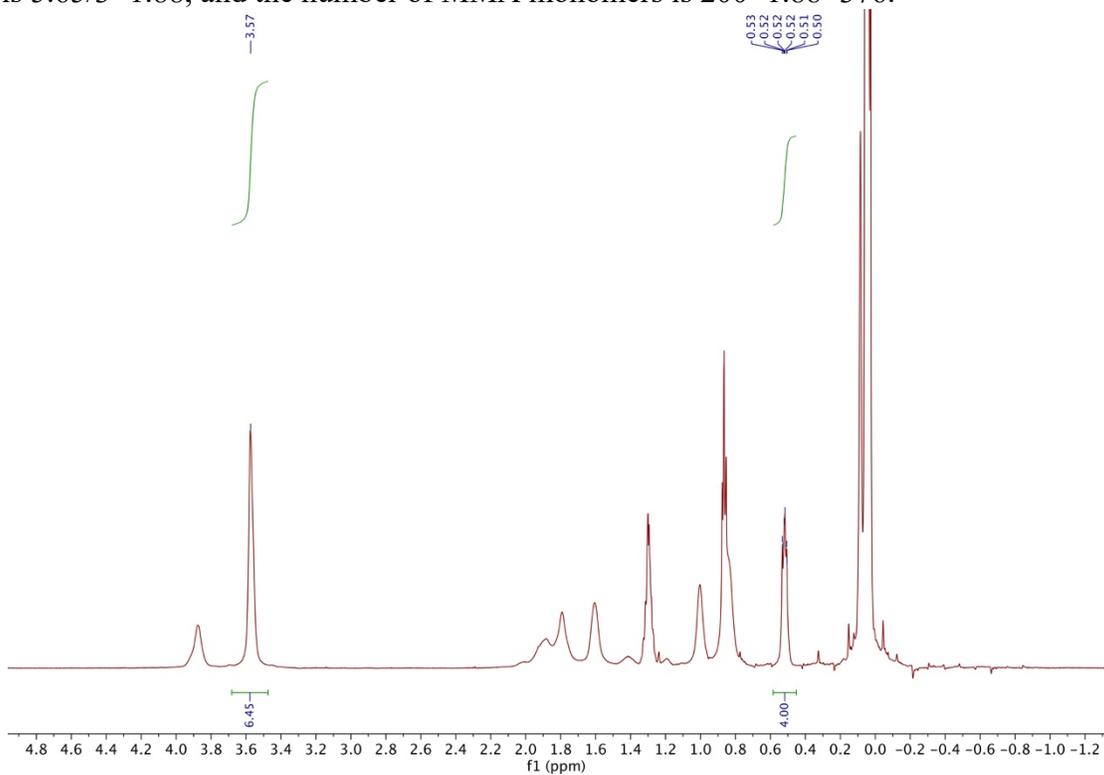

**Fig. S42.** $^1$H NMR of $(MMA_{2.15}\text{-}r\text{-}PDMS^1)_{200}$. The number of PDMS side chains is 200, the spacer ratio is 6.45/3=2.15, and the number of MMA monomers is 200×2.15=430.

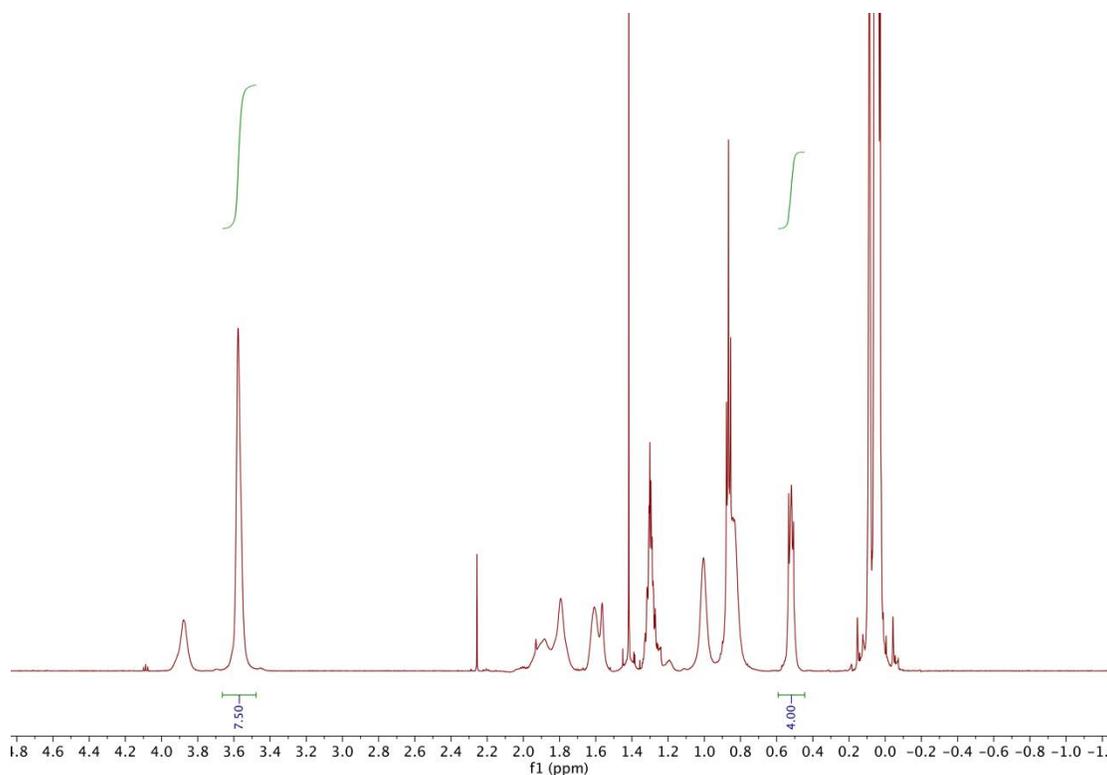

**Fig. S43.** $^1$H NMR of (MMA$_{2.50}$-$r$-PDMS$^1$)$_{200}$. The number of PDMS side chains is 200, the spacer ratio is 7.50/3=2.50, and the number of MMA monomers is 200×2.50=500.

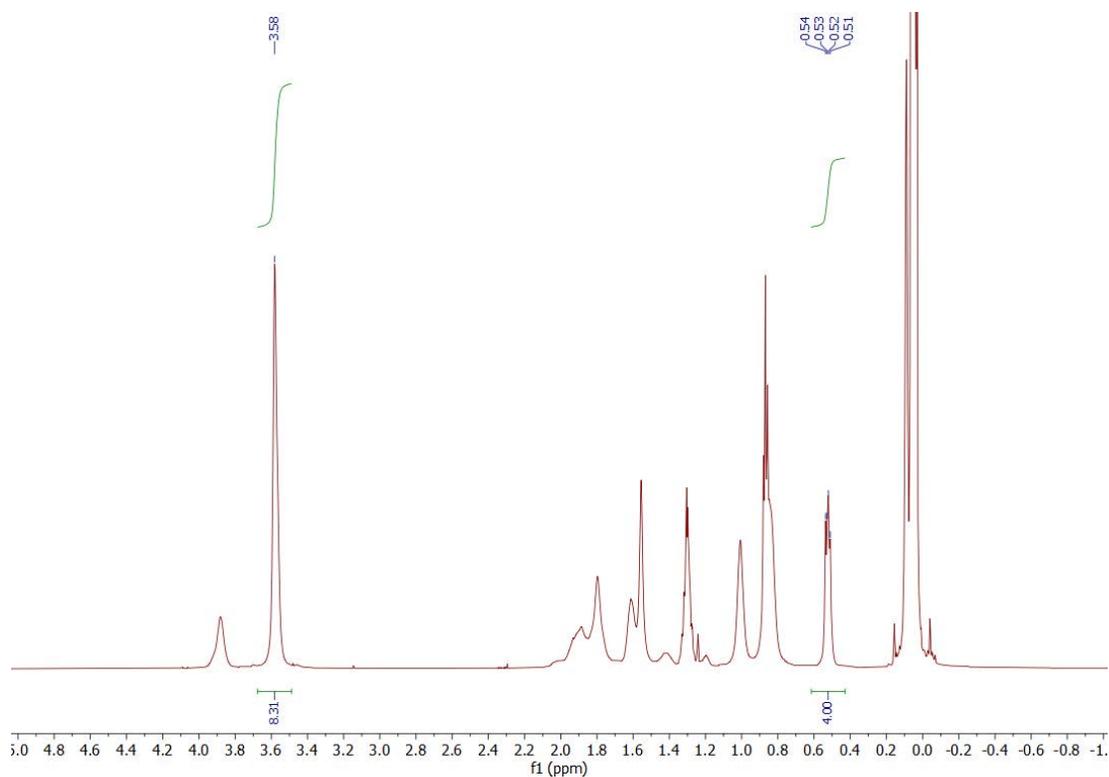

**Fig. S44.** $^1$H NMR of (MMA$_{2.77}$-$r$-PDMS$^1$)$_{200}$. The number of PDMS side chains is 200, the spacer ratio is 8.31/3=2.77, and the number of MMA monomers is 200×2.77=554.

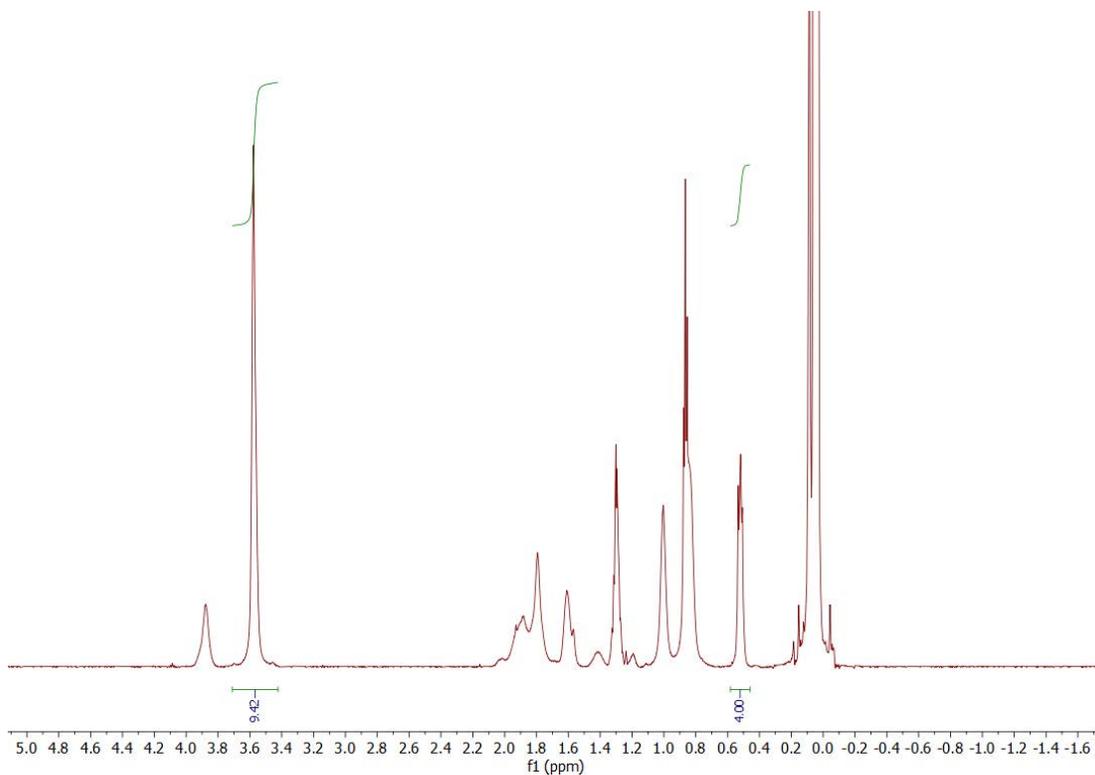

**Fig. S45.** $^1$H NMR of (MMA$_{3.14}$-$r$-PDMS$^1$)$_{200}$. The number of PDMS side chains is 200, the spacer ratio is 9.42/3=3.14, and the number of MMA monomers is 200×3.14=628.

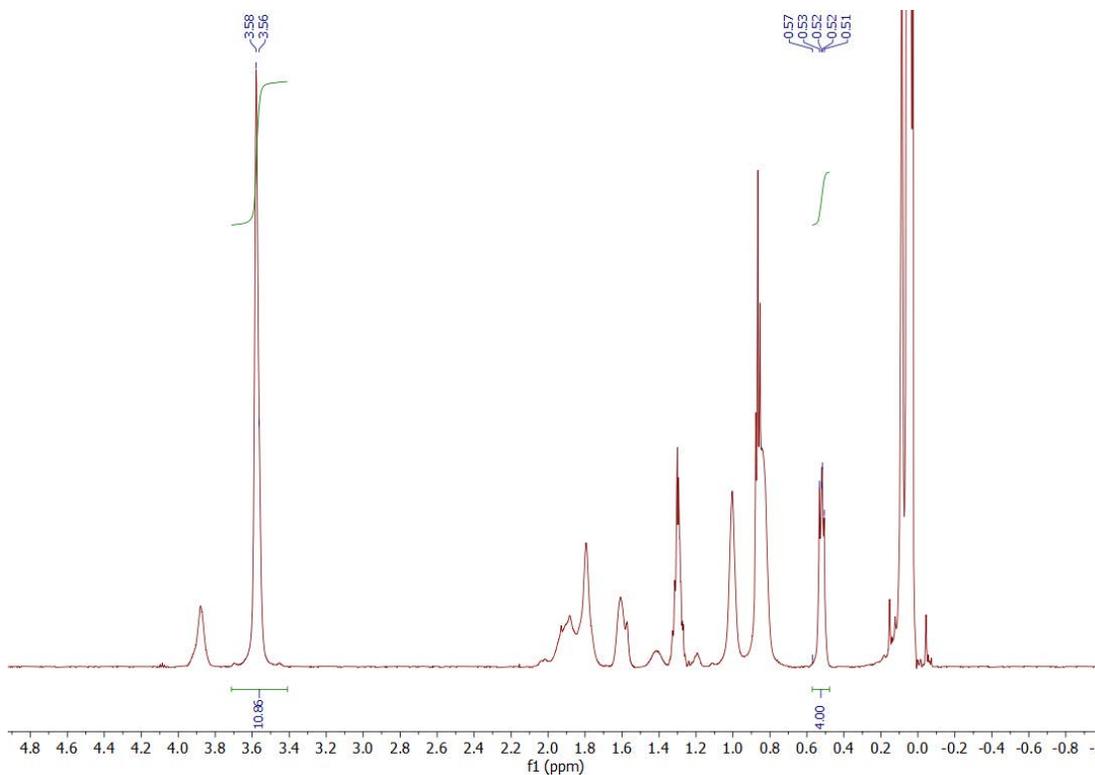

**Fig. S46.** $^1$H NMR of (MMA$_{3.62}$-$r$-PDMS$^1$)$_{192}$. The number of PDMS side chains is 192, the spacer ratio is 10.86/3=3.62, and the number of MMA monomers is 192×3.62=695.

2.2. Foldable bottlebrush polymer networks: [~200, ~60, 0-3.62]

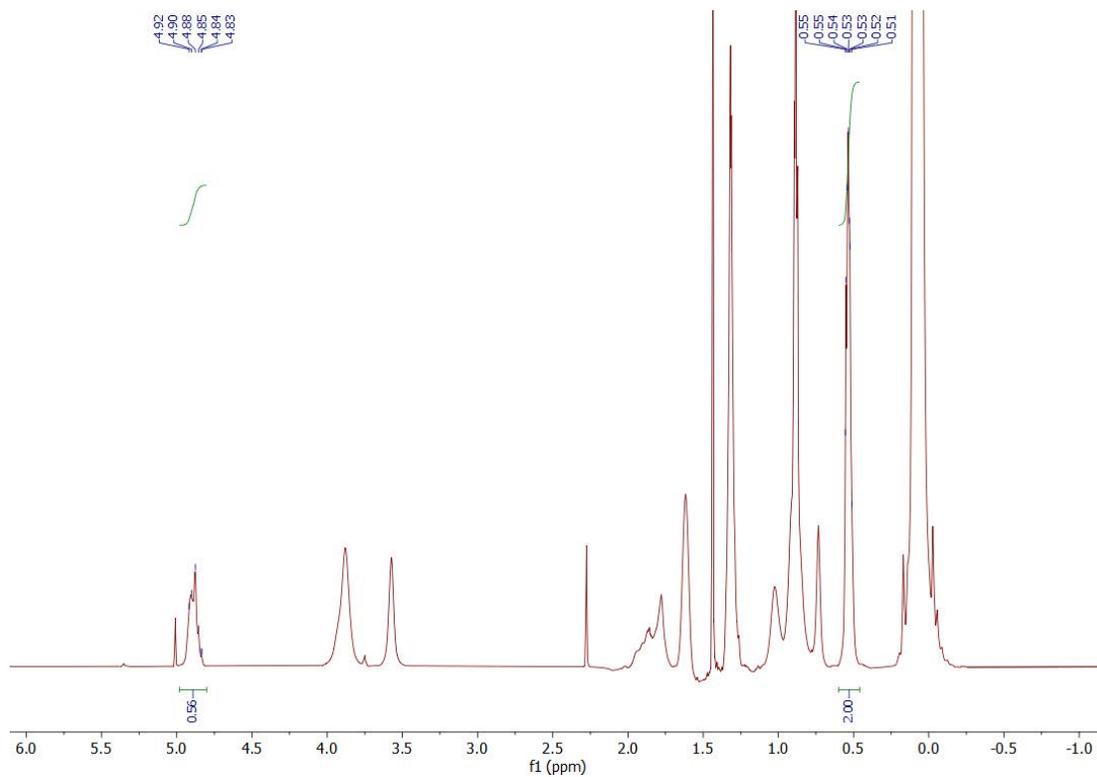

**Fig. S47.** $^1$H NMR of BnMA$_{56}$-$b$-(MMA$_{0.30}$-$r$-PDMS$^1$)$_{200}$-$b$-BnMA$_{56}$. The number of total BnMA monomers is 200×0.56=112. The number of BnMA monomers on each end block is 112/2=56.

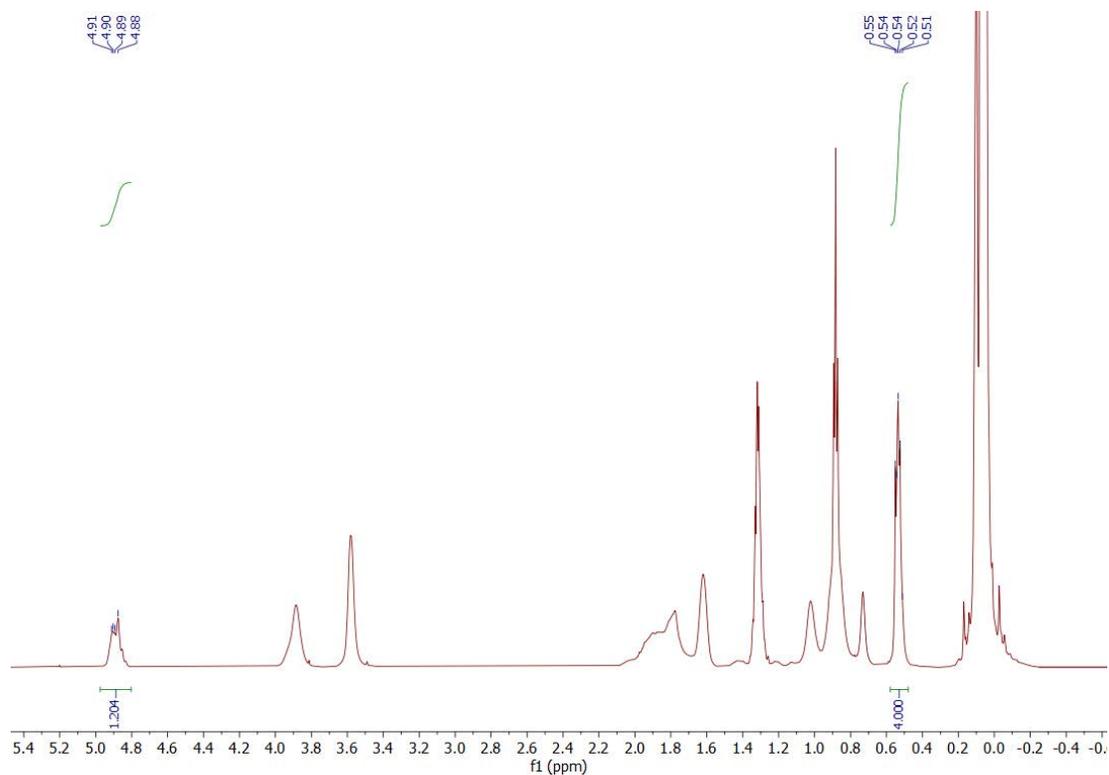

**Fig. S48.** $^1$H NMR of BnMA$_{60}$-*b*-(MMA$_{0.80}$-*r*-PDMS$^1$)$_{200}$-*b*-BnMA$_{60}$. The number of total BnMA monomers is 200×1.204/2=120. The number of BnMA monomers on each end block is 120/2=60.

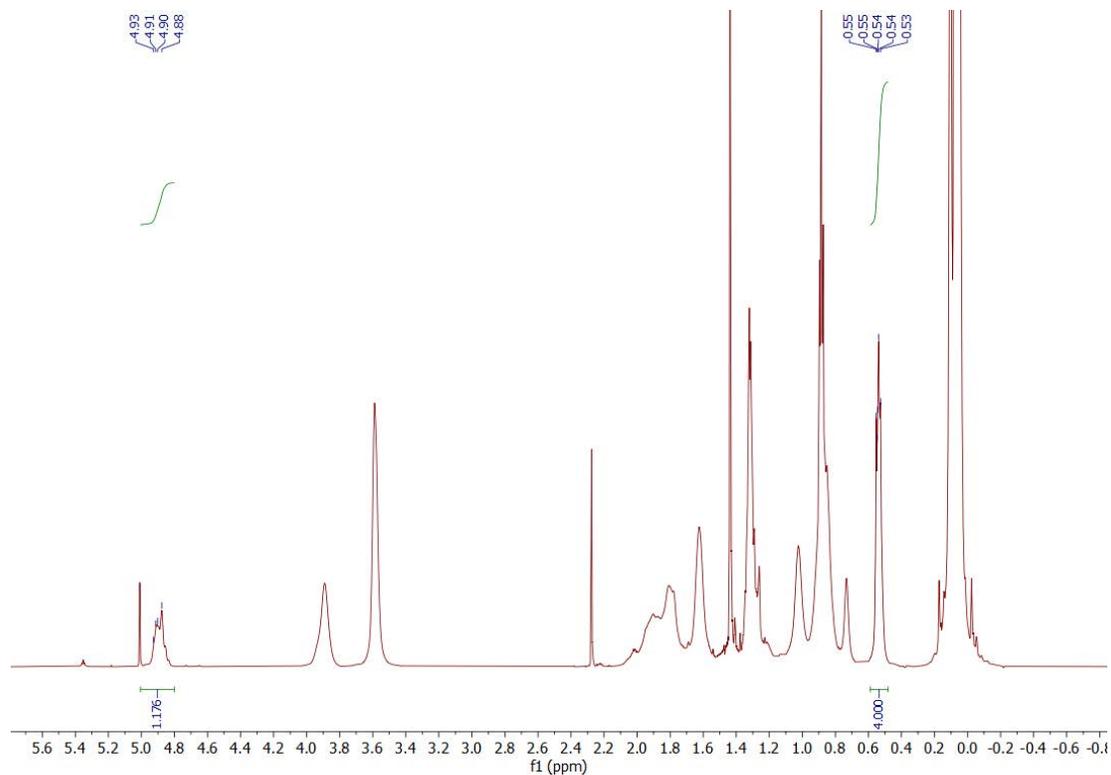

**Fig. S49.** $^1$H NMR of BnMA$_{58}$-*b*-(MMA$_{1.24}$-*r*-PDMS$^1$)$_{198}$-*b*-BnMA$_{58}$. The number of total BnMA monomers is 198×1.176/2=116. The number of BnMA monomers on each end block is 116/2=58.

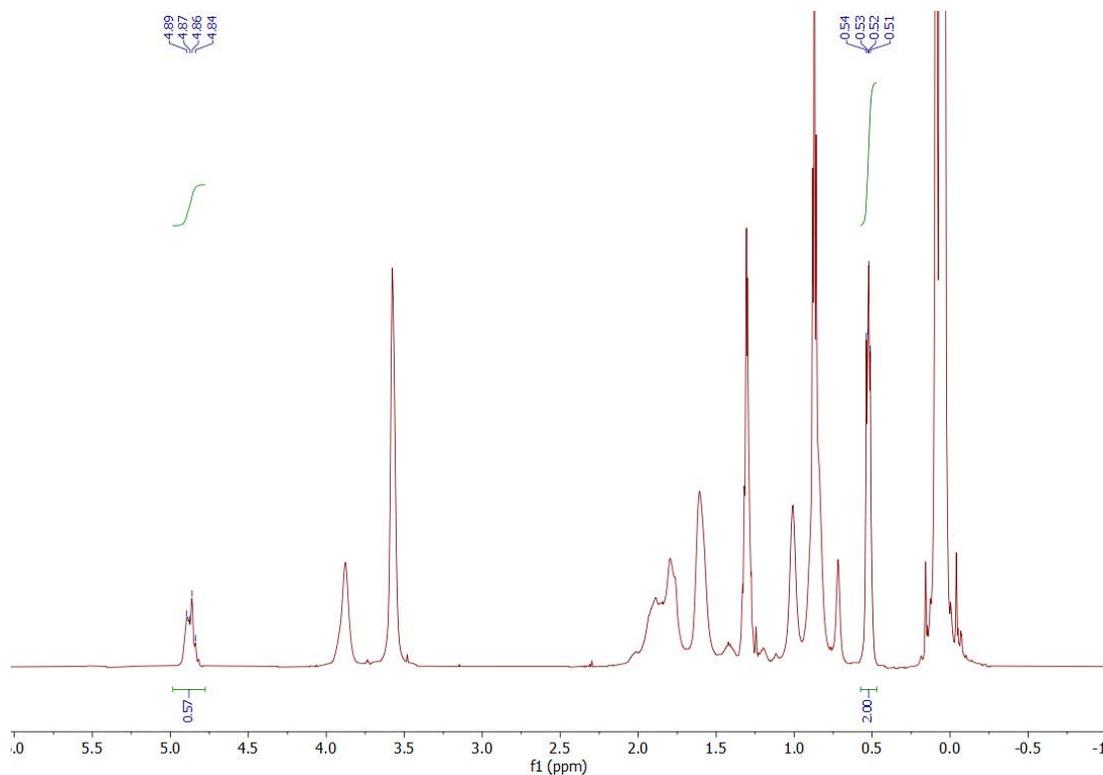

**Fig. S50.** $^1$H NMR of BnMA$_{57}$-b-(MMA$_{1.50}$-r-PDMS$^1$)$_{200}$-b-BnMA$_{57}$. The number of total BnMA monomers is 200×0.57=114. The number of BnMA monomers on each end block is 114/2=57.

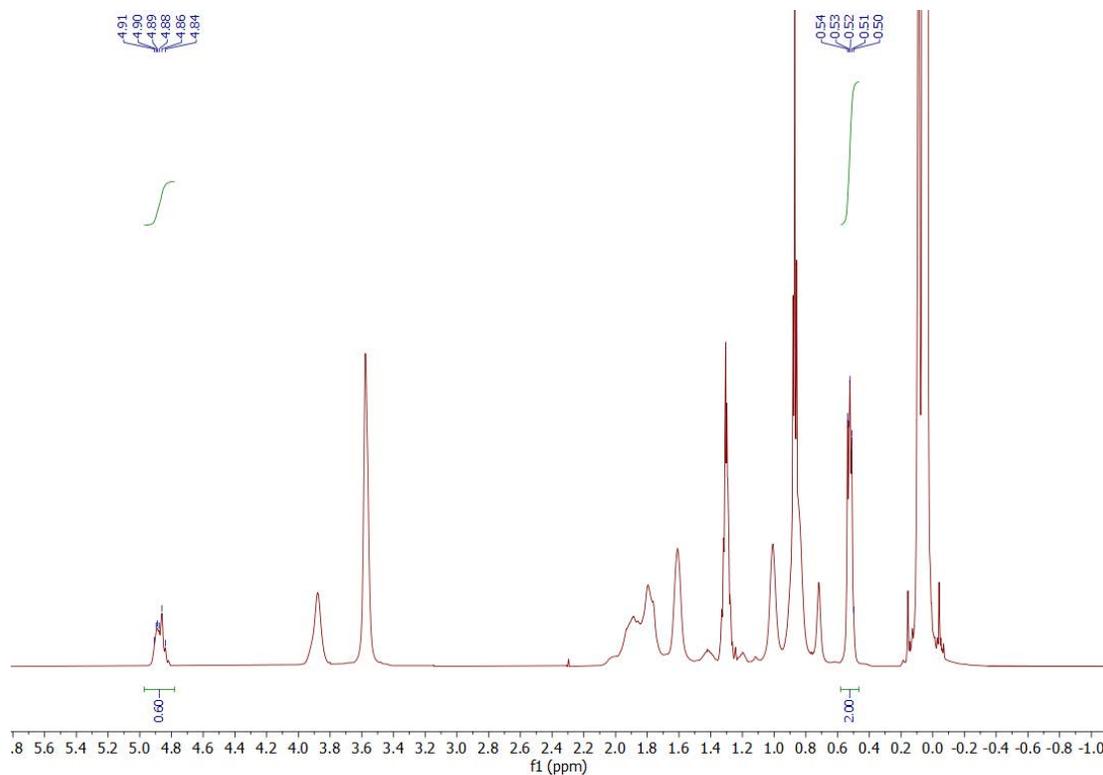

**Fig. S51.** $^1$H NMR of BnMA$_{60}$-b-(MMA$_{1.70}$-r-PDMS$^1$)$_{200}$-b-BnMA$_{60}$. The number of total BnMA monomers is 200×0.60=120. The number of BnMA monomers on each end block is 120/2=60.

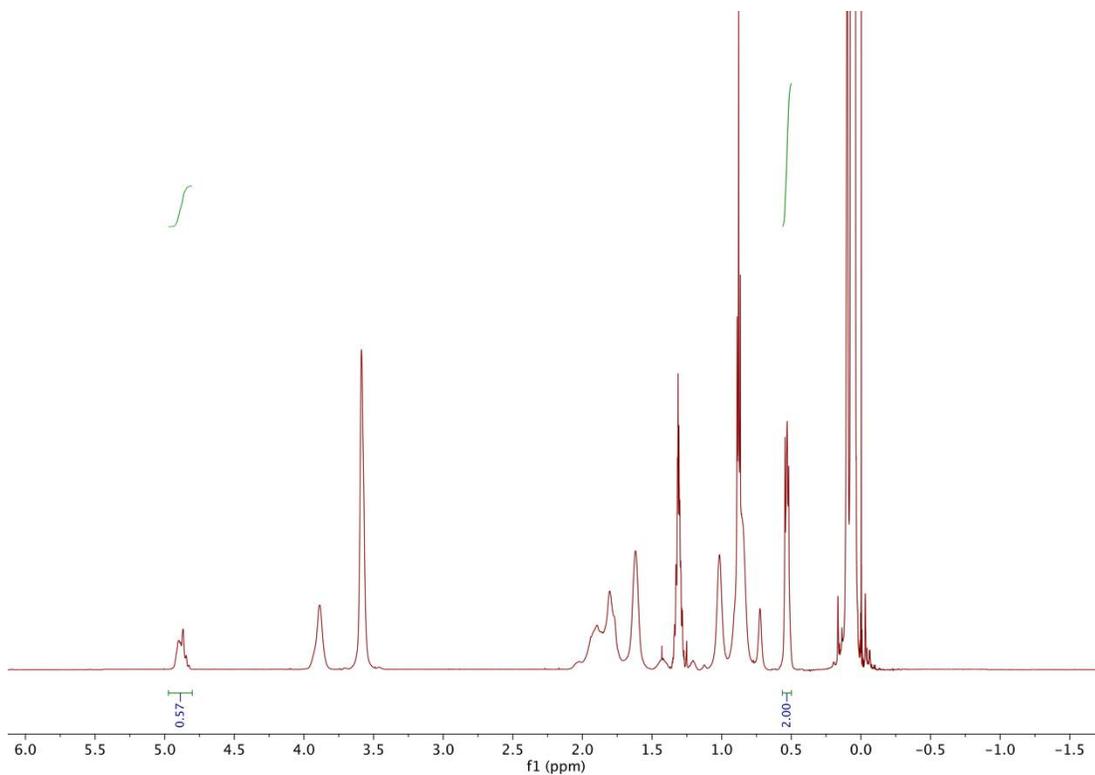

**Fig. S52.** $^1$H NMR of BnMA$_{57}$-b-(MMA$_{1.88}$-r-PDMS$^1$)$_{200}$-b-BnMA$_{57}$. The number of total BnMA monomers is 200×0.57=114. The number of BnMA monomers on each end block is 114/2=57.

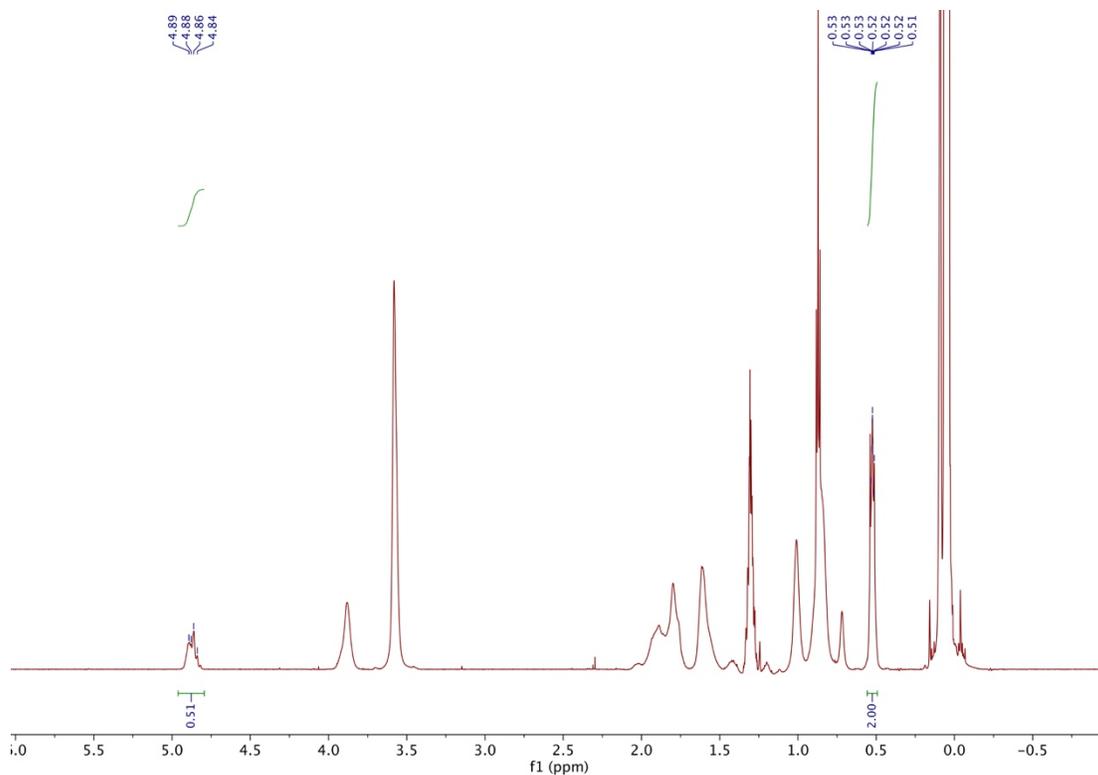

**Fig. S53.** $^1$H NMR of BnMA$_{51}$-b-(MMA$_{2.15}$-r-PDMS$^1$)$_{200}$-b-BnMA$_{51}$. The number of total BnMA monomers is 200×0.51=102. The number of BnMA monomers on each end block is 102/2=51.

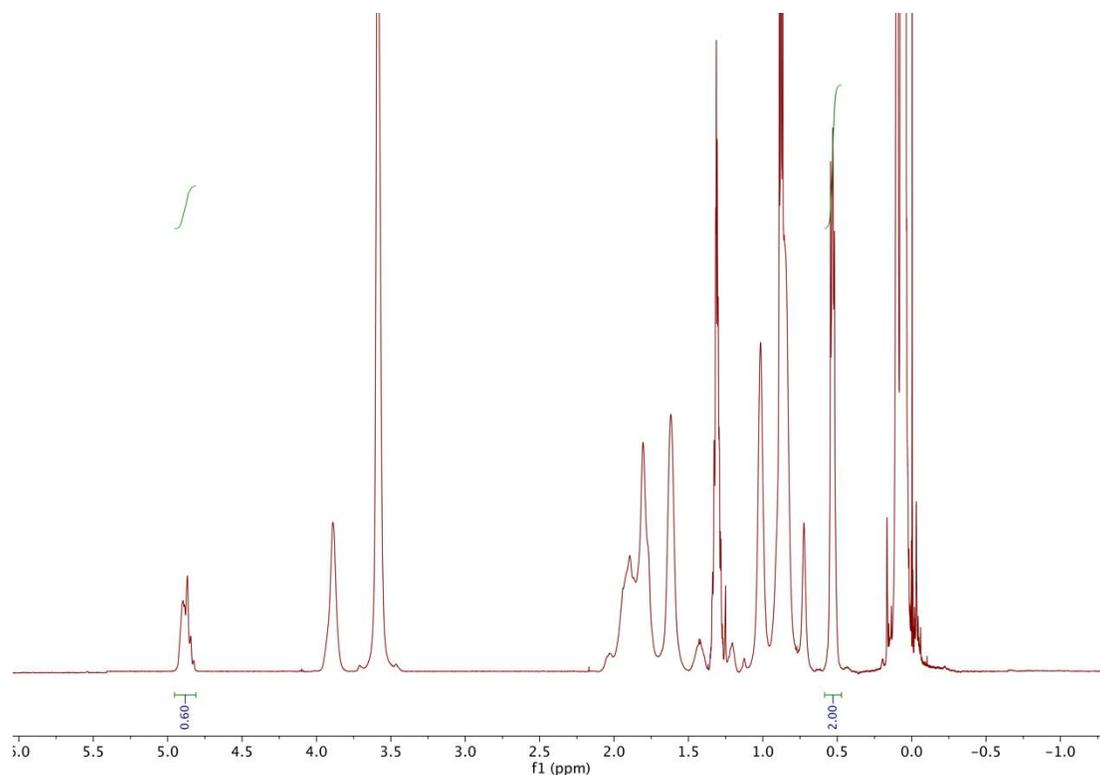

**Fig. S54.** $^1$H NMR of BnMA$_{60}$-$b$-(MMA$_{2.50}$-$r$-PDMS$^1$)$_{200}$-$b$-BnMA$_{60}$. The number of total BnMA monomers is 200×0.60=120. The number of BnMA monomers on each end block is 120/2=60.

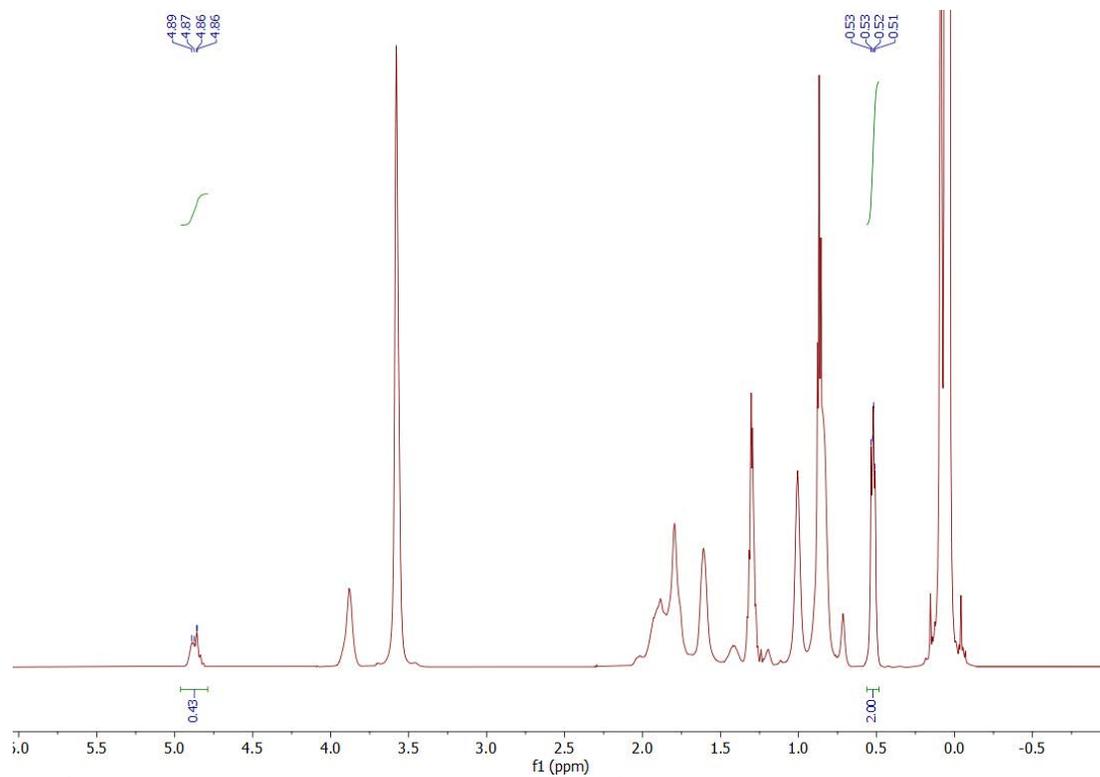

**Fig. S55.** $^1$H NMR of BnMA$_{43}$-$b$-(MMA$_{2.77}$-$r$-PDMS$^1$)$_{200}$-$b$-BnMA$_{43}$. The number of total BnMA monomers is 200×0.43=86. The number of BnMA monomers on each end block is 86/2=43.

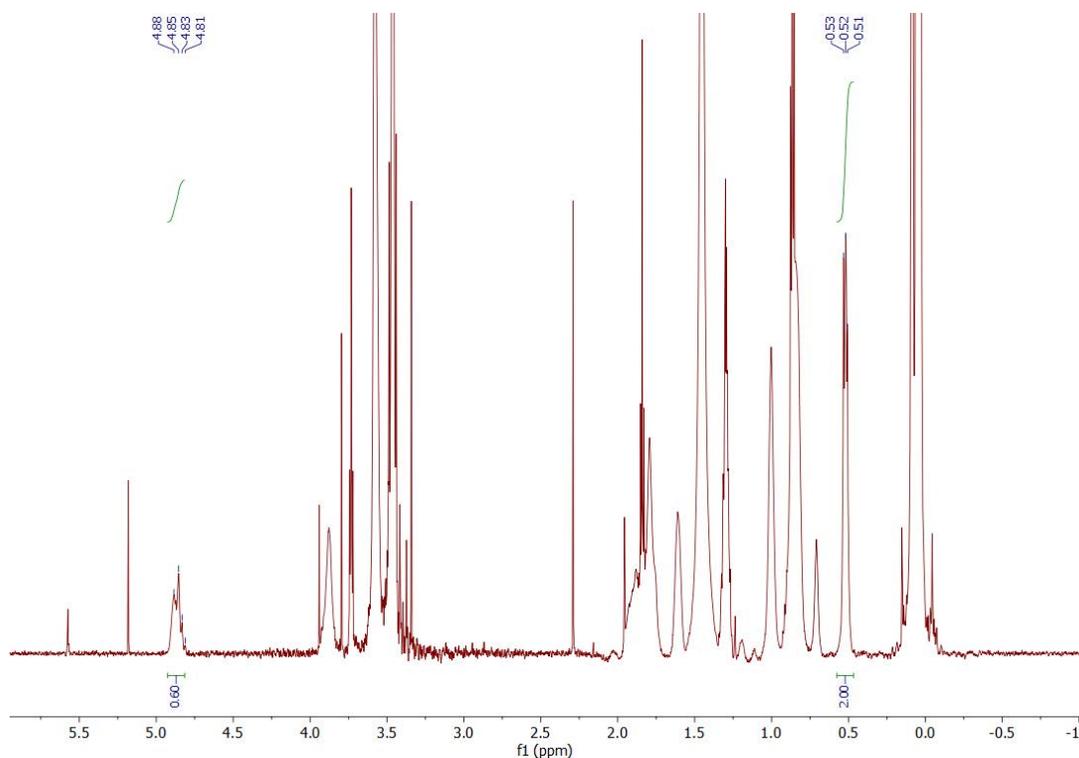

**Fig. S56.** $^1$H NMR of BnMA$_{60}$-$b$-(MMA$_{3.14}$-$r$-PDMS$^1$)$_{200}$-$b$-BnMA$_{60}$. The number of total BnMA monomers is 200×0.60=120. The number of BnMA monomers on each end block is 120/2=60.

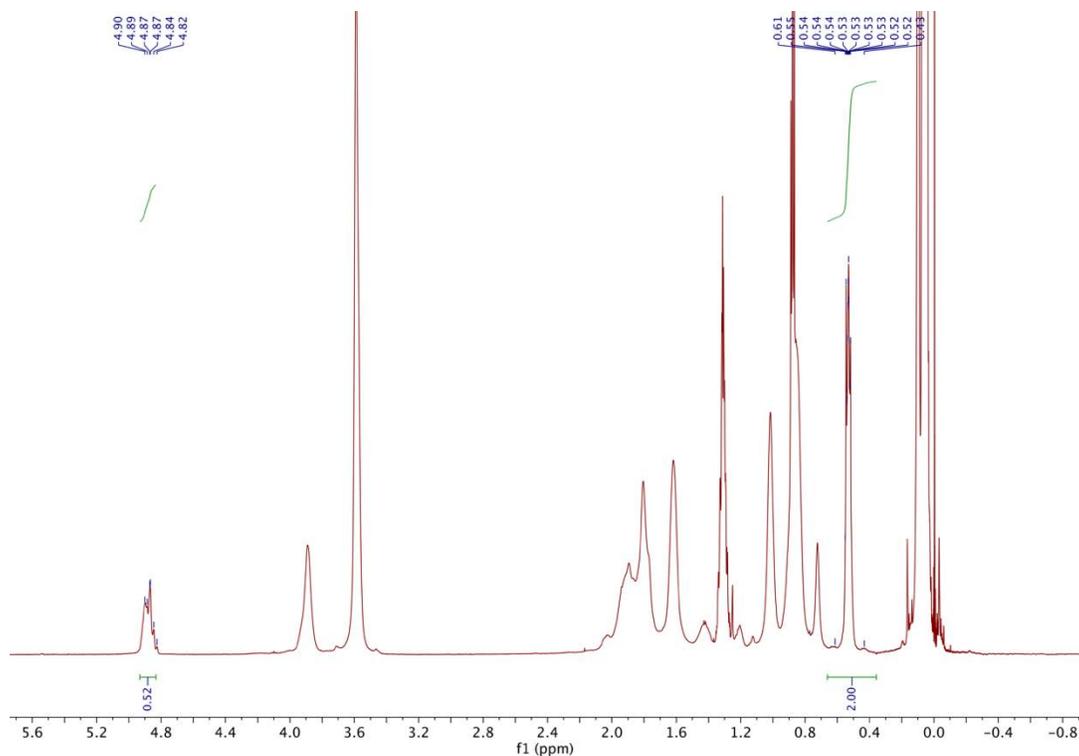

**Fig. S57.** $^1$H NMR of BnMA$_{50}$-$b$-(MMA$_{3.62}$-$r$-PDMS$^1$)$_{192}$-$b$-BnMA$_{50}$. The number of total BnMA monomers is 192×0.52=110. The number of BnMA monomers on each end block is 110/2=50.